\documentclass[12 pt]{article}
\usepackage{amsfonts,lscape,inputenc,hyperref}
\usepackage{amsmath,subfigure,ragged2e}
\usepackage{amssymb}
\usepackage{graphicx}
\usepackage[onehalfspacing]{setspace}
\usepackage[top=1.0in, bottom=1.0in, left=1in, right=1in]{geometry}

\providecommand{\U}[1]{\protect\rule{.1in}{.1in}}

\begin{document}

\title{{\Huge Misspecification and Weak Identification in Asset Pricing}}
\author{Frank Kleibergen\thanks{%
Email: f.r.kleibergen@uva.nl. Amsterdam School of Economics, University of
Amsterdam, Roetersstraat 11, 1018 WB Amsterdam, The Netherlands. }\ \ \ \ \
\ \ \ \ Zhaoguo Zhan\thanks{%
Email: zzhan@kennesaw.edu. Department of Economics, Finance and Quantitative
Analysis, Coles College of Business, Kennesaw State University, GA 30144,
USA.}}
\date{\today}
\maketitle

\begin{abstract}
The widespread co-existence of misspecification and weak identification in
asset pricing has led to an overstated performance of risk factors. Because
the conventional Fama and MacBeth (1973) methodology is jeopardized by
misspecification and weak identification, we infer risk premia by using a
double robust Lagrange multiplier test that remains reliable in the presence
of these two empirically relevant issues. Moreover, we show how the
identification, and the resulting appropriate interpretation, of the risk
premia is governed by the relative magnitudes of the misspecification $J$%
-statistic and the identification $IS$-statistic. We revisit several
prominent empirical applications and all specifications with one to six
factors from the factor zoo of Feng, Giglio, and Xiu (2020) to emphasize the
widespread occurrence of misspecification and weak identification.

\bigskip

\bigskip

\noindent \textbf{JEL Classification}: G12

\noindent \textbf{Keywords:} misspecification, weak identification, risk
factor, asset pricing
\end{abstract}

\doublespace

\newpage

\doublespace

\section{Introduction}

Over the past decades, hundreds of risk factors have been proposed to
explain the cross-section of asset returns, making up the, so-called, zoo of
factors; see, e.g., Cochrane (2011) and Harvey, Liu, and Zhu (2016).\nocite%
{harvey2016and} \nocite{cochrane2011presidential} Much of the empirical
support for these factors is based on the beta representation, where risk
premia are identified by projecting expected returns on the betas, and the
betas are the factor loadings in the time-series regression of asset returns
on risk factors; see, e.g., Fama and MacBeth (FM, 1973).\nocite{fm73} The
resulting FM $t$-statistic on risk premia, \nocite{han82} Hansen (1982)'s $J$%
-statistic for misspecification and the cross-sectional $R^{2}$, are
popularly used in the asset pricing literature to show support for proposed
risk factors. The reliability of these conventional statistics has, however,
recently been brought into question by two empirically relevant issues:
misspecification and weak identification.

Up till now, misspecification has been widely acknowledged as an inherent
feature of asset pricing models; see, e.g., Gospodinov, Kan, and Robotti
(2014). \nocite{gkr14} For the beta representation, misspecification results
in expected returns that are not fully explained by the betas of the
specified factors. Thus, pricing errors generally exist and should be taken
into account when conducting inference on risk premia. Kan, Robotti, and
Shanken (KRS, 2013) \nocite{krs13} therefore propose the KRS $t$-test, which
explicitly allows for the existence of misspecification while the FM $t$%
-test does not. As warned by Kan, Robotti, and Shanken (2013), failure to
account for misspecification tends to enlarge the $t$-statistic on risk
premia, leading to overstated pricing performance of risk factors.

Weak (or no) identification, on the other hand, is driven by poor quality
risk factors or more generally, limited information contained in the data.
Kan and Zhang (1999)\nocite{kz99}, for example, warn that risk factors
proposed in the literature could be useless factors with zero betas. When
betas are zero, risk premia are unidentified in the beta representation and
paired with misspecification, the FM $t$-statistic can be spuriously large
to support useless factors as shown by Kan and Zhang (1999). In addition,
Kleibergen (2009) \nocite{kf09} illustrates the malfunction of the FM $t$%
-statistic caused by factors that are only weakly correlated with asset
returns and proposes robust tests to address weak identification of risk
premia. Moreover, Kleibergen and Zhan (2015, 2021) \nocite{kz15} show that
the cross-sectional $R^{2}$ and $J$-statistic may also spuriously favor
statistically weak factors, respectively.

Despite that misspecification and weak identification are well recognized,
there is currently no unified approach for identification and inference that
can deal with these empirically relevant issues. We therefore advocate such
an approach and apply it to show the widespread occurrence of
misspecification and weak identification in applied asset pricing studies.

When misspecification is present, there is no longer a (true) value of the
risk premia at which the pricing error is zero. Different estimation
procedures then lead to distinct pseudo-true values of risk premia, which
are the minimizers of the population objective function of the respective
estimation procedure. We show that, in case of generic misspecification, the
difference between the pseudo-true value of two estimation procedures, i.e.
the FM two-pass approach and the continuous updating estimator (CUE)\ of
Hansen, Heaton, and Yaron (1996),\nocite{hhy96} and the baseline risk premia
that would apply in case of correct specification, depends on the strength
of misspecification compared to the strength of identification. The strength
of misspecification is reflected by the population equivalent of the $J$%
-statistic, while the strength of identification is revealed by the,
so-called, identification strength ($IS$) statistic. The $IS$-statistic is a
rank test statistic on the $\beta $ matrix and is, by construction, always
larger than or equal to the $J$-statistic. The difference between these two
measures is indicative of whether we can interpret the pseudo-true value as
a risk premium. When this difference is small, the pseudo-true value
basically results from the close to reduced rank value of the $\beta $
matrix, so it can be far off from the baseline risk premium under correct
specification, and does not reflect a risk premium. The sample equivalents
of the $J$ and $IS$ statistics are therefore important in gauging whether we
can interpret risk premia estimates resulting from empirical studies as
genuine risk premia of interest.

For eight prominent empirical studies in asset pricing: Fama and French
(1993), Jagannathan and Wang (1996), Yogo (2006), Lettau and Ludvigson
(2001), Savov (2011), Adrian, Etula, and Muir (2014), Kroencke (2017) and
He, Kelly, and Manela (2017) and for all (14 billion+) one to six factor
specifications resulting from the factor zoo of Feng, Giglio, and Xiu
(2020), we compute the $J$ and $IS$ statistics to illustrate the widespread
co-existence of misspecification and weak identification. The resulting $J$%
-statistics are mostly large and significant, and if they are not, their
smaller values are often induced by an accompanying small value of the $IS$%
-statistic. The smallish $J$-statistic then basically results from a close
to reduced rank value of the $\beta $ matrix, so the risk premia estimates
do so as well and do not reflect risk premia. When we do not incorporate the
zero-$\beta $ return or use more time series observations, these
identification issues become less problematic for some settings, such as the
three-factor model from Fama and French (1993), but remain for many others.

The conventional test statistics for conducting inference on risk premia,
such as the FM two-pass $t$-test, also with the Shanken (1992)\nocite{sh92}
correction of the standard errors, and the KRS $t$-test, become unreliable
in the presence of misspecification and/or weak identification. We
illustrate this by showing that the limit behavior of the FM\ two-pass
estimator consists of four components. Two of these components are
negligible in case of strong factors and correct specification, but lead to
non-standard behavior in case of misspecification and/or weak
identification. It is also not possible to correct for them by using the
bootstrap. We therefore use the double robust Lagrange multiplier (DRLM)
statistic from Kleibergen and Zhan (2021) that provenly remains reliable in
case of misspecification and/or weak identification. For ease of exposition,
we conduct a small simulation experiment to illustrate the pros and cons of
these different statistics for testing hypotheses on risk premia.

We illustrate the practical usage of the DRLM test by using the influential
Fama and French (1993) three-factor model \nocite{ff93} and the conditional
consumption capital asset pricing model in Lettau and Ludvigson (2001). 
\nocite{letlud01} For the Fama and French (1993) three-factor model, we show
that for commonly used data sets, such as those resulting from Lettau and
Ludvigson (2001) and Lettau, Ludvigson, and Ma (2019), identification of the
risk premia can be problematic when incorporating the zero-$\beta $ return
because of the near constancy of the $\beta $'s associated with the market
return. This identification issue is alleviated when using more time series
observations or by removing the zero-$\beta $ return. The joint 95\%
confidence set for the risk premia resulting from the DRLM\ test is then
bounded (ellipsoid), while it is unbounded in the direction of the risk
premia on the market return when using fewer observations and also
incorporating the zero-$\beta $ return. For the conditional capital asset
pricing model stemming from Lettau and Ludvigson (2001), the 95\% confidence
sets of the risk premia resulting from the DRLM\ test are all unbounded,
indicating limited information in the data for precisely identifying the
risk premia. The 95\% confidence sets from the DRLM\ test are, for all of
these settings, fully in line with the $J$ and $IS$ statistics. When the $J$
and $IS$ statistics are relatively close, we obtain unbounded 95\%
confidence sets from the DRLM test, in contrast with the bounded ones
resulting from the FM two-pass and KRS $t$-tests which are then unreliable.
Also the CUE differs considerably from the FM two-pass estimator in this
scenario. When the $J$ and $IS$ statistics are substantially apart, the
results from all these procedures are, as expected, rather similar, but the
95\% (projected) confidence sets from the DRLM\ test are notably narrower
than those from the FM two-pass and KRS $t$-tests. It all shows the
importance of jointly using the $J$ and $IS$ statistics to gauge the
identification of risk premia while using the DRLM\ test to conduct
inference on risk premia.

Overall, our study adds to an emerging body of research that aims to bring
discipline to the zoo of factors. Harvey, Liu, and Zhu (2016), for example,
propose a higher hurdle such as a $t$-statistic greater than 3.0 instead of
the commonly used 1.96, since they are concerned that significant $t$%
-statistics documented in the existing literature could result from data
mining. Unlike Harvey, Liu, and Zhu (2016) and the follow-up work, we do not
provide new hurdles of $t$-statistics; instead, we suggest the DRLM, $J$ and 
$IS$ statistics for gauging the quality of risk factors. Feng, Giglio, and
Xiu (2020) \nocite{feng2020taming} focus on whether a proposed risk factor
adds explanatory power beyond the existing zoo of factors, so their
methodology builds on a high-dimensional set of existing factors. In
contrast, our approach is suitable for evaluating the explanatory power of
each factor model individually. In order to jointly address misspecification
and no identification, Gospodinov, Kan, and Robotti (2014) propose a model
selection procedure to eliminate potentially useless factors. Unlike
Gospodinov, Kan, and Robotti (2014), our proposed DRLM test aims to infer
risk premia regardless of whether misspecification and weak identification
are present.

Finally, we note that misspecification and weak identification are not
limited to the beta representation; see, e.g., Stock and Wright (2000) and
Hansen and Lee (2021), \nocite{sw00} \nocite{hl21} who have studied weak
identification and misspecification in the generalized method of moments
framework, respectively. Given that asset pricing models are at best
approximations of reality while lots of risk factors have little explanatory
power for asset returns, misspecification and weak identification are a
likely common threat to empirical asset pricing studies.

The rest of the paper is organized as follows. Section \ref{prelim} starts
with discussing the setup of misspecification and the consequences it has
for commonly used risk premia estimators. It illustrates that the commonly
used FM $t$-test is jeopardized by both misspecification and weak
identification, while the DRLM test takes both issues into account. Section %
\ref{j_is} uses the $J$ and $IS$ statistics to highlight the prevalence of
misspecification and weak identification in existing studies. Section \ref%
{section_app} contains the empirical applications for conducting inference
on risk premia, while Section \ref{conclusion} concludes. Technical details
are relegated to the Appendix.

\section{Misspecification in the beta representation}

\label{prelim}

Let $R_{i,t}$ be the return on the $i$-th asset at time $t,$ with $%
i=1,...,N, $ and $t=1,...,T.$ The beta representation of expected returns
models it as linear in the beta vector of factor loadings: 
\begin{equation}
E(R_{i,t})=\beta _{i}^{\prime }\lambda _{F},  \label{beta_representation1}
\end{equation}%
where $\lambda _{F}$ is the $K\times 1$ vector of risk premia, and $\beta
_{i}$ is the $K\times 1$ vector of factor loadings: 
\begin{equation}
\beta _{i}=var(F_{t})^{-1}cov(F_{t},R_{i,t}),
\end{equation}%
with $F_{t}$ the $K\times 1$ vector of the specified risk factors, and $%
K<N+1 $. We can as well represent the beta representation jointly for all
assets by stacking the $N$ equations of (\ref{beta_representation1}) to get 
\begin{equation}
\mu _{R}=E(R_{t})=\beta \lambda _{F},  \label{matrixvector_representation2}
\end{equation}%
with $R_{t}=(R_{1,t},...,R_{N,t})^{\prime },$ $\beta =(\beta _{1},...,\beta
_{N})^{\prime }.$

In both (\ref{beta_representation1}) and its misspecification extension
provided later on (i.e., Equation (\ref{beta_representation2})),
identification of $\lambda _{F}$ relies on the quality of $\beta _{i}$. For
instance, if some of the specified risk factors are just useless noise with
zero betas, then $\lambda _{F}$ is unidentified in the beta representation.
This reasoning generalizes to the full rank condition of the $N\times K$%
-dimensional matrix $\beta ,$ i.e. for $\lambda _{F}$ to be identified, $%
\beta $ needs to have full rank. In the single factor case with $K=1$, this
rank condition then just requires that $\beta $ should be non-zero. If the
full rank condition of $\beta $ is jeopardized by risk factors of poor
quality, then $\lambda _{F}$ is potentially weakly identified; see, e.g.,
Kan and Zhang (1999),\nocite{kz99} Kleibergen (2009),\nocite{kf09} and Kan,
Robotti, and Shanken (2013)\nocite{krs13} for a related discussion.

\textbf{Remark 1: The zero-}$\beta ,$ \textbf{$\lambda _{0}=0,$ restriction.}
A scalar $\lambda _{0}$ is often added to (\ref{matrixvector_representation2}%
) so 
\begin{equation}
E(R_{t})=\iota _{N}\lambda _{0}+\beta \lambda _{F},
\label{beta_representation_intercept}
\end{equation}%
where $\iota _{N}$ is the $N\times 1$ vector of ones, and $\lambda _{0}$ is
the, so-called, zero-beta return, or the expected return to an asset with no
exposure to priced risks. Since our interest mainly lies in $\lambda _{F}$,
we opt to focus on (\ref{matrixvector_representation2}) instead of (\ref%
{beta_representation_intercept}) for ease of exposition. This treatment is
related to the $\lambda _{0}=0$ restriction, so that (\ref%
{beta_representation_intercept}) reduces to (\ref%
{matrixvector_representation2}). The zero restriction can be achieved by
considering $R_{t}$ as the excess return. Our discussion on (\ref%
{matrixvector_representation2}), however, can be straightforwardly extended
to incorporate $\lambda _{0}$. For example, the full rank condition of $%
\beta $ for (\ref{matrixvector_representation2}) extends to the full rank
condition of $(\iota _{N}$ $\vdots $ $\beta )$ for (\ref%
{beta_representation_intercept}) once $\lambda _{0}$ is allowed for.

\textbf{Remark 2: Incorporating the zero-}$\beta $\textbf{\ return.} If $%
\lambda _{0}=0$ is not assumed, we can still use (\ref%
{matrixvector_representation2}) to focus on $\lambda _{F}$ by considering $%
R_{t}$ as the return in deviation of a reference asset. To illustrate, let $%
\mathcal{R}_{t}=(\mathcal{R}_{1,t}\ldots \mathcal{R}_{N+1,t})^{\prime }$ be
the $(N+1)\times 1$ vector of returns such that 
\begin{equation}
E(\mathcal{R}_{t})=\iota _{N+1}\lambda _{0}+\mathcal{B}\lambda _{F},
\label{matrixvector_extra}
\end{equation}%
where $\mathcal{B}$ is the $(N+1)\times K$-dimensional matrix of factor
loadings. By subtracting the $(N+1)$-th asset return, we obtain the $N\times
1$ vector $R_{t}:$ $R_{t}=(\mathcal{R}_{1,t}\ldots \mathcal{R}%
_{N,t})^{\prime }-\iota _{N}\mathcal{R}_{N+1,t}$ such that (\ref%
{matrixvector_extra}) implies (\ref{matrixvector_representation2}), with $%
R_{t}=J_{N}\mathcal{R}_{t}$, $\beta =J_{N}\mathcal{B}$, and $J_{N}=(I_{N}$ $%
\vdots $ $-\iota _{N})$. Thus, the full rank condition of $(\iota _{N+1}$ $%
\vdots $ $\mathcal{B)}$ in (\ref{matrixvector_extra}) is equivalent to the
full rank condition of $\beta $ in (\ref{matrixvector_representation2}). Our
subsequent analysis is invariant with respect to the choice of the $(N+1)$%
-th asset; see Kleibergen and Zhan (2020).\nocite{kz20}

Remarks 1 and 2 above show that we can focus on (\ref%
{matrixvector_representation2}) to illustrate inference on $\lambda _{F}$
regardless of whether a zero-$\beta $ return is incorporated or not.%
\footnote{%
It is worth noting that whether $\lambda _{0}$ should be excluded already
sheds light on misspecification and weak identification in asset pricing. On
the one hand, if $\lambda _{0}$ is non-zero, then incorrectly imposing the
zero restriction leads to a misspecified condition. On the other hand, if $%
\lambda _{0}$ is zero, then including the redundant intercept term in (\ref%
{beta_representation_intercept}) potentially weakens the identification of
risk premia. In particular, when there is little cross-sectional variation
in $\beta $, including $\iota _{N}$ causes near-multicollinearity in $(\iota
_{N}$ $\vdots $ $\beta )$, which will further induce the weak identification
of $\lambda _{F}$. Misspecification and weak identification are, however,
more general than whether to impose $\lambda _{0}=0$ or not.}

\subsection{Misspecification}

Under misspecification, the expected returns are not fully explained by the
specified $\beta $, so $\mu _{R}\neq \beta \lambda _{F}$. We therefore
consider the correctly specified setting above as the baseline, and
introduce the pricing error $\tilde{e}$ in (\ref{beta_representation2}) to
reflect misspecification. In particular, we assume the pricing error to be
potentially of a different order of magnitude:%
\begin{equation}
\begin{array}{rl}
\mu _{R}= & \beta \lambda _{F}+\tilde{e} \\ 
\tilde{e}= & O(\tilde{e})\cdot a \\ 
\beta = & O(\beta )\cdot b,%
\end{array}
\label{beta_representation2}
\end{equation}%
where $a$ is a normalized $N$-dimensional vector, so $O(\tilde{e})$ reflects
the magnitude of misspecification; similarly, $b$ is a normalized $N\times K$%
-dimensional matrix, so $O(\beta )$ reflects the magnitude of identification.

The specification in (\ref{beta_representation2}) is without loss of
generality and, for example, allows for misspecification which is much
smaller than the expected return resulting from the beta representation, or
proportional to it which would result in case of weak identification. To
capture this, the magnitudes can be modelled to be proportional to the
number of time series observations as is common in the literature on weak
identification, for example, $O(\tilde{e})=O(T^{c_{\tilde{e}}}),$ $O(\beta
)=O(T^{c_{\beta }})$, $c_{\tilde{e}}$ and $c_{\beta }$ are finite constants;
see, e.g., Staiger and Stock (1997)\nocite{stst97} and Kleibergen (2009).%
\nocite{kf09}

The misspecification is generic, so the difference from the baseline pricing
error, $\tilde{e},$ lies both in the space spanned by $\beta $ and outside
of it. It captures the notion of the correctly specified setting as a
baseline, on top of which there is an unstructured misspecification
component. It resembles the estimation errors which the sample moment
equations add to the population moment equations. These errors are also
unstructured, and lie both in the space spanned by $\beta $ and outside of
it. The magnitude components, $O(\tilde{e})$ and $O(\beta ),$ can then
further be such that the estimation errors in the sample moment equations
are of the same order of magnitude as the misspecification and
identification strengths.

Next, we show that the relative magnitudes of misspecification and
identification, $O(\tilde{e})$ and $O(\beta ),$ drive the identification of
risk premia in the FM two-pass methodology and the CUE. Since the
misspecification $J$-statistic gauges $O(\tilde{e})$ while the
identification $IS$-statistic reflects $O(\beta ),$ the comparison of these
two statistics plays a crucial role in our later discussion.

\subsection{Population objective function and pseudo-true value}

\subsubsection{FM two-pass estimator}

The population objective functions of different risk premia estimators are
all quadratic forms of the pricing error but weigh it differently. The
population objective function of the FM two-pass estimator involves no
weight function and is therefore just the quadratic form of the pricing
error:%
\begin{equation}
Q_{FM}(l)=(\mu _{R}-\beta l)^{\prime }(\mu _{R}-\beta l),  \label{fm objf}
\end{equation}%
having as its minimizer the, so-called, pseudo-true value of risk premia.
Unlike the true value in case of correct specification, the pseudo-true
value denoted by $\lambda _{F,FM}^{\ast }$ does not set the objective
function to zero as $\mu _{R}=\beta \lambda _{F}+\tilde{e},$ with $\tilde{e}%
\neq 0:$%
\begin{equation}
\begin{array}{c}
\lambda _{F,FM}^{\ast }=\arg \min_{l\in \mathbb{R}^{K}}\text{ }%
Q_{FM}(l)=\left( \beta ^{\prime }\beta \right) ^{-1}\beta ^{\prime }\mu _{R}.%
\end{array}
\label{fm pseudo true}
\end{equation}%
Using the specification of the expected asset returns in (\ref%
{beta_representation2}), the pseudo-true value becomes:%
\begin{equation}
\begin{array}{c}
\lambda _{F,FM}^{\ast }=\lambda _{F}+\frac{O(\tilde{e})}{O(\beta )}\cdot
(b^{\prime }b)^{-1}b^{\prime }a,%
\end{array}
\label{pseudo misspriced}
\end{equation}%
where $\frac{O(\tilde{e})}{O(\beta )}\cdot (b^{\prime }b)^{-1}b^{\prime }a$
reflects the difference between the pseudo-true value $\lambda _{F,FM}^{\ast
}$ and the generic risk premia $\lambda _{F}.$

The specification of the pseudo-true value (\ref{pseudo misspriced})
therefore shows that it depends on the strength of misspecification, $O(%
\tilde{e}),$ compared to the identification strength, $O(\beta )$, unless $%
(b^{\prime }b)^{-1}b^{\prime }a=0$ so the misspecification error is outside
of the space spanned by $\beta $. For example, when we use the specification
from the weak instrument/factor literature: $O(\tilde{e})=O(T^{c_{\tilde{e}%
}})=O_{c_{\tilde{e}}}\times T^{c_{\tilde{e}}},$ $O(\beta )=O(T^{c_{\beta
}})=O_{c_{\beta }}\times T^{c_{\beta }},$ with $c_{\tilde{e}}=c_{\beta }=-%
\frac{1}{2}$ and $O_{c_{\tilde{e}}},$ $O_{c_{\beta }}$ non-zero finite
constants:%
\begin{equation}
\begin{array}{c}
\lambda _{F,FM}^{\ast }=\lambda _{F}+\frac{O_{c_{\tilde{e}}}}{O_{c_{\beta }}}%
\cdot (b^{\prime }b)^{-1}b^{\prime }a,%
\end{array}
\label{labfm miss}
\end{equation}%
so depending on $O_{c_{\tilde{e}}}$ being larger or smaller than $%
O_{c_{\beta }},$ there can be a considerable difference between $\lambda
_{F,FM}^{\ast }$ and $\lambda _{F}.$ Hence, it is important to compare the
strength of misspecification reflected by $O(\tilde{e})$ to the strength of
identification reflected by $O(\beta )$, in order to gauge the difference
between $\lambda _{F,FM}^{\ast }$ and $\lambda _{F}.$

Under correct specification, the pricing error is zero at the true value of
risk premia, and remains so when repackaging the assets. The minimizers of
the population objective functions underlying different estimators all
therefore have the true value of risk premia as their minimizer, and are all
invariant to repackaging. Under the standard conditions, these estimators
are also consistent for the true value. This is no longer the case in
misspecified settings where the minimizers of the population objective
functions associated with various estimators (\textit{i.e. }the pseudo-true
values) differ, since there is no longer a value of risk premia at which the
population objective function is equal to zero.

Kandel and Stambaugh (1995)\nocite{kanstam95} show that the pseudo-true
value for the FM two-pass estimator is not invariant to repackaging of the
assets, while the population cross-section generalized least squares (GLS)\
estimator is. The FM two-pass estimator provides a consistent estimator of
the pseudo-true value, which makes Kandel and Stambaugh (1995) skeptical
about how much of interest the cross-section ordinary least squares (OLS)
estimator, or FM two-pass estimator, is under misspecification. They thus
show a preference for the cross-section GLS estimator.

\paragraph{Theorem 1.}

For $A$ an invertible $N\times N$ matrix of weights, $A\iota _{N}=\iota _{N}$
with $\iota _{N}$ the $N$-dimensional vector of ones, the FM two-pass
pseudo-true value for the orginal test assets, $R_{t},$ does not equal that
of the repackaged test assets, $A\times R_{t}.$\smallskip

\noindent \textbf{Proof: } See the Appendix and Kandel and Stambaugh (1995).
\smallskip

\subsubsection{Continuous updating estimator}

Kandel and Stambaugh (1995)'s analysis primarily concerns the population
cross-section OLS and GLS objective functions. These population objective
functions treat the expected asset returns, $\mu _{R},$ and $\beta $'s as
known while we replace them by estimators in the sample objective functions.
When extending these population objective functions to sample ones, an
important and empirically relevant issue occurs if the proximity of the true 
$\beta $'s to a reduced rank value is comparable to its estimation error.
The resulting risk premia estimators are then exposed to multi-collinearity
issues and test statistics, like, for example, the FM\ two-pass $t$-test and
its misspecification robust extension by Kan, Robotti, and Shanken (2013),%
\nocite{krs13} are no longer reliable for conducting tests on the
pseudo-true value of the FM two-pass estimator. Test statistics which remain
reliable for conducting tests on the pseudo-true value do, however, exist
and are based on the continuous updating estimator (CUE) of Hansen, Heaton,
and Yaron (1996).\nocite{hhy96} The population objective function of the CUE
provides an extension of the GLS population objective function by accounting
for the estimation error of all estimable components of the pricing error, $%
i.e.$ $\mu _{R}$ and $\beta :$

\begin{equation}
\begin{array}{rl}
Q_{CUE}(l)= & (\mu _{R}-\beta l)^{\prime }\left[ \text{Var(}\sqrt{T}(\hat{\mu%
}_{R}-\hat{\beta}l))\right] ^{-1}(\mu _{R}-\beta l) \\ 
= & \frac{1}{1+l^{\prime }Q_{FF}^{-1}l}(\mu _{R}-\beta l)^{\prime }\Omega
^{-1}(\mu _{R}-\beta l),%
\end{array}
\label{cue objf}
\end{equation}%
where $\hat{\mu}_{R}=\bar{R}=\frac{1}{T}\sum_{t=1}^{T}R_{t},$ $\hat{\beta}=%
\frac{1}{T}\sum_{t=1}^{T}\bar{R}_{t}\bar{F}_{t}^{\prime }\left( \frac{1}{T}%
\sum_{j=1}^{T}\bar{F}_{j}\bar{F}_{j}^{\prime }\right) ^{-1},$ $\bar{R}%
_{t}=R_{t}-\bar{R},$ $\bar{F}_{t}=F_{t}-\bar{F},$ $\bar{F}=\frac{1}{T}%
\sum_{t=1}^{T}F_{t}.$ The specification in the last line of (\ref{cue objf})
is for a setting of i.i.d. data, so $\Omega =$Var($R_{t}-\beta F_{t})$ and $%
Q_{FF}=$Var($F_{t}).$ The GLS population objective function results if we
remove the first part of the expression on the bottom line of (\ref{cue objf}%
).

The CUE population objective function normalizes the pricing error, so it is
invariant under repackaging and transformations of the factors. The
(normalized) sample value of the pricing error used in (\ref{cue objf}) has
unit variance, which implies that, under mild conditions (see, e.g., Shanken
(1992)),\nocite{sh92} the sample CUE objective function has a non-central $%
\chi ^{2}$ distribution in large samples. The pseudo-true value associated
with the CUE then results as 
\begin{equation}
\begin{array}{c}
\lambda _{F,CUE}^{\ast }=\arg \min_{l\in \mathbb{R}^{K}}Q_{CUE}(l),%
\end{array}
\label{cue pseudo true}
\end{equation}%
which is also invariant to repackaging.

The population CUE objective function is multi-modal, and the pseudo-true
value results from the smallest mode. For example, in an i.i.d. setting, it
results from an eigenvalue problem so the number of modes equals the number
of eigenvalues/characteristic roots.\footnote{%
For the setting of i.i.d. data, a closed-form expression of the pseudo-true
value can be provided as resulting from the eigenvector associated with the
smallest root of the characteristic polynomial $\left\vert \tau \left( 
\begin{array}{cc}
1 & 0 \\ 
0 & Q_{FF}^{-1}%
\end{array}%
\right) -\left( 
\begin{array}{cc}
\mu _{R} & \beta%
\end{array}%
\right) ^{\prime }\Omega ^{-1}\left( 
\begin{array}{cc}
\mu _{R} & \beta%
\end{array}%
\right) \right\vert =0$ by specifying that eigenvector as $c\binom{1}{%
-\lambda _{F,CUE}^{\ast }}$ with $c$ a scalar.} To guarantee that the
characteristic root from which the pseudo-true value results, identifies
risk premia, the identification strength has to exceed the misspecification
strength. These identification properties result because the CUE\ population
objective function is the optimized function in the stepwise optimization of
a generalized reduced rank objective function, which imposes a reduced rank
value on the $N\times (K+1)\ $matrix $\left( \mu _{R}\text{ }\vdots \text{ }%
\beta \right) ,$ see Kleibergen (2007)\nocite{kf04} and Kleibergen and Zhan
(2021):\nocite{kz21}%
\begin{equation}
\begin{array}{rl}
Q_{CUE}(l)= & \min_{B\in \mathbb{R}^{N\times K}}Q_{CUE}(l,B) \\ 
Q_{CUE}(l,B)= & \left[ \text{vec}\left( \left( \mu _{R}\text{ }\vdots \text{ 
}\beta \right) -B\left( l\text{ }\vdots \text{ }I_{K}\right) \right) \right]
^{\prime }\left[ \text{Var}\left( \sqrt{T}\left( \hat{\mu}^{\prime }\text{ }%
\vdots \text{ vec(}\hat{\beta})^{\prime }\right) ^{\prime }\right) \right]
^{-1} \\ 
& \left[ \text{vec}\left( \left( \mu _{R}\text{ }\vdots \text{ }\beta
\right) -B\left( l\text{ }\vdots \text{ }I_{K}\right) \right) \right] \\ 
= & \text{tr}\left[ Q_{FF}^{-1}\left( \left( \mu _{R}\text{ }\vdots \text{ }%
\beta \right) -B\left( l\text{ }\vdots \text{ }I_{K}\right) \right) ^{\prime
}\Omega ^{-1}\left( \left( \mu _{R}\text{ }\vdots \text{ }\beta \right)
-B\left( l\text{ }\vdots \text{ }I_{K}\right) \right) \right]%
\end{array}
\label{rank}
\end{equation}%
where the expression on the last line is for the setting of i.i.d. data.%
\footnote{%
vec($A)$ is the column vectorization of a matrix $A$ that results from
stacking its columns, so vec($A)=(a_{1}^{\prime }\ldots a_{m}^{\prime
})^{\prime }$ for $A=(a_{1}\ldots a_{m}).$ tr($A)$ is the trace, or sum of
its diagonal elements, of the square matrix $A.$} The second expression in (%
\ref{rank}), $Q_{CUE}(l,B),$\ is a normalized distance measure between the $%
N\times (K+1)$ matrix $\left( \mu _{R}\text{ }\vdots \text{ }\beta \right) ,$
which is at most of rank $K+1,$ and the $N\times (K+1)$ matrix $B\left( l%
\text{ }\vdots \text{ }I_{K}\right) ,$ which is at most of rank $K.$%
\footnote{%
The rank of $B\left( l\text{ }\vdots \text{ }I_{K}\right) $ is at most $K$
because $B$ is an $N\times K$ matrix and $\left( l\text{ }\vdots \text{ }%
I_{K}\right) $ a $K\times (K+1)$ matrix, so the ranks of both of these are
at most $K.$} The sample analog of the CUE population objective function is
therefore a rank test on $\left( \mu _{R}\text{ }\vdots \text{ }\beta
\right) $. Its minimal value is the $J$-statistic for misspecification.

Since the pseudo-true value of the CUE results from a generic rank test on $%
\left( \mu _{R}\text{ }\vdots \text{ }\beta \right) ,$ it does not
necessarily represent risk premia when there is misspecification. Consider,
for example, a setting where the strength of misspecification, $O(\tilde{e}%
), $ is considerably larger than the strength of identification, $O(\beta ),$
which is tiny. The minimization over $(l,B)$ then leads to a tiny value of $%
B $ and a very large value of $l$ that does not reflect the generic risk
premia. This reasoning is thus also in line with (\ref{pseudo misspriced})
for the FM two-pass approach.

The above point is further illustrated in Figure 1. It shows, for a single
factor setting, the contour lines of the pseudo-true values of the FM
two-pass estimator and the CUE in deviation from the risk premium in case of
correct specification, as a function of the misspecification and
identification strengths. Both for FM two-pass (left panel) and CUE (right
panel), Figure 1 shows that the pseudo-true values deviate considerably from
the baseline risk premium when the misspecification strength exceeds the
identification strength. The pseudo-true value of CUE then no longer
represents a risk premium, because the closest proximity of $\left( \mu _{R}%
\text{ }\vdots \text{ }\beta \right) $ from a reduced rank value results
mainly from the small value of $\beta $, and much less so from the
combination of $\mu _{R}$ and $\beta .$ It is therefore important to be able
to diagnose if the pseudo-true value results from such a setting. \smallskip

\begin{figure}[htp]
\centering
\includegraphics[width=450 pt,  height=150 pt]{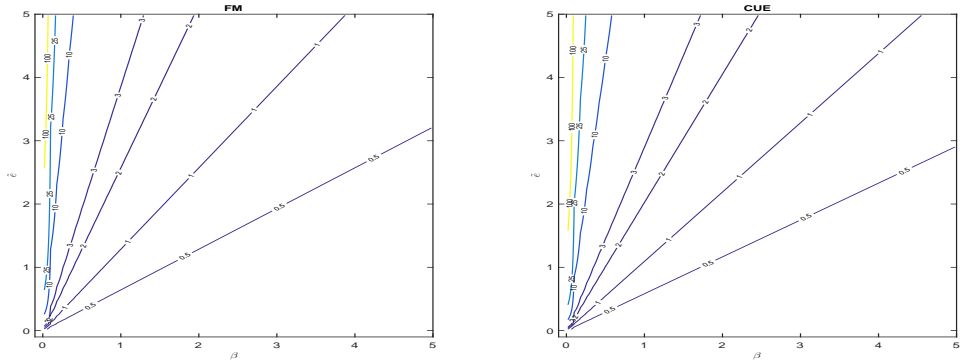}
\caption{\textbf{Contour lines that show the deviation of pseudo-true values
of FM and CUE from the baseline risk premium in case of correct specification as a
function of the strengths of\ misspecification $\tilde{e}$ and identification $\beta$.}  }
\end{figure}
\singlespace{\small \noindent Notes: The pseudo-true values of FM and CUE
are defined in (\ref{fm pseudo true}) and (\ref{cue pseudo true}),
respectively. The baseline risk premium is set to 2, so the contour lines
show the deviation of pseudo-true values from 2. The parameters used for
Figure 1 are calibrated to the data from Kroencke (2017). The magnitudes of $\tilde{e}$ and $\beta $ vary from 0 to 5.}

\doublespace

\subsection{$IS$-statistic versus $J$-statistic}

To obtain a diagnostic for interpreting the pseudo-true value as a risk
premium, we use that the CUE population objective function evaluated at the
pseudo-true value equals a rank test on $(\mu _{R}$ $\vdots $ $\beta )$ as
shown by (\ref{rank}), which is also the population equivalent of a $J$%
-statistic: 
\begin{equation}
\begin{array}{c}
J=Q_{CUE}(\lambda _{F,CUE}^{\ast }).%
\end{array}
\label{jstat}
\end{equation}%
It is then always smaller than or equal to an appropriately specified rank
test statistic conducted on any sub-matrix of $\left( \mu _{R}\text{ }\vdots 
\text{ }\beta \right) .$ An example of a rank test on a sub-matrix of $(\mu
_{R}$ $\vdots $ $\beta )$ is a rank test on just $\beta ,$ whose sample
analog is typically used to test for the identification of the risk premia
under correct specification; see, e.g., Cragg and Donald (1997), Kleibergen
and Paap (2006) and Robin and Smith (2000).\nocite{cradon97}\nocite{kpaap02}%
\nocite{robsmit00} This rank test statistic is thus always larger than or
equal to the CUE objective function evaluated at the pseudo-true value (i.e.
the $J$-statistic in (\ref{jstat})), and provides a measure of the
identification strength of $\lambda _{F},$ see Kleibergen and Zhan (2021): 
\begin{equation}
\begin{array}{rl}
IS= & \min_{d\in \mathbb{R}^{(K-1)}}Q_{\beta }(d) \\ 
Q_{\beta }(d)= & \binom{1}{-d}^{\prime }\beta ^{\prime }\left[ \left( \binom{%
1}{-d}\otimes I_{N}\right) ^{\prime }\text{Var}\left( \sqrt{T}\text{vec(}%
\hat{\beta})\right) \left( \binom{1}{-d}\otimes I_{N}\right) \right]
^{-1}\beta \binom{1}{-d} \\ 
= & \min_{G\in \mathbb{R}^{N\times (K-1)}}Q_{p,r}(d,G) \\ 
Q_{p,r}(d,G)= & \left[ \text{vec}\left( \beta -G\left( d\text{ }\vdots \text{
}I_{K-1}\right) \right) \right] ^{\prime }\left[ \text{Var}\left( \sqrt{T}%
\text{vec(}\hat{\beta})\right) \right] ^{-1}\left[ \text{vec}\left( \beta
-G\left( d\text{ }\vdots \text{ }I_{K-1}\right) \right) \right] .%
\end{array}
\label{rank 3}
\end{equation}

When the $J$ misspecification measure is close to the $IS$ identification
measure, the pseudo-true value basically results from a close to reduced
rank value of $\beta .$ It implies that the pseudo-true value $\lambda
_{F,CUE}^{\ast }$ is very large and does not represent risk premia. Hence,
only when the $J$ misspecification measure is considerably less than the $IS$
identification strength measure does the pseudo-true value represent risk
premia. It shows that when misspecification is present, the identification
of risk premia is no longer just reflected by the rank strength value of $%
\beta ,$ as is the case under correct specification, but by the difference
between a measure of the degree of misspecification and a measure of the
rank strength of $\beta $. In case of correct specification, the
misspecification measure equals zero, so the identification then solely
results from the rank value of $\beta ,$ but not so when misspecification is
present. The cut-off for identification is when the measures of
misspecification and rank strength of $\beta $ are equal in which case $%
\lambda _{F}$ is not identified, while it can be identified as risk premia
when the latter exceeds the former.

To summarize, the sample analog of the $IS$ identification measure is a rank
test statistic on $\beta $, while the sample analog of the minimal value of
the population CUE objective function is the $J$-statistic for
misspecification. It is thus important to compare the $IS$-statistic on $%
\beta $ with the $J$-statistic for misspecification. When using aligning
specifications for the $J$-statistic and the $IS$-statistic on $\beta ,$ the
former is always less than or equal to the latter. Close values, however,
indicate an issue with identifying risk premia. The estimated values of the
risk premia are then also typically very large, which sheds further doubt on
whether they can be interpreted as risk premia.\footnote{%
It is important to note that we cannot test for the equality of the $J$-test
and rank test for $\beta $, since they involve the same estimators. For
example, in case $\beta $ is of reduced rank while the expected returns, $%
\mu _{R},$ are different from zero, the population values of these two test
statistics would be the same, and the joint sampling distribution of their
estimators would be degenerate, so we cannot establish a test for their
equality.}

\subsection{Tests of risk premia}

To conduct inference on the pseudo-true values of risk premia, it is
important to have tests that remain reliable for a wide range of values of
the misspecification and identification strengths. We show that this does
not hold for the FM $t$-test and its misspecification robust extension by
Kan, Robotti, and Shanken (2013), which are size distorted when the
identification strength of risk premia is minor. We conduct a small
simulation exercise to highlight this sensitivity. Thereafter we state the
double robust Lagrange multiplier (DRLM) test from Kleibergen and Zhan
(2021), which is size correct for all settings of the misspecification and
identification strengths.

\subsubsection{Large sample behavior of $t$-tests based on the FM estimator
under misspecification and weak identification}

We use a single factor setting, so $K=1,$ to show that the FM two-pass $t$%
-test and its misspecification robust extension by Kan, Robotti, and Shanken
(2013),\nocite{krs13} for conducting inference on the FM two-pass
pseudo-true value become unreliable, when the value of the $\beta $-matrix
is close to rank deficient. For the single factor setting, a close to rank
deficient value of $\beta $ implies that $\beta $ is close to zero. Because
of the commonality of misspecification paired with weak identification, we
construct the large sample behavior of the FM\ two-pass estimator for a
setting where both the $\beta $'s and the misspecification are small. We
model this using the weak factor/small $\beta $ and misspecification
assumption (\ref{beta_representation2}) and (\ref{labfm miss}), and further
postulate that both $\beta $ and the misspecification $\mu _{R}-\beta
\lambda _{F}$ are drifting to zero at rate $1/\sqrt{T}:$%
\begin{equation}
\begin{array}{c}
\beta =\beta _{T}=\frac{b}{\sqrt{T}},\text{ }\mu _{R}-\beta \lambda _{F}=%
\frac{a}{\sqrt{T}},\text{ }\lambda _{F,FM}^{\ast }=\lambda _{F}+(b^{\prime
}b)^{-1}b^{\prime }a,%
\end{array}
\label{small beta}
\end{equation}%
with $b$ and $a$ $N$-dimensional vectors of constants. The small $\beta $
assumption (\ref{small beta}) is common in the weak identification
literature; see, e.g., Staiger and Stock (1997),\nocite{stst97} Kleibergen
(2005, 2009) and Kleibergen and Zhan (2018, 2021).\nocite%
{kleibergen2018identification}\nocite{kf09} \nocite{kf00a} \nocite{kz20} It
leads to the small values of the $F$-statistic testing the joint
significance of the $\beta $'s that we often observe; see Kleibergen and
Zhan (2015).\nocite{kz15} The small misspecification assumption further
accommodates the small but significant values of the $J$-statistic that are
regularly seen.

\paragraph{Theorem 2:}

For i.i.d. data and under the weak $\beta $ and misspecification assumption (%
\ref{small beta}), the large sample distribution of the FM two-pass
estimator, $\hat{\lambda}_{F}=(\hat{\beta}^{\prime }\hat{\beta})^{-1}\hat{%
\beta}^{\prime }\hat{\mu}_{R}$, consists of four components:%
\begin{equation}
\begin{array}{rl}
\hat{\lambda}_{F}\underset{d}{\rightarrow } & \underset{1}{\underbrace{%
\lambda _{F,FM}^{\ast }}}+\underset{2}{\underbrace{\frac{\psi _{\mu
}^{\prime }(b+\psi _{\beta })}{(b+\psi _{\beta })^{\prime }(b+\psi _{\beta })%
}}}-\underset{3}{\underbrace{\lambda _{F,FM}^{\ast }\frac{\psi _{\beta
}^{\prime }(b+\psi _{\beta })}{(b+\psi _{\beta })^{\prime }(b+\psi _{\beta })%
}}}+\underset{4}{\underbrace{\frac{e^{\prime }\psi _{\beta }}{(b+\psi
_{\beta })^{\prime }(b+\psi _{\beta })}}},%
\end{array}
\label{large sample beta}
\end{equation}%
with $e=a-b(b^{\prime }b)^{-1}b^{\prime }a$; $\psi _{\mu }$ and $\psi
_{\beta }$ are independent normally distributed $N$-dimensional random
vectors with mean zero and covariance matrices var($R_{t})=\Omega +\beta
Q\beta ^{\prime }$ and $\Omega Q^{-1},$ $\Omega =$Var($R_{t}-\beta F_{t})$
and $Q_{FF}=$Var($F_{t}).$

\noindent \textbf{Proof: }See the Appendix. \smallskip

The four different components of the large sample behavior of the FM
two-pass estimator in Theorem 2 are characterized by:

\begin{enumerate}
\item The object of interest: $\lambda _{F,FM}^{\ast },$ the pseudo-true
value of the FM two-pass estimator.

\item $\frac{\psi _{\mu }^{\prime }(b+\psi _{\beta })}{(b+\psi _{\beta
})^{\prime }(b+\psi _{\beta })}:$ Under i.i.d. data and assumption (\ref%
{small beta}), $\sqrt{T}\hat{\beta}\underset{d}{\rightarrow }b+\psi _{\beta
} $ and $\sqrt{T}\hat{\mu}_{R}\underset{d}{\rightarrow }e+b\lambda
_{F,FM}^{\ast }+\psi _{\mu }$, so it shows the large sample behavior of $%
\frac{(\hat{\mu}_{R}-\mu_R)^{\prime }\hat{\beta}}{\hat{\beta}^{\prime }\hat{%
\beta}}$ , which is such that $\frac{\psi _{\mu }^{\prime }(b+\psi _{\beta })%
}{\sqrt{(b+\psi _{\beta })^{\prime }(b+\psi _{\beta })}}\sim N(0,\Omega
+\beta Q\beta ^{\prime }),$ since $\psi _{\mu }$ is independent of $\psi
_{\beta }.$

\item $-\lambda _{F,FM}^{\ast }\frac{\psi _{\beta }^{\prime }(b+\psi _{\beta
})}{(b+\psi _{\beta })^{\prime }(b+\psi _{\beta })}:$ Since the $\psi
_{\beta }$ elements in the numerator are positively correlated, it creates a
negative bias in the FM two-pass estimator for small values of $b.$ It also
implies that the large sample distribution of the FM two-pass estimator is
not a normal one for such values of $b.$ When $b$ is much larger than $\psi
_{\beta },$ so we are in a setting of sizeable $\beta $'s, $\psi _{\beta }$
becomes negligible compared to $b$ in the $(b+\psi _{\beta })$ elements.
This is the setting covered by Shanken (1992),\nocite{sh92} who provides the
correction for the standard errors of the FM two-pass estimator to
incorporate the contribution of this component for the variance of the FM
two-pass estimator.

\item $\frac{e^{\prime }\psi _{\beta }}{(b+\psi _{\beta })^{\prime }(b+\psi
_{\beta })}$: It appears in the large sample distribution of the FM two-pass
estimator because of misspecification. When $e=0$ or the identification
strength, $b^{\prime }b,$ is much larger than the length of $e,$ or the
amount of misspecification, it has little effect on the large sample
distribution of the FM two-pass estimator. Because of the dependence between
the $\psi _{\beta }$ elements in the numerator and denominator, this
component only has a normal distribution for large values of $b,$ so $\psi
_{\beta }$ becomes negligible in the $(b+\psi _{\beta })$ elements. This is
the setting covered by Kan, Robotti, and Shanken\ (2013), who provide a
correction of the standard errors to incorporate this component.
\end{enumerate}

The third and fourth components lead to a large sample distribution for the
FM two-pass estimator which is not a normal one for the empirically relevant
setting of small values of $b.$ Test statistics based on the FM two-pass
estimator, like the FM $t$-test and the KRS $t$-test, therefore become size
distorted for such small values. To illustrate this, Figure 2 shows the
simulated rejection frequencies of 5\% significance FM and KRS $t$-tests on $%
\lambda _{F,FM}^{\ast }$ for a range of values of the misspecification and
identification strengths.\footnote{%
Throughout the paper, simulated data are generated from the linear factor
model using parameters calibrated to the data from Kroencke (2017).} Figure
2.1 shows the rejection frequencies for H$_{0}:\lambda _{F,FM}^{\ast }=0$,
and Figure 2.2 is for $\lambda _{F,FM}^{\ast }$ corresponding with the
values resulting from Figure 1.

\begin{figure}[htp]
\subfigure[$H_{0}:\lambda _{F,FM}^{\ast }=0$ ]{\includegraphics[width=450 pt,  height=160 pt]{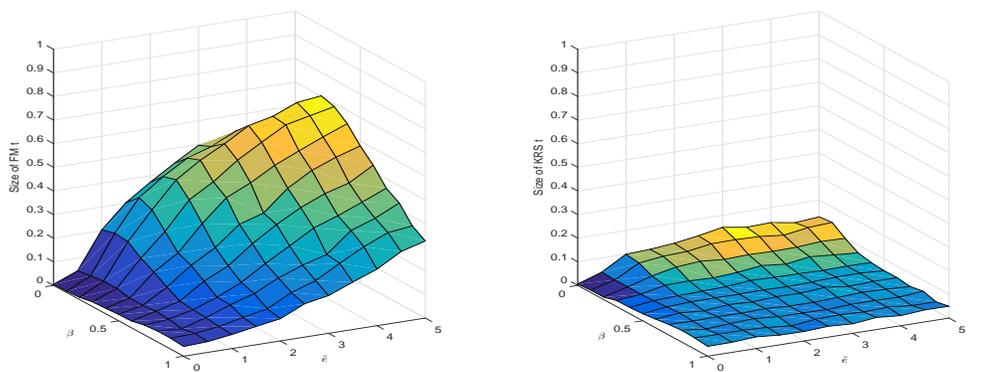}}
\subfigure[$H_{0}:\lambda _{F,FM}^{\ast }$ results from Figure 1]{\includegraphics[width=450 pt,  height=160 pt]{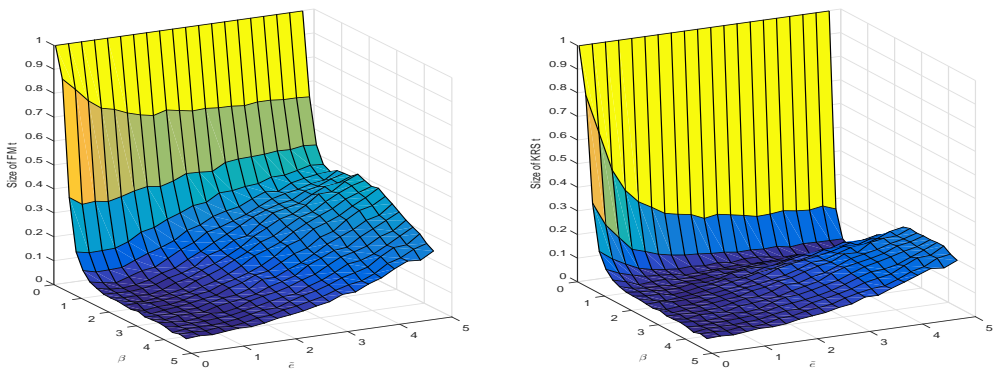}}
\caption{\textbf{Rejection frequencies of 5\% significance FM and KRS t-tests for varying identification strengths $\beta$ and misspecification $\tilde{e}$.}}
\end{figure}

Figure 2.1 shows that the FM $t$-test and the KRS $t$-test are conservative
or correctly sized when testing H$_{0}:\lambda _{F,FM}^{\ast }=0$ and there
is no misspecification so $\tilde{e}=0.$ For increasing values of $\tilde{e}$%
, however, the FM $t$-test and the KRS $t$-test over-reject for small values
of $\beta $, i.e. the rejection frequencies are larger than the nominal 5\%.
For the FM $t$-test, this over-rejection also extends to larger values of $%
\beta $.

For testing H$_{0}:\lambda _{F,FM}^{\ast }$ corresponding with the
pseudo-true values resulting from Figure 1, Figure 2.2 shows that the FM $t$%
-test and the KRS $t$-test severely over-reject for small values of $\beta $
which are less than $\tilde{e}$. When $\beta$ approaches zero, the rejection
frequencies are even equal to one, which is in line with Kan and Zhang
(1999).\nocite{kz99}

\subsubsection{The DRLM\ test which remains size correct under
misspecification and weak identification}

The decomposition in Theorem 2 can also be conducted for the GLS risk premia
estimator, so test statistics based on it, such as the GLS $t$-test, become
similarly size distorted. We therefore use a statistic based on the CUE: the
DRLM statistic proposed in Kleibergen and Zhan (2021), whose asymptotic
distribution is bounded by the $\chi _{K}^{2}$ distribution. When using
appropriate critical values from the $\chi _{K}^{2}$ distribution, the DRLM\
test of hypotheses specified on the pseudo-true value of the CUE remains
size correct for general levels of misspecification and identification. In
the Appendix, we state the DRLM statistic, and provide a brief discussion of
its implementation.

Figure 3 presents the rejection frequencies of a 5\% significance DRLM\ test
of H$_{0}:\lambda _{F,CUE}^{\ast }=0$. We note that, unlike the large sample
distributions of the FM and KRS $t$-tests, the limiting distributions of the
statistics underlying the DRLM test do not depend on the tested parameter or
the covariance matrices $\Omega $ and $Q_{FF}$. In contrast with Figure 2
for the simulated sizes of the FM and KRS $t$-tests, Figure 3 shows that the
DRLM test is size correct (i.e. the rejection frequencies do not exceed the
nominal 5\%), and is conservative for combined small identification and
misspecification strengths.

\begin{figure}[htp]
\centering
\includegraphics[width=320 pt,  height=260 pt]{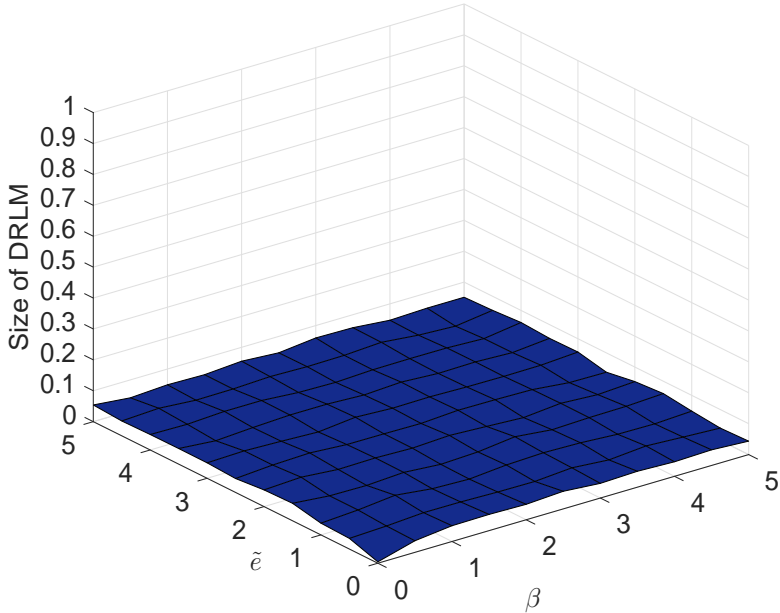}
\caption{\textbf{Rejection frequencies of the 5\% significance DRLM test of $H_{0}:\lambda _{F,CUE}^{\ast }=0$  for varying identification strengths $\beta$ and
misspecification $\tilde{e}$.}  }
\end{figure}

\subsection{Power comparison}

We briefly illustrate the power of the FM and KRS $t$-tests compared to the
DRLM test. A more extensive power study of the DRLM test is conducted in
Kleibergen and Zhan (2021). We first show the power for a correct
specification, so all the aforementioned tests target the same value of the
risk premium, while there is also strong identification. Thus, we have the
ideal setting that $\beta $ is sizeable while $\tilde{e}=0$ in the data
generation process. All the examined tests are therefore size correct in
Figure 4a-b at the hypothesized value, i.e. all the rejection frequencies
are near the nominal level of 5\% when the distance to the tested risk
premium is zero. Since there is no misspecification, the power curves of FM $%
t$ (dotted black) and KRS $t$ (dashed green) largely overlap in Figure 4a.
The comparison of Figure 4a and 4b, however, also shows that the DRLM test
can have more power than FM and KRS $t$-tests. This results since the DRLM
test is based on the GLS framework, while the FM and KRS $t$-tests for
Figure 4 are based on OLS.

\begin{figure}[h]\centering
\subfigure[FM $t$ (black) and KRS $t$ (green)]{\includegraphics[width=180 pt,  height=170 pt]{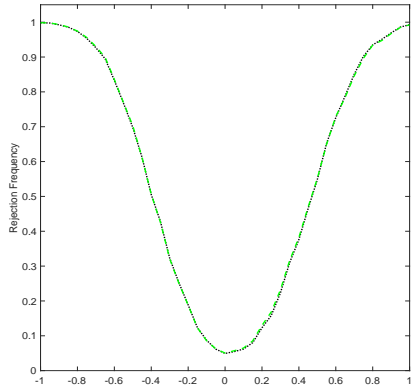}}
\subfigure[DRLM]{\includegraphics[width=180 pt,  height=170 pt]{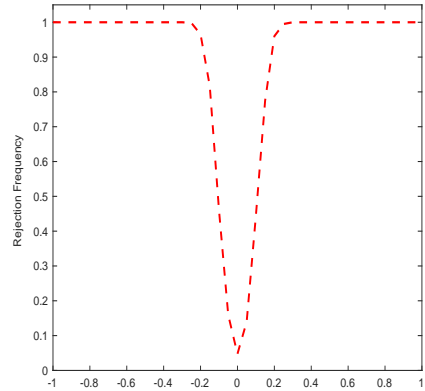}}
\subfigure[FM $t$ (black) and KRS $t$ (green) ]{\includegraphics[width=180 pt,  height=170 pt]{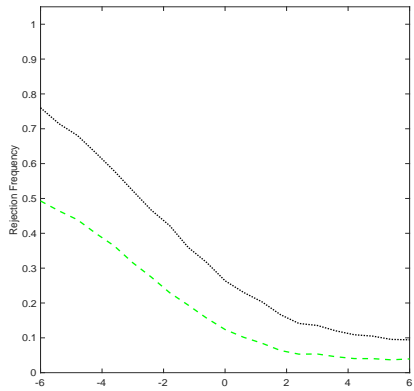}}
\subfigure[DRLM]{\includegraphics[width=180 pt,  height=170 pt]{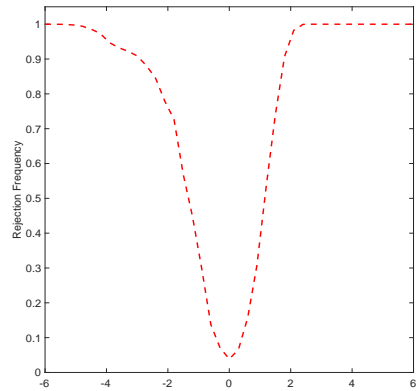}}
\caption{\textbf{Power comparison of FM, KRS $t$-tests and DRLM at the 5\% level.}  a-b: Correct specification, strong identification; c-d:
Misspecification, weak identification.}
\end{figure}

In contrast with Figure 4a-b, we allow for misspecification as well as weak
identification in Figure 4c-d. This is achieved by considering smallish $%
\beta $ and nonzero $\tilde{e}$ in the data generation process. For this
scenario, Figure 4c shows that the FM and KRS $t$-tests are no longer size
correct, while their power curves substantially differ from those in Figure
4a. In contrast, the DRLM test, which involves the power improvement rule
mentioned in the Appendix, remains size correct in Figure 4d.

Overall, Figure 4 illustrates that it is appealing to use the DRLM test for
conducting inference on risk premia. In contrast, the conventional FM $t$%
-test, together with the KRS $t$-test, is jeopardized by joint
misspecification and weak identification, both of which are prevalent in
empirical studies as we show next.

\section{\protect\bigskip Diagnostic statistics}

\label{j_is}

Before we turn to empirical applications of the tests on the risk premia, we
first apply the diagnostic statistics which help to gauge whether we can
interpret the risk premia accordingly. The $IS$-statistic, which tests for a
reduced rank value of $\beta ,$ is commonly used to test for identification
of the risk premia in correctly specified settings. In misspecified
settings, it is, however, no longer just the $IS$-statistic which governs
the identification and interpretation of the pseudo-true value of the risk
premia, but its difference with the $J$-statistic, which equals the minimal
value of the sample CUE objective function. This is, for example, shown by
the deviation of the FM two-pass pseudo-true value in (\ref{pseudo
misspriced}) from its counterpart under correct specification, and further
illustrated in Figure 1. The $J$ and $IS$ statistics have well established
limiting distributions under their hypotheses of interest, which are $\chi
_{N-K}^{2}$ for the $J$-statistic under correct specification, and $\chi
_{N-K+1}^{2}$ for the $IS$-statistic under a reduced rank $\beta $ matrix;
see the Appendix for their explicit expressions. We are, however, mainly
interested in these statistics because of the inequality between them, $J$%
-statistic $\leq $ $IS$-statistic, and because a close proximity between
these two statistics shows that the pseudo-true value does not identify risk
premia as discussed in Section 2. We therefore compute these two statistics
first for eight well known specifications of the linear asset pricing model,
and second for the specifications resulting from the factor zoo of Feng,
Giglio, and Xiu (2020).\nocite{feng2020taming}

\subsection{Empirical identification of risk premia}

Figure 5 shows a scatter plot of the $J$ and $IS$ statistics for eight well
known specifications of the linear asset pricing model: Fama and French
(1993), Jagannathan and Wang (1996), Yogo (2006), Lettau and Ludvigson
(2001), Savov (2011), Adrian, Etula, and Muir (2014), Kroencke (2017) and
He, Kelly, and Manela (2017).\footnote{\label{footnote_data}We thank the
authors of Jagannathan and Wang (1996), Yogo (2006), Lettau and Ludvigson
(2001), Savov (2011), and Kroencke (2017) for sharing their data. For the
models of Fama and French (1993), Adrian, Etula, and Muir (2014), and He,
Kelly, and Manela (2017), we use the extended data of risk factors and test
assets as in Lettau, Ludvigson, and Ma (2019).\nocite{llm19}} In line with
common practice, we incorporate the zero-$\beta $ return, $\lambda _{0},$
while the factors used in the eight different specifications are:

\begin{enumerate}
\item Fama and French (1993), the prominent three, so-called Fama-French,
factors: the market return $R_{m}$, SMB (small minus big), and HML (high
minus low). We use the quarterly data from Lettau, Ludvigson, and Ma (2019)
over 1963Q3 to 2013Q4, so $T=202,$ for the three factors, and the
twenty-five size and book-to-market sorted portfolios as test assets.

\item Jagannathan and Wang (1996), three factors: $R_{m}$, corporate bond
yield spread, and per capita labor income growth. We use their monthly data
from July 1963 to December 1990 so $T=330$, while one hundred size and beta
sorted portfolios are used as test assets. \nocite{jw96}

\item Yogo (2006), three factors: $R_{m}$, durable and nondurable
consumption growth. The sample period is from 1951Q1 to 2001Q4 so $T=204$,
with twenty-five size and book-to-market sorted portfolios as test assets. 
\nocite{yogo06}

\item Lettau and Ludvigson (2001), three factors: consumption-wealth ratio,
consumption growth, and their interaction. We use the quarterly data from
1963Q3 to 1998Q3 so $T=141$, while the test assets are the twenty-five
Fama-French portfolios.

\item Savov (2011), one factor: garbage growth. We use the same annual data,
1960 - 2006, while the test assets are the twenty-five Fama-French
portfolios augmented by the ten industry portfolios, as suggested by
Lewellen, Nagel, and Shanken (2010).\nocite{lns2009} \nocite{sav11}

\item Adrian, Etula, and Muir (2014), one factor: leverage. Following
Lettau, Ludvigson, and Ma (2019), we extend the time period to 1963Q3 -
2013Q4, and use twenty-five size and book-to-market sorted portfolios as
test assets. \nocite{aem14}

\item Kroencke (2017), one factor: unfiltered annual consumption growth. We
use the postwar 1960 - 2014 sample from Kroencke (2017), while thirty
portfolios, sorted by size, value and investment alongside the market
portfolio, are used as test assets. \nocite{Kro17}

\item He, Kelly, and Manela (2017), two factors: banking equity-capital
ratio and $R_{m}$. The data are also taken from Lettau, Ludvigson, and Ma
(2019) for the period 1963Q3 - 2013Q4, and twenty-five size and
book-to-market sorted portfolios are the test assets.\nocite{hkm17}
\end{enumerate}

\begin{figure}[htp]
\centering
\includegraphics[width=320 pt,  height=250 pt]{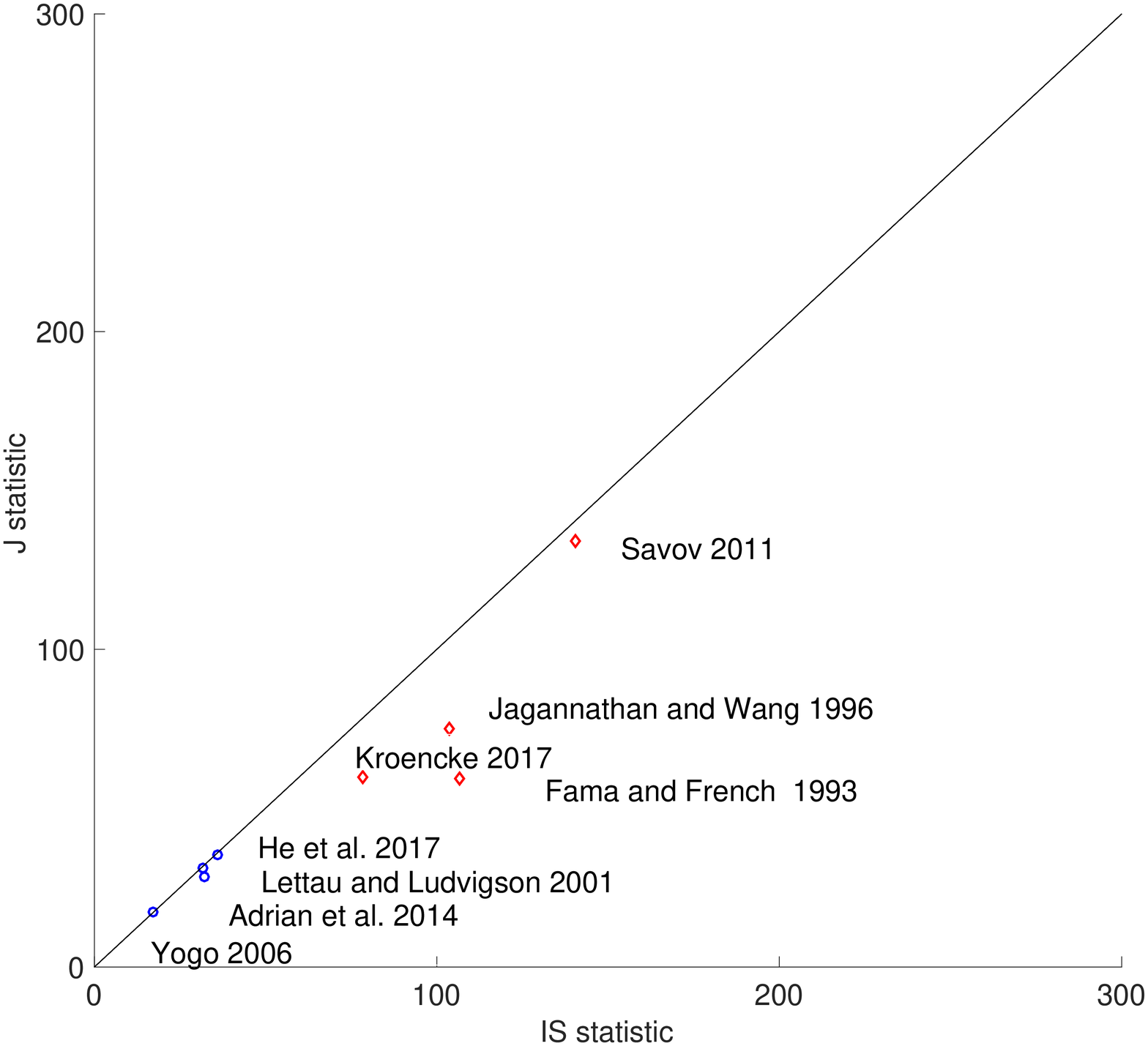}
\caption{\textbf{Scatter plot of $J$ and $IS$ statistics for different
specifications.}  }
\justify{\small Notes:  The zero-$\beta$ return is incorporated. We revisit eight models  and their associated factors. Fama and French
(1993): $R_{m}$, SMB, and HML; Jagannathan and  Wang (1996): $R_{m}$, corporate bond yield spread, and
per capita labor income growth; Yogo (2006):  $R_{m}$, durable and nondurable consumption growth; Lettau and Ludvigson (2001): consumption growth, consumption wealth ratio and their interaction; Savov
(2011): garbage growth; Adrian, Etula, and Muir(2014): leverage; Kroencke (2017): unfiltered consumption growth;
He, Kelly, and Manela (2017): $R_{m}$ and the banking  equity-capital ratio. For detailed descriptions of
risk factors and test assets, we refer to the published articles. }
\end{figure}

\doublespace

The plotted points in Figure 5 roughly exhibit two patterns, which we
illustrate by using different colors (red and blue). For Fama and French
(1993), Jagannathan and Wang (1996), Savov (2011) and Kroencke (2017), we
observe both large $J$ and $IS$ statistics, so the corresponding models
appear to be misspecified. In contrast, for Lettau and Ludvigson (2001),
Yogo (2006), Adrian, Etula, and Muir (2014), and He, Kelly, and Manela
(2017), we encounter small $J$ and $IS$ statistics, so these models are
likely to be weakly identified.

The $IS$-statistics for the first set of four specifications are large and
mostly significant but they are also close to their respective $J$%
-statistics, which is revealed by the proximity of these points to the
45-degree line. We can therefore not conclude from these $IS$-statistics
that we identify risk premia. The most likely setting for identification is
for Fama and French (1993), whose ($IS$, $J$) point is most distant from the
45-degree line.

For the second set of four points, the $J$-statistics are small and mostly
insignificant, indicating that the models might not be misspecified.
However, their corresponding $IS$-statistics are also similarly small, which
implies that the low values of the $J$-statistics are induced by the low
values of the $IS$-statistics, since $J$-statistics are necessarily exceeded
by $IS$-statistics. Hence, we cannot identify risk premia for these
specifications.

\begin{table}[tph]
\caption{\textbf{$J$ and $IS$ statistics}}
\label{j_is_models}\justify Panel A contains the $J$ and $IS$ statistics
plotted in Figure 5, for which the zero-$\beta $ return is incorporated. In
Panel B, the zero-$\beta $ return is removed so $\lambda _{0}=0.$
Significance at 1\%, ***; 5\%, **; 10\%, *.
\par
\bigskip \centering
\par
\begin{tabular}{lccccc}
\hline
& \multicolumn{2}{c}{(A) Impose $\lambda _{0}=0$: No} &  & 
\multicolumn{2}{c}{(B) Impose $\lambda _{0}=0$: Yes} \\ 
\cline{2-3}\cline{5-6}
& $J$-statistic & $IS$-statistic &  & $J$-statistic & $IS$-statistic \\ 
\hline
Fama and French (1993) & 59.34*** & 106.81*** &  & 87.47*** & 974.39*** \\ 
Jagannathan and Wang (1996) & 75.07 & 103.54 &  & 86.46 & 103.56 \\ 
Lettau and Ludvigson (2001) & 31.11* & 31.75* &  & 37.15** & 40.90** \\ 
Yogo (2006) & 17.14 & 17.34 &  & 19.42 & 19.60 \\ 
Savov (2011) & 134.27*** & 140.68*** &  & 268.60*** & 296.78*** \\ 
Adrian, Etula, and Muir (2014) & 28.42 & 31.97 &  & 30.41 & 42.03** \\ 
Kroencke (2017) & 59.84*** & 78.47*** &  & 60.03*** & 102.77*** \\ 
He, Kelly, and Manela (2017) & 35.32** & 35.88** &  & 44.44*** & 59.74*** \\ 
\hline
\end{tabular}%
\end{table}

\subsection{Removing the zero-$\protect\beta $ return}

Panel A in Table \ref{j_is_models} states the values of the $J$ and $IS$
statistics plotted in Figure 5. Panel B states these statistics when the
zero-$\beta $ return is removed, so $\lambda _{0}=0.$ All statistics in
Panel B therefore exceed their corresponding counterparts in Panel A.
Removing the zero-$\beta $ return thus increases both the misspecification
and identification. The misspecification has increased since we removed a
parameter from the pricing equation, while identification has improved since
removing the zero-$\beta $ return allows to identify risk premia when the $%
\beta $'s are constant over the assets, which was not so when the zero-$%
\beta $ return was included. For some of the specifications, the increase of
the $J$ and $IS$ statistics is disproportional. This is most notably so for
the Fama and French (1993) specification, where the $IS$-statistic increases
ninefold while the $J$-statistic does so only minorly. For the other
specifications, the increase of the $IS$-statistic typically exceeds that of
the $J$-statistic, but not by an amount which makes it clear that the risk
premia become well identified as is the case for the Fama and French (1993)
specification. For more than half of the specifications, the removal of the
zero-$\beta $ return has therefore little effect on the identification of
risk premia.

We note that the $J$ and $IS$ statistics presented in Figure 5 are just for
illustrative purposes. We do not aim to use these statistics to strictly
reject or favor any model in Figure 5, since there are many other issues
involved. These issues include, for example, that the models used for Figure
5 contain various numbers of risk factors; in addition, their corresponding
empirical studies have used non-identical test assets; furthermore, the
considered time periods and frequencies also vary to a large extent. To
address these concerns, we next use the factor zoo data from Feng, Giglio,
and Xiu (2020) to investigate misspecification and weak identification in a
more systematic manner.

\subsection{Misspecification and weak identification in the factor zoo}

How prevalent are misspecification and weak identification in empirical
asset pricing studies? With this question in mind, we extend our analysis to
the factor zoo collected by Feng, Giglio, and Xiu (2020),\nocite%
{feng2020taming} which covers one hundred and fifty risk factors from July
1976 to December 2017, so $T$ = 498. We use the twenty-five Fama and French
size and book-to-market portfolios as test assets, which have been widely
used as the default choice.

\subsubsection{Prevalence of misspecification and weak identification}

In line with existing studies, we evaluate all possible specifications of
linear factor models resulting from the factor zoo with $K=1$, $2$, $3$, $4$%
, $5$ and $6$ factors. For $K=1$, we therefore consider $C_{150}^{1}=150$
single factor models. Similarly, we examine $C_{150}^{2}=11,175$ $%
(=150\times 149/2)$ two-factor models; $C_{150}^{3}=551,300$ three-factor
models; $C_{150}^{4}=20,260,275$ four-factor models; $%
C_{150}^{5}=591,600,030 $ five-factor models; and $%
C_{150}^{6}=14,297,000,725 $ six-factor models.\footnote{%
We used high-performance computers to conduct this study.} For each model,
we compute its $J$ and $IS$ statistics while incorporating the zero-$\beta $
return.

Table \ref{t_k1} states the frequencies of, at the 5\% level, significant
values of the $J$-statistics, and insignificant values of the $IS$%
-statistics, signaling misspecification and weak identification,
respectively. Table \ref{t_k1} shows that, when we increase the number of
factors, the resulting factor models tend to be less misspecified, while
becoming weaker identified. This results naturally since on the one hand,
adding more factors helps to better explain asset returns, so the resulting
models are less likely to be rendered misspecified; while on the other hand,
since some factors could be closely related to others, including more
factors decreases the distance of the $\beta $ matrix from a reduced rank
value, which leads to weaker identified models. We note that the reported
percentages in Table \ref{t_k1} likely understate the severeness of
misspecification, because the misspecification $J$-statistic tends to be
insignificant under weak identification, as we have discussed for Figure 5.

Overall, Table \ref{t_k1} shows that the majority of the examined models
seem to suffer from misspecification and/or weak identification. Apparently,
there is also a trade-off between the reported percentages of
misspecification and weak identification in Table \ref{t_k1}. We previously
have, however, shown that we cannot analyze the $J$ and $IS$ statistics in
isolation, as in Table \ref{t_k1}, to determine whether risk premia are
identified, so we next analyze them jointly.

\begin{table}[tph]
\caption{\textbf{Prevalence of misspecification and weak identification}}
\label{t_k1}\justify The data of one hundred and fifty risk factors are
taken from Feng, Giglio, and Xiu (2020). The test assets are the twenty-five
Fama and French size and book-to-market portfolios from July 1976 to
December 2017. Models are deemed misspecified at the 5\% level, if the $p$%
-value of the $J$-statistic does not exceed 5\%. Models are deemed weakly
identified at the 5\% level, if the $p$-value of the $IS$-statistic exceeds
5\%. The $\lambda _{0}=0$ restriction is not imposed. \bigskip
\par
\centering{\small \centering%
\begin{tabular}{lcccccc}
\hline
& \multicolumn{6}{c}{Number of Factors, $K$} \\ \cline{2-7}
& {\ $1$} & {\ $2$} & {\ $3$} & {\ $4$} & {\ $5$} & {\ $6$} \\ \hline
Number of Models ($C_{150}^K$) & 150 & 11175 & 551300 & 20260275 & 591600030
& 14297000725 \\ 
&  &  &  &  &  &  \\ 
Misspecified (\%) & 98.67\% & 95.84\% & 87.11\% & 68.65\% & 43.58\% & 18.37\%
\\ 
&  &  &  &  &  &  \\ 
Weakly Identified (\%) & 0.67\% & 2.08\% & 5.49\% & 15.40\% & 33.35\% & 
50.28\% \\ \hline
\end{tabular}
}
\end{table}

\subsubsection{Joint empirical density of $J$ and $IS$ statistics resulting
from the different specifications}

Because the $J$ and $IS$ statistics jointly indicate if the risk premia are
identified, the six pictures in Figure 6 show the bivariate empirical
density functions (histograms) of the $J$ and $IS$ statistics resulting from
all the specifications having one to six factors presented in Table 2.
Figure 7 shows the contour lines of these six empirical density functions.

\begin{figure}[htp]\centering
\subfigure[K=1]{\includegraphics[width=150 pt,  height=100 pt]{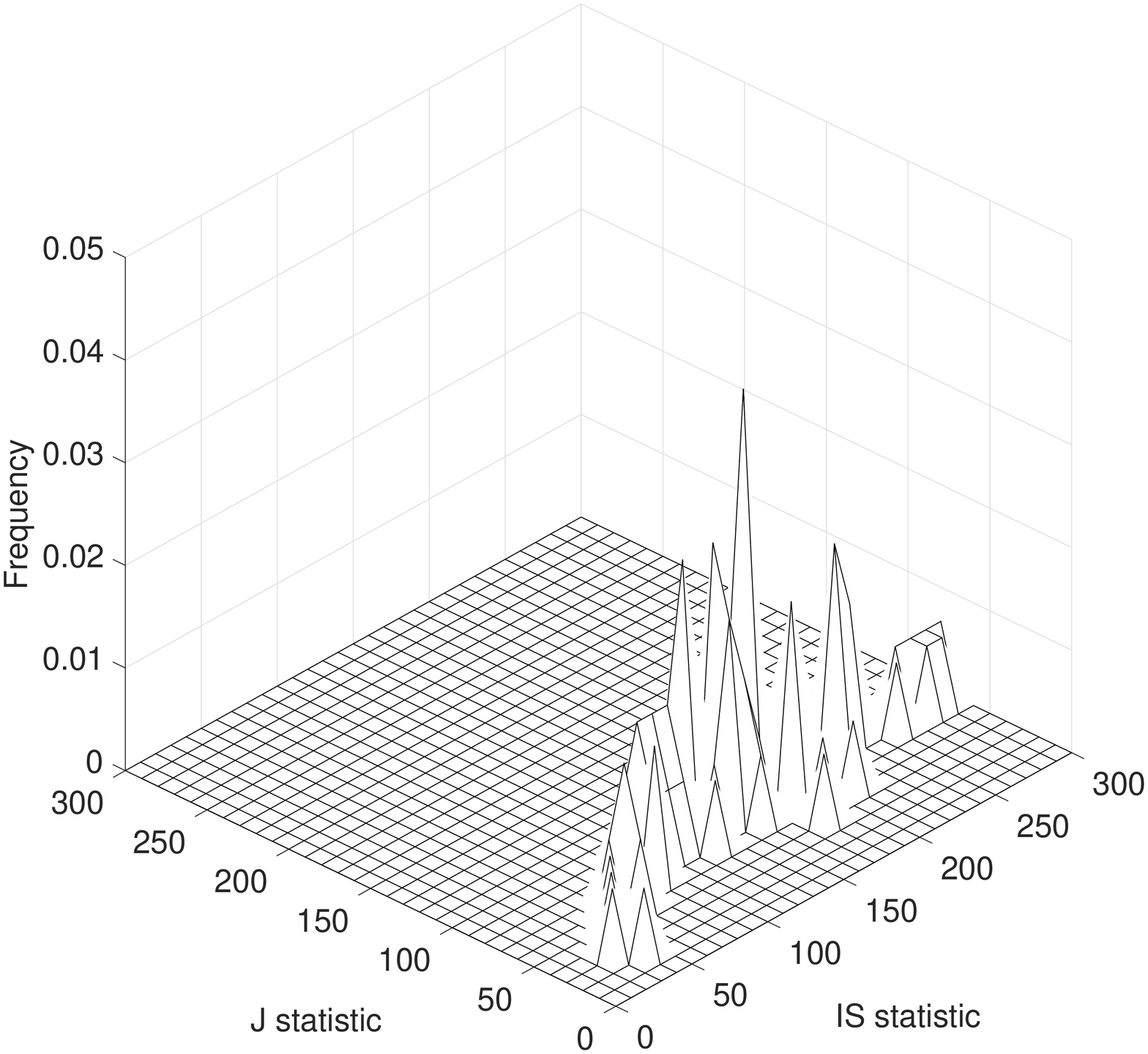}}
\subfigure[K=2]{\includegraphics[width=150 pt,  height=100 pt]{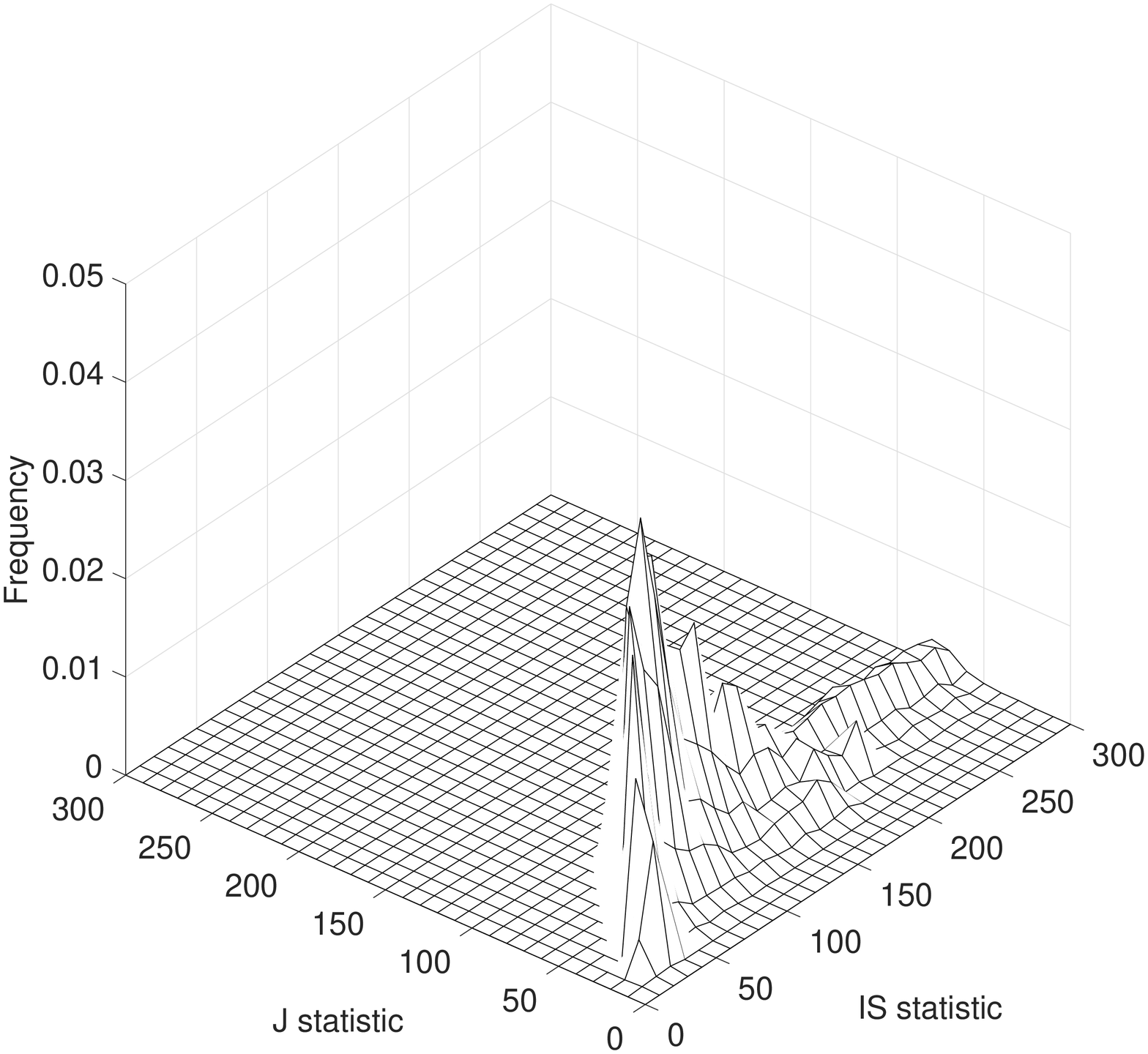}}
\subfigure[K=3]{\includegraphics[width=150 pt,  height=100 pt]{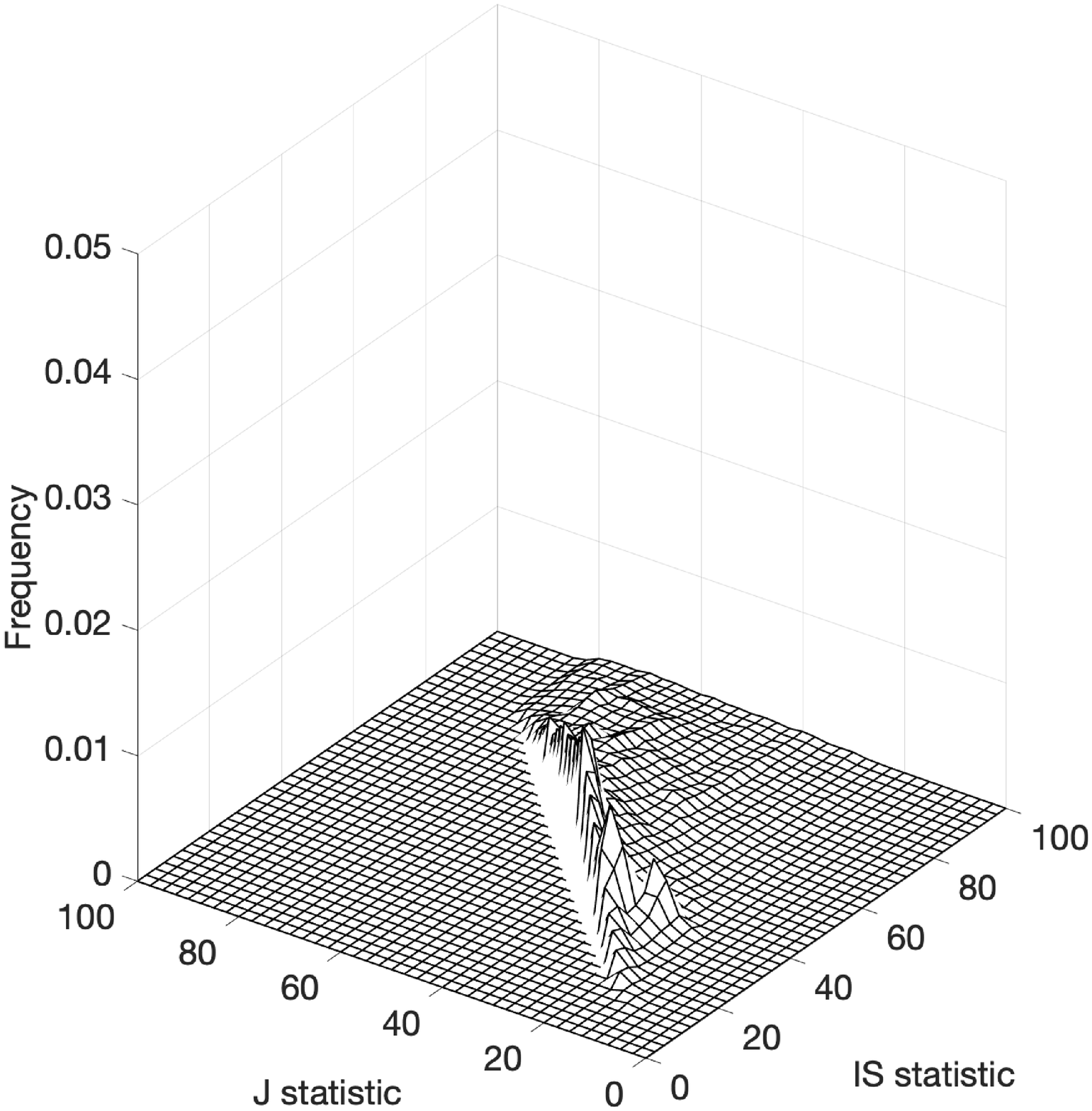}}
\subfigure[K=4]{\includegraphics[width=150 pt,  height=100 pt]{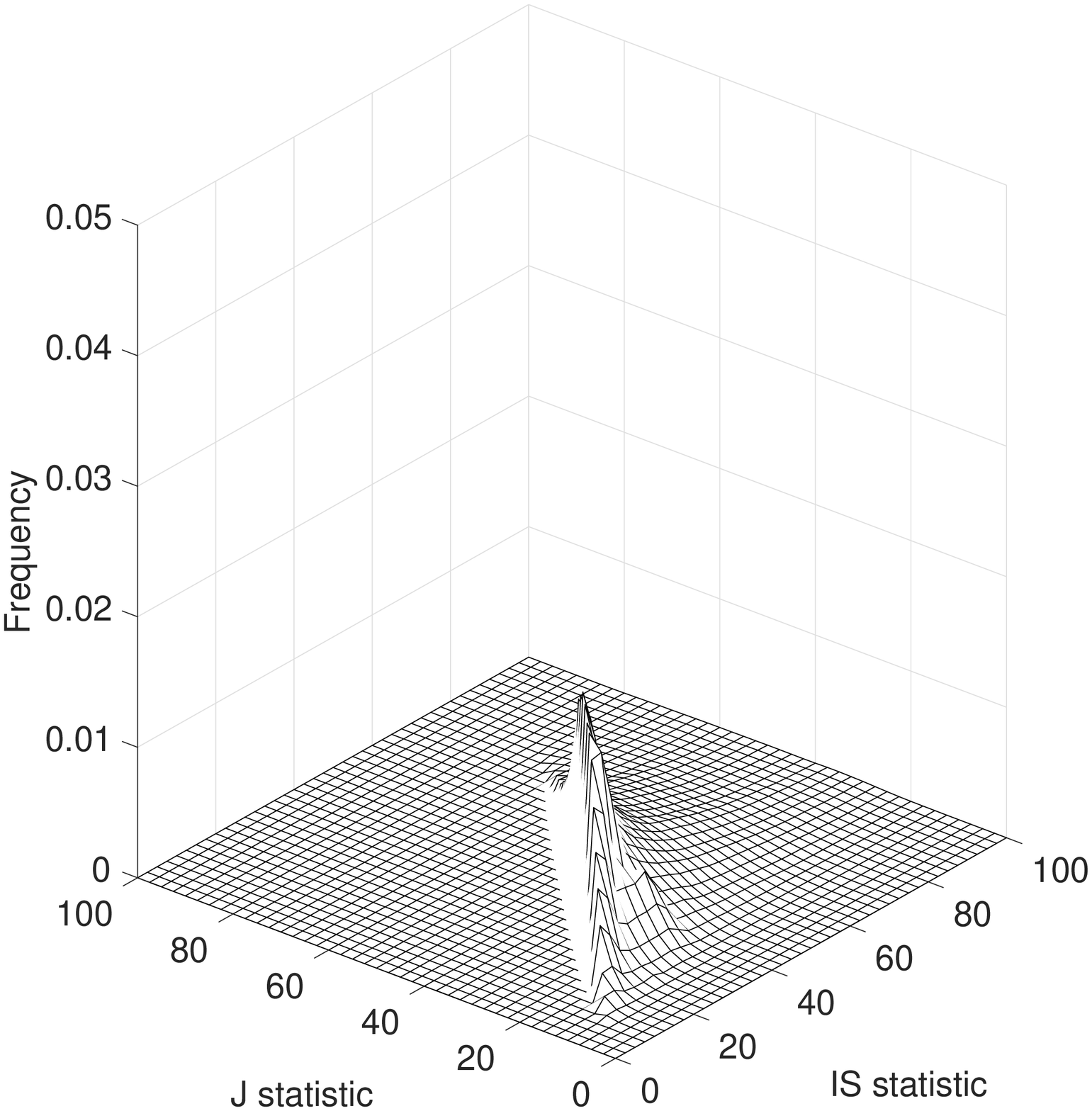}}
\subfigure[K=5]{\includegraphics[width=150 pt,  height=100 pt]{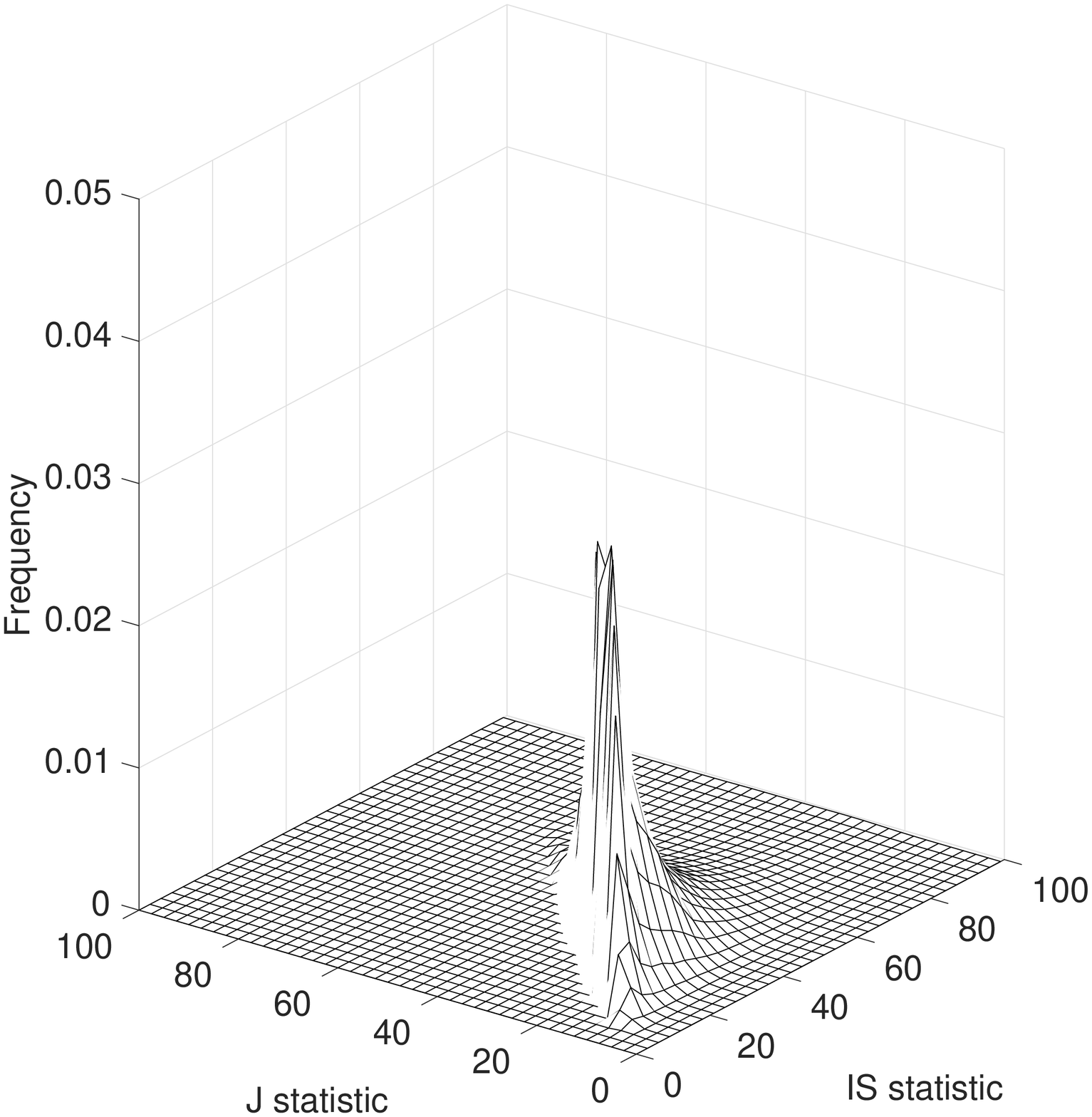}}
\subfigure[K=6]{\includegraphics[width=150 pt,  height=100 pt]{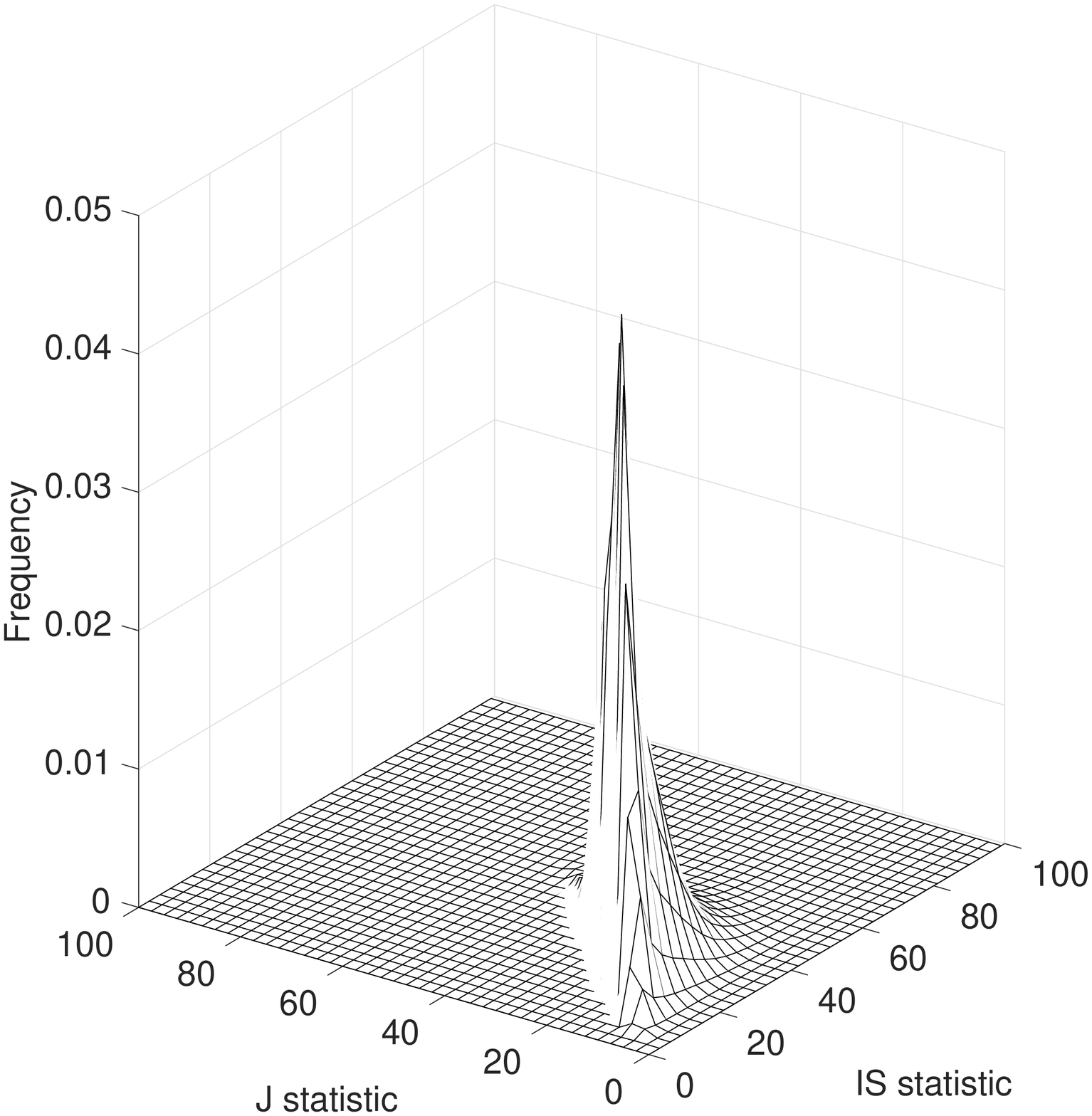}}
\caption{\textbf{Joint empirical density of $J$ and $IS$ statistics over the specifications with $K$ factors from the factor zoo.}}
\end{figure}

\begin{figure}[htp]\centering
\subfigure[K=1]{\includegraphics[width=150 pt,  height=100 pt]{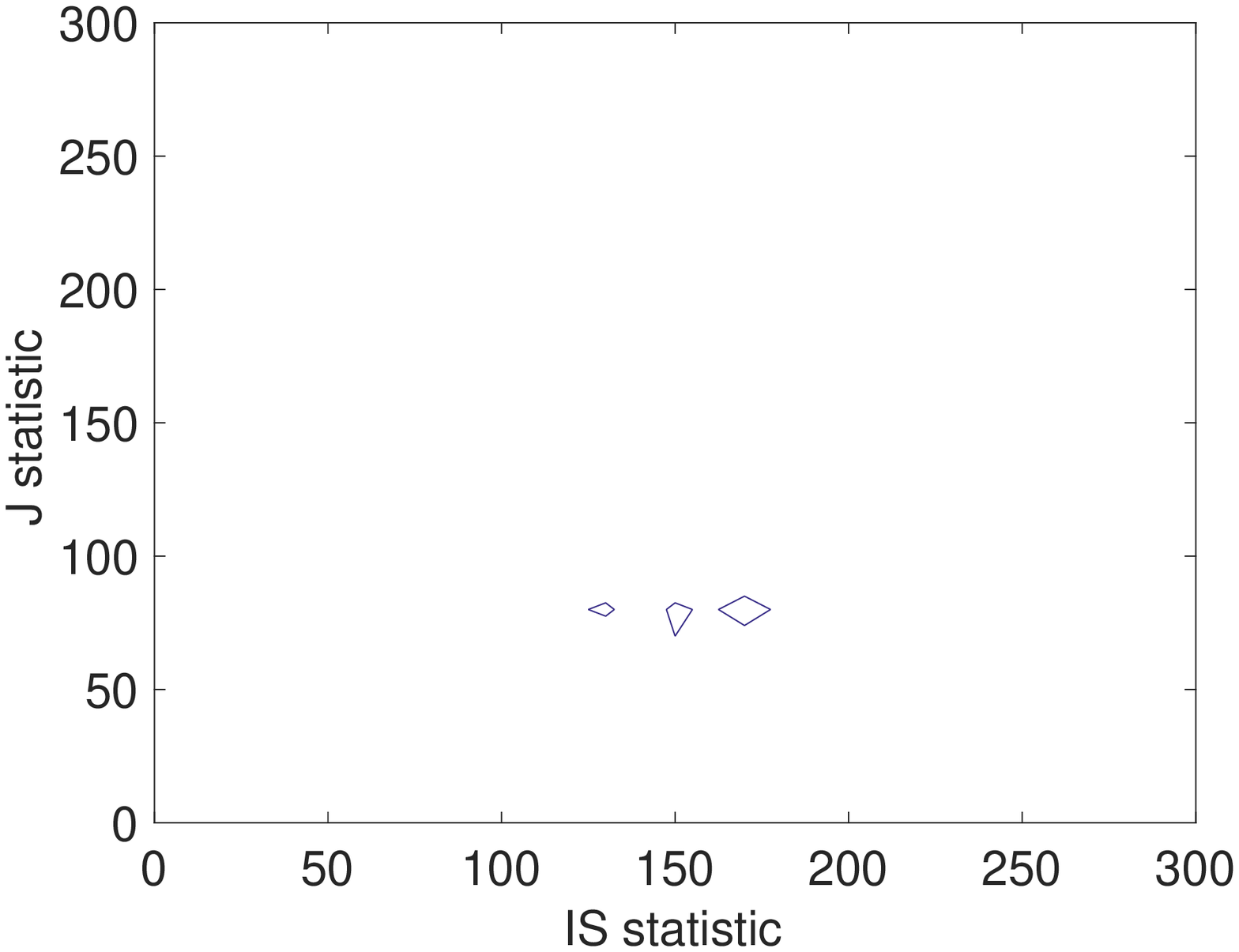}}
\subfigure[K=2]{\includegraphics[width=150 pt,  height=100 pt]{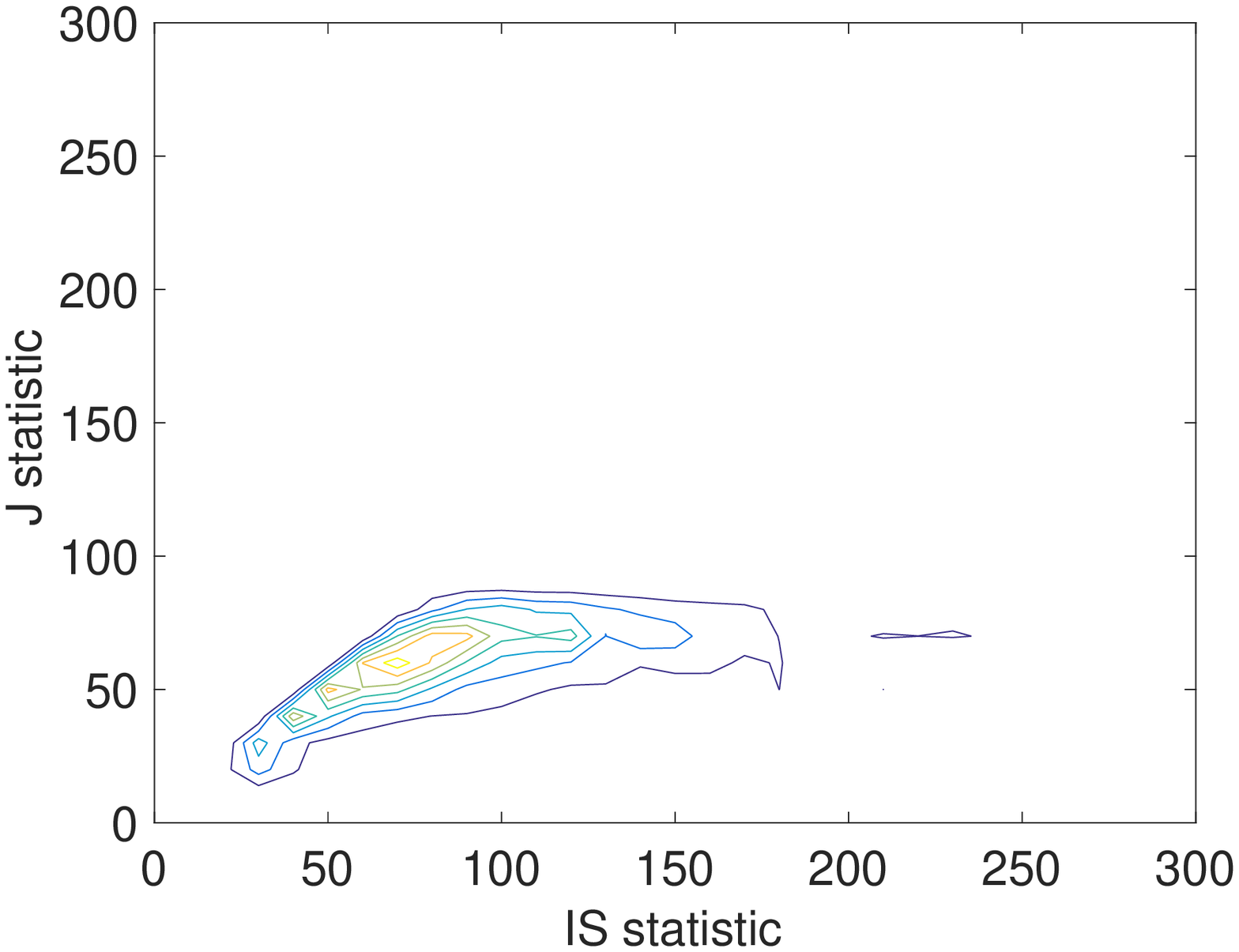}}
\subfigure[K=3]{\includegraphics[width=150 pt,  height=100 pt]{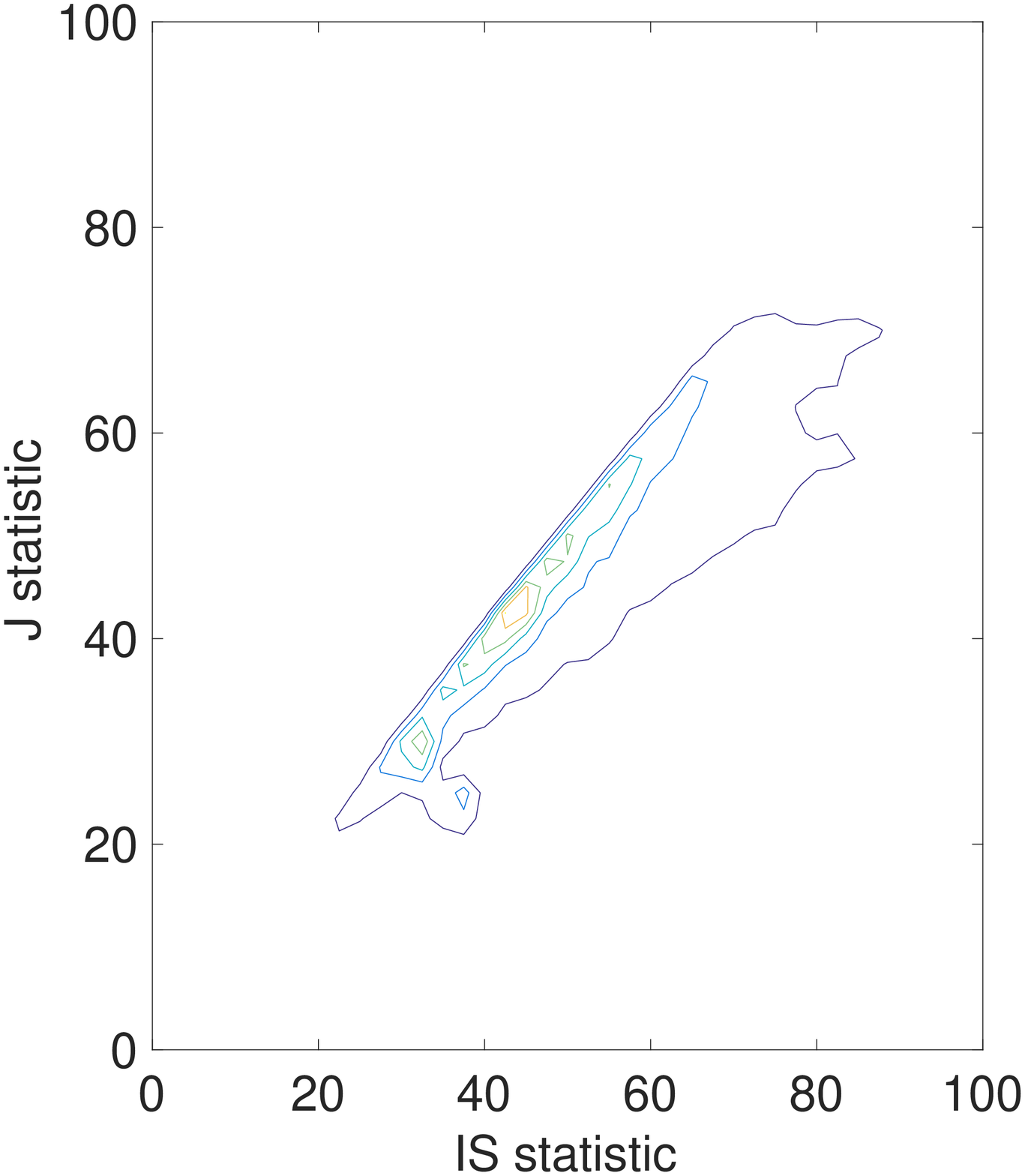}}
\subfigure[K=4]{\includegraphics[width=150 pt,  height=100 pt]{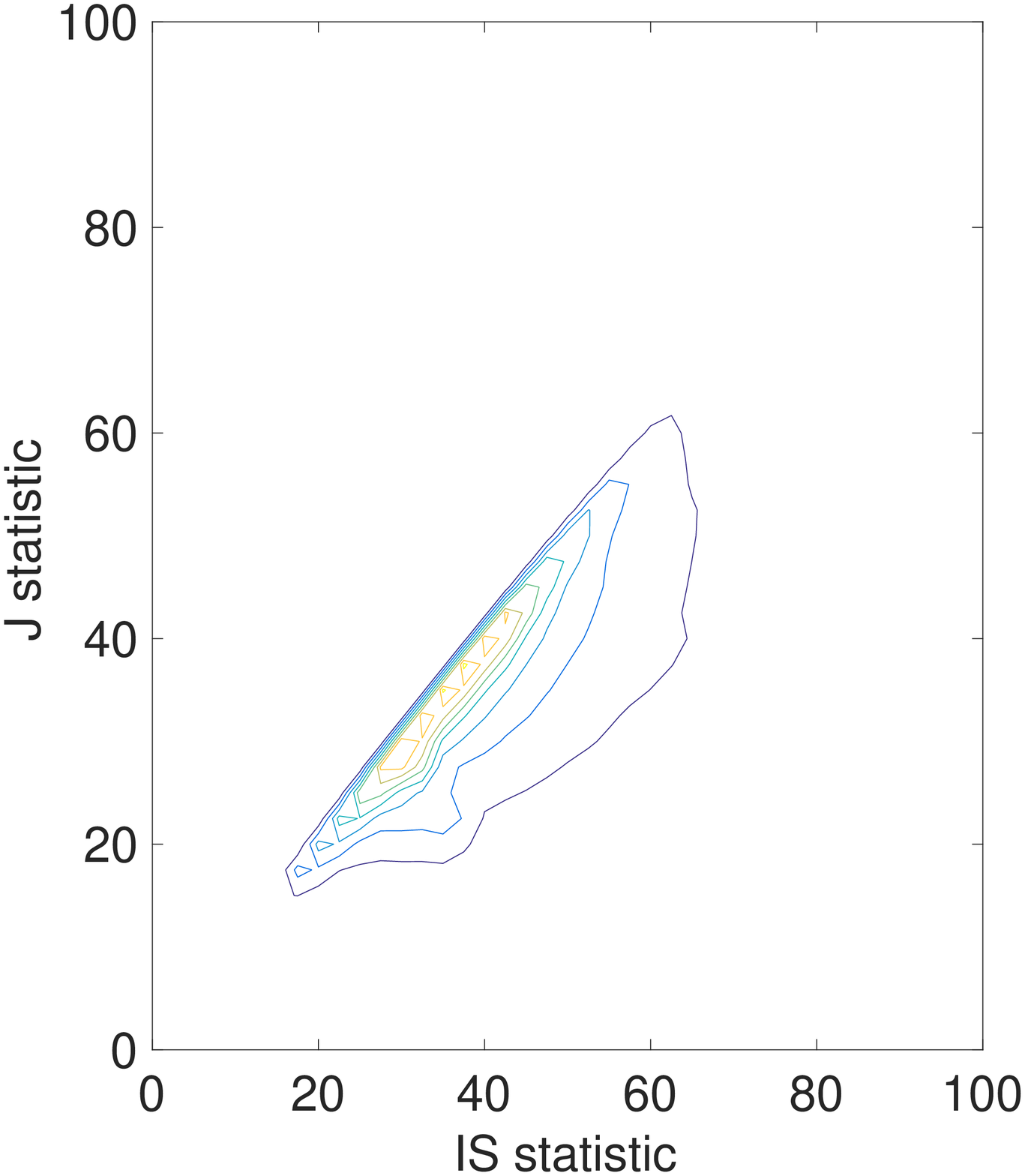}}
\subfigure[K=5]{\includegraphics[width=150 pt,  height=100 pt]{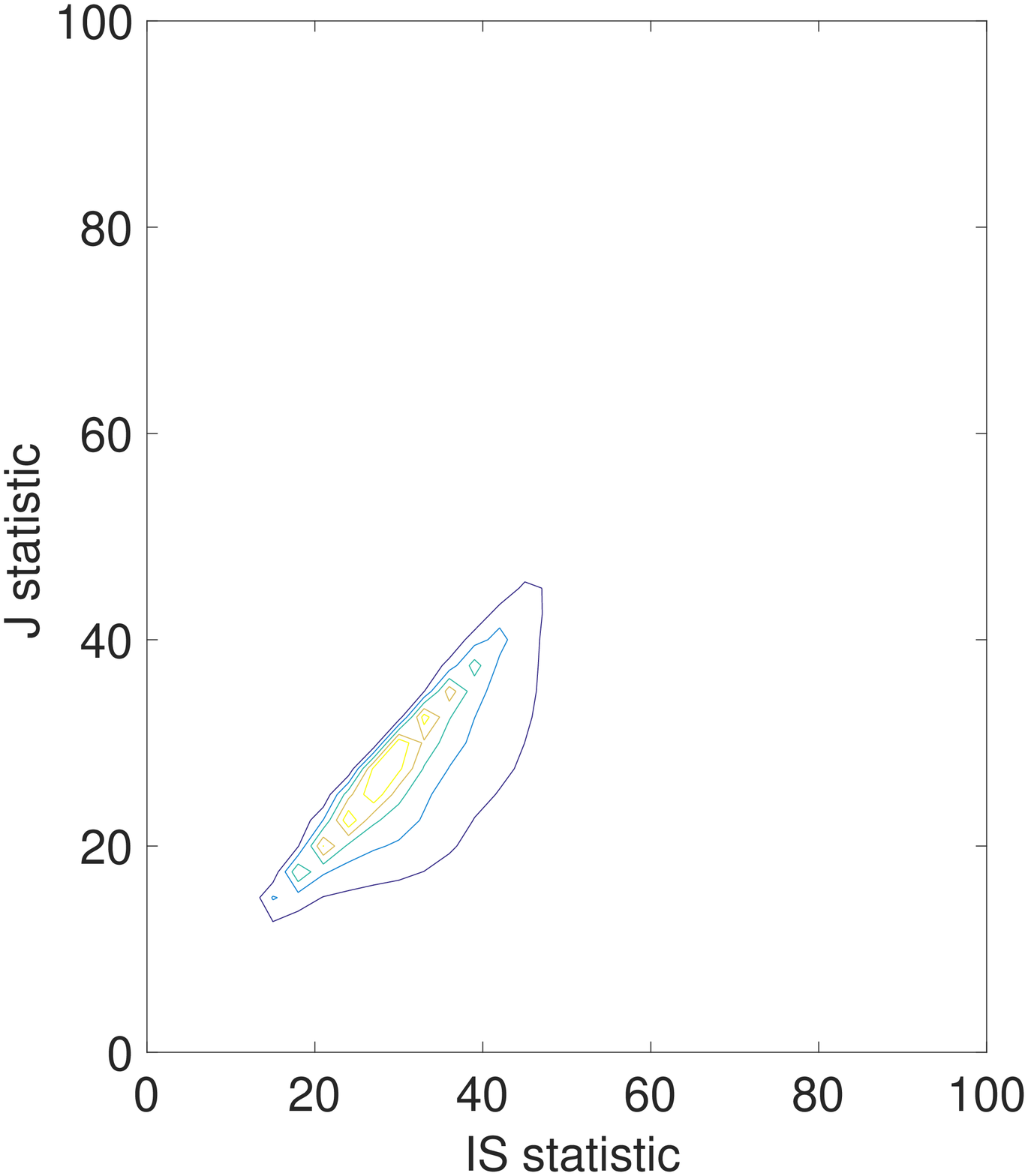}}
\subfigure[K=6]{\includegraphics[width=150 pt,  height=100 pt]{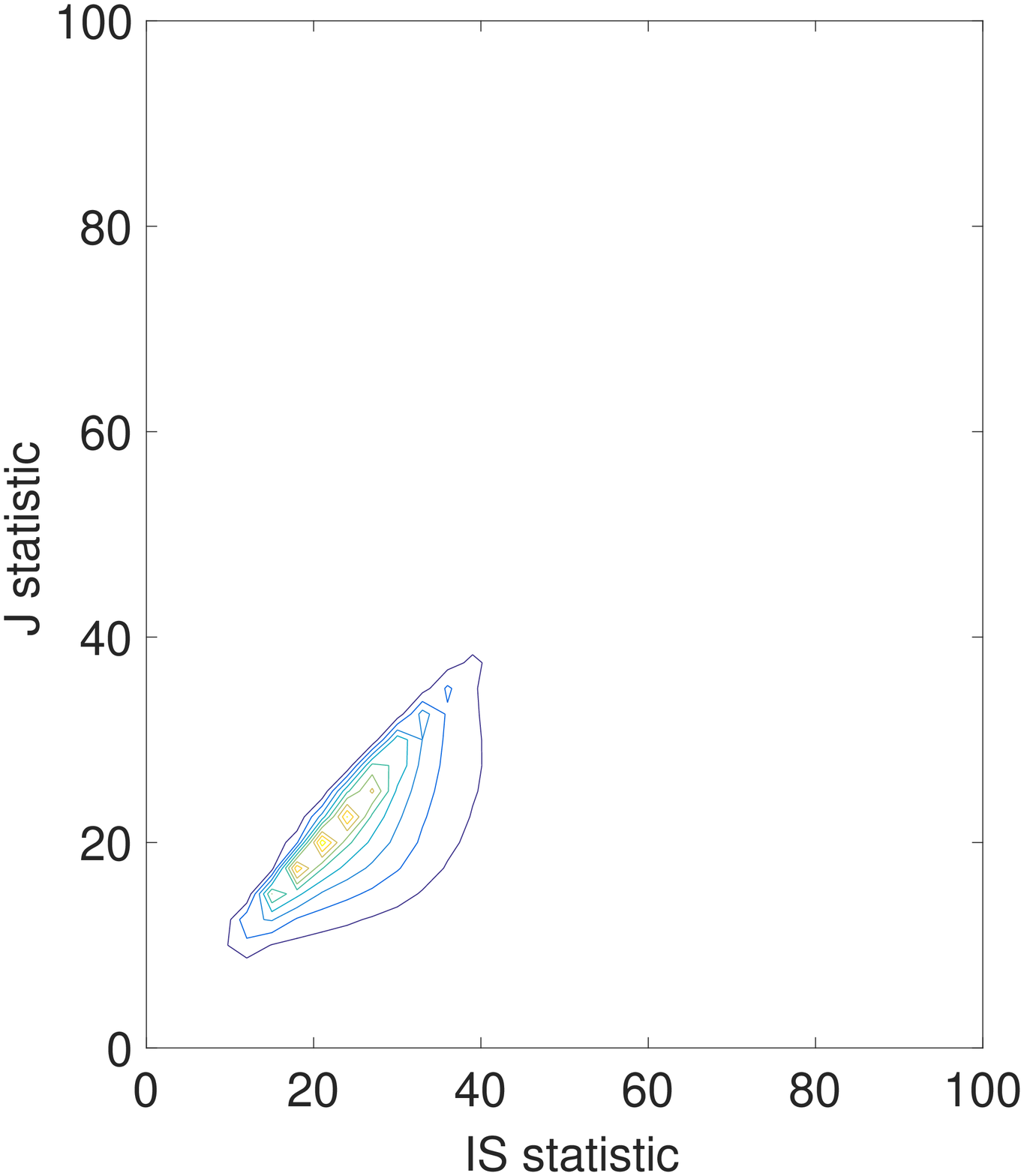}}
\caption{\textbf{Contour lines of the joint empirical density of $J$ and $IS$
statistics over the specifications with $K$  factors from the factor zoo.}}
\end{figure}

When $K=1$, Figures 6a and 7a show that almost all single factor models are
associated with large significant (at the 5\% level) $J$-statistics. This is
as expected, since it is unlikely that a single factor explains all the
variation in the cross-section of expected asset returns. Specifically, 148
out of 150 single factor models are deemed misspecified by their significant 
$J$-statistics at the 5\% level, leading to $148/150\approx 98.67\%$ in
Table \ref{t_k1}. The remaining two single factor models have small $J$%
-statistics (see the bottom left of Figure 6a), but their $IS$-statistics
are also small. One of these two $IS$-statistics is even insignificant at
the 5\% level, leading to the $1/150\approx 0.67\%$ for weak identification
in Table \ref{t_k1}, while the other is insignificant at the 1\% level.
Thus, these two factors are of poor quality, and their resulting single
factor models should also be deemed misspecified. The $J$-statistic fails to
signal misspecification for these two models, because the small $IS$%
-statistic forces it to be very small, given that the $J$-statistic is less
than or equal to the $IS$-statistic. Joint misspecification and weak
identification therefore occur in these two single factor models.

When we increase the number of factors so $K=2$, $3$, $4$, $5$ and $6,$
Figures 6b-f and 7b-f show two clear patterns:

\begin{enumerate}
\item The $J$ and $IS$ statistics are overall decreasing.

\item The empirical bivariate density of the $J$ and $IS\ $statistics moves
closer to the 45-degree line.
\end{enumerate}

While Table \ref{t_k1} shows that for $K=6,$ more than 50\% of the examined
specifications are weakly identified, the second pattern listed above
implies that many more specifications are weakly identified, and similarly
so for specifications involving fewer factors. This is analogous to what we
observed for Figure 5. Taken all together, it all shows that joint
misspecification and weak identification is a common problem that needs to
be addressed.

Encouragingly, Figures 6 and 7 also show that there are specifications for
which the ($IS$ , $J$) combination is distant from the 45-degree line. For
these specifications, the $IS$-statistics exceed their $J$-statistics to a
larger extent when compared to those in Figure 5. From a purely statistical
point of view, the specifications with $J$ much smaller than $IS$ are worth
investigation, since their pricing errors tend to be small while the risk
premia are likely to be well identified. It would be interesting if
researchers could further relate these models to economic theories. Yet
given the large number of such models as shown by Figure 6, we leave them
for future research.

\section{Application}

\label{section_app}

We show the importance of $J,$ $IS$ and DRLM statistics for applied asset
pricing by revisiting two prominent examples: the Fama and French (1993)
model and the conditional consumption capital asset pricing model from
Lettau and Ludvigson (2001). Table \ref{lambdaf_app} therefore reports the
estimation results for the three-factor specification used in Lettau and
Ludvigson (2001), and for Fama and French (1993) using three different data
sets. Accordingly, Figure 8 shows the joint 95\% confidence sets resulting
from DRLM\ for all four specifications in Table \ref{lambdaf_app}.

\begin{table}[tph]
\caption{\textbf{Risk Premia $\protect\lambda _{F}$ for the three-factor
models of Fama and French (1993) and Lettau and Ludvigson (2001)}}
\label{lambdaf_app}\justify The test assets are the 25 Fama-French
portfolios from 1963Q3 - 2013Q4 used in Lettau, Ludvigson, and Ma (2019) for
Panel A, from 1963Q3 - 1998Q3 used in Lettau and Ludvigson (2001) for Panel
B and Panel C, and from July 1963 - June 2021 downloaded from French's
website for Panel D, respectively. The estimates of $\lambda _{F, FM}$
result from the Fama and MacBeth (1973) two-pass procedure, and are
identical to those reported in Lettau, Ludvigson, and Ma (2019) and Lettau
and Ludvigson (2001) when using the same data. The presented FM $t$%
-statistic and its resulting 95\% confidence interval (C.I.) of risk premia
do not use the Shanken (1992) correction, while Shanken $t$ does. The KRS $t$
is computed by using the programs provided by Kan, Robotti, and Shanken
(2013). The zero-$\beta $ return is incorporated.
\par
\bigskip
\par
{\small \centering%
\begin{tabular}{lccccccc}
\hline
& \multicolumn{3}{c}{\footnotesize A: Lettau, Ludvigson, and Ma (2019)} &  & 
\multicolumn{3}{c}{\footnotesize B: Lettau and Ludvigson (2001)} \\ 
\cline{2-4}\cline{6-8}
& $R_{m}$ & SMB & HML &  & $R_{m}$ & SMB & HML \\ \hline
$\lambda _{F,FM}$ & -1.96 & 0.70 & 1.35 &  & 1.33 & 0.47 & 1.46 \\ 
$\lambda _{F,CUE}$ & -4.15 & 0.82 & 0.86 &  & -11.26 & 0.69 & 1.52 \\ 
&  &  &  &  &  &  &  \\ 
FM $t$ & -1.72 & 1.67 & 2.64 &  & 0.83 & 0.94 & 3.24 \\ 
95\% C.I. & {\footnotesize (-4.18, 0.27)} & {\footnotesize (-0.12, 1.52)} & 
{\footnotesize (0.35, 2.35)} &  & {\footnotesize (-1.81, 4.46)} & 
{\footnotesize (-0.51, 1.45)} & {\footnotesize (0.58, 2.34)} \\ 
Shanken $t$ & -1.64 & 1.66 & 2.60 &  & 0.78 & 0.94 & 3.22 \\ 
95\% C.I. & {\footnotesize (-4.29, 0.38)} & {\footnotesize (-0.13, 1.52)} & 
{\footnotesize (0.33, 2.37)} &  & {\footnotesize (-2.02, 4.68)} & 
{\footnotesize (-0.51, 1.45)} & {\footnotesize (0.57, 2.35)} \\ 
KRS $t$ & -1.33 & 1.65 & 2.52 &  & 0.63 & 0.96 & 3.26 \\ 
95\% C.I. & {\footnotesize (-4.83, 0.92)} & {\footnotesize (-0.13, 1.53)} & 
{\footnotesize (0.30, 2.40)} &  & {\footnotesize (-2.78, 5.44)} & 
{\footnotesize (-0.49, 1.43)} & {\footnotesize (0.58, 2.34)} \\ 
&  &  &  &  &  &  &  \\ 
$R^{2}$ & \multicolumn{3}{c}{0.73} &  & \multicolumn{3}{c}{0.80} \\ 
$J$-statistic & \multicolumn{3}{c}{59.34} &  & \multicolumn{3}{c}{45.38} \\ 
$IS$-statistic & \multicolumn{3}{c}{106.81} &  & \multicolumn{3}{c}{47.01}
\\ 
&  &  &  &  &  &  &  \\ 
& \multicolumn{3}{c}{\footnotesize C: Lettau and Ludvigson (2001)} &  & 
\multicolumn{3}{c}{\footnotesize D: Monthly data from July 1963 - June 2021}
\\ \cline{2-4}\cline{6-8}
& $\triangle c$ & cay & $\triangle c\times cay$ &  & $R_{m}$ & SMB & HML \\ 
\hline
$\lambda _{F,FM}$ & 0.02 & -0.13 & 0.06 &  & -0.53 & 0.13 & 0.34 \\ 
$\lambda _{F,CUE}$ & -1.45 & -3.80 & 0.01 &  & -0.52 & 0.19 & 0.32 \\ 
&  &  &  &  &  &  &  \\ 
FM $t$ & 0.20 & -0.43 & 3.12 &  & -1.75 & 1.09 & 2.93 \\ 
95\% C.I. & {\footnotesize (-0.20, 0.25)} & {\footnotesize (-0.70, 0.45)} & 
{\footnotesize (0.02, 0.09)} &  & {\footnotesize (-1.11, 0.06)} & 
{\footnotesize (-0.11, 0.37)} & {\footnotesize (0.11, 0.57)} \\ 
Shanken $t$ & 0.15 & -0.31 & 2.25 &  & -1.74 & 1.09 & 2.92 \\ 
95\% C.I. & {\footnotesize (-0.29, 0.34)} & {\footnotesize (-0.93, 0.68)} & 
{\footnotesize (0.01, 0.11)} &  & {\footnotesize (-1.12, 0.07)} & 
{\footnotesize (-0.11, 0.38)} & {\footnotesize (0.11, 0.57)} \\ 
KRS $t$ & 0.08 & -0.27 & 1.95 &  & -1.67 & 1.09 & 2.94 \\ 
95\% C.I. & {\footnotesize (-0.53, 0.58)} & {\footnotesize (-1.03, 0.78)} & 
{\footnotesize (-0.00, 0.11)} &  & {\footnotesize (-1.14, 0.09)} & 
{\footnotesize (-0.11, 0.37)} & {\footnotesize (0.11, 0.57)} \\ 
&  &  &  &  &  &  &  \\ 
$R^{2}$ & \multicolumn{3}{c}{0.70} &  & \multicolumn{3}{c}{0.75} \\ 
$J$-statistic & \multicolumn{3}{c}{31.11} &  & \multicolumn{3}{c}{52.25} \\ 
$IS$-statistic & \multicolumn{3}{c}{31.75} &  & \multicolumn{3}{c}{425.55}
\\ \hline
\end{tabular}
}
\end{table}

\subsection{Fama and French (1993)}

The Fama and French (1993) three-factor model has been widely used as a
benchmark in the asset pricing literature; see, e.g., Lettau and Ludvigson
(2001) and Lettau, Ludvigson, and Ma (2019). In line with the large,
significant (at the 1\% level) $J$-statistics in Table \ref{lambdaf_app},
the Fama and French (1993) three-factor model is well acknowledged to be
misspecified. Yet this model appears able to explain the cross-section of
asset returns, as reflected by the reported large cross-sectional $R^{2}$.

Using quarterly data from Lettau and Ludvigson (2001) and Lettau, Ludvigson,
and Ma (2019), the two-pass procedure of Fama and MacBeth (1973) suggests
that the risk premia on SMB and HML are both positive; in contrast, the
point estimate of the market premium is negative in Lettau, Ludvigson, and
Ma (2019), but positive in Lettau and Ludvigson (2001); see Panel A versus
Panel B in Table \ref{lambdaf_app}. Because Lettau, Ludvigson, and Ma (2019)
use longer time series than Lettau and Ludvigson (2001), its $IS$-statistic
is considerably larger (106.81 vs. 47.01), reflecting that more information
is available for identification in Lettau, Ludvigson, and Ma (2019). On the
other hand, the larger $J$-statistic for Lettau, Ludvigson, and Ma (2019)
(59.34 vs. 45.38) implies more severe misspecification. A question thus
arises: how do we reconcile the seemingly conflicting findings in Lettau,
Ludvigson, and Ma (2019) and Lettau and Ludvigson (2001) for $R_{m}$ in the
Fama and French (1993) three-factor model?

\begin{figure}[htp]\centering
\subfigure[Data: Lettau, Ludvigson, and Ma (2019)]{\includegraphics[width=200 pt,  height=150 pt]{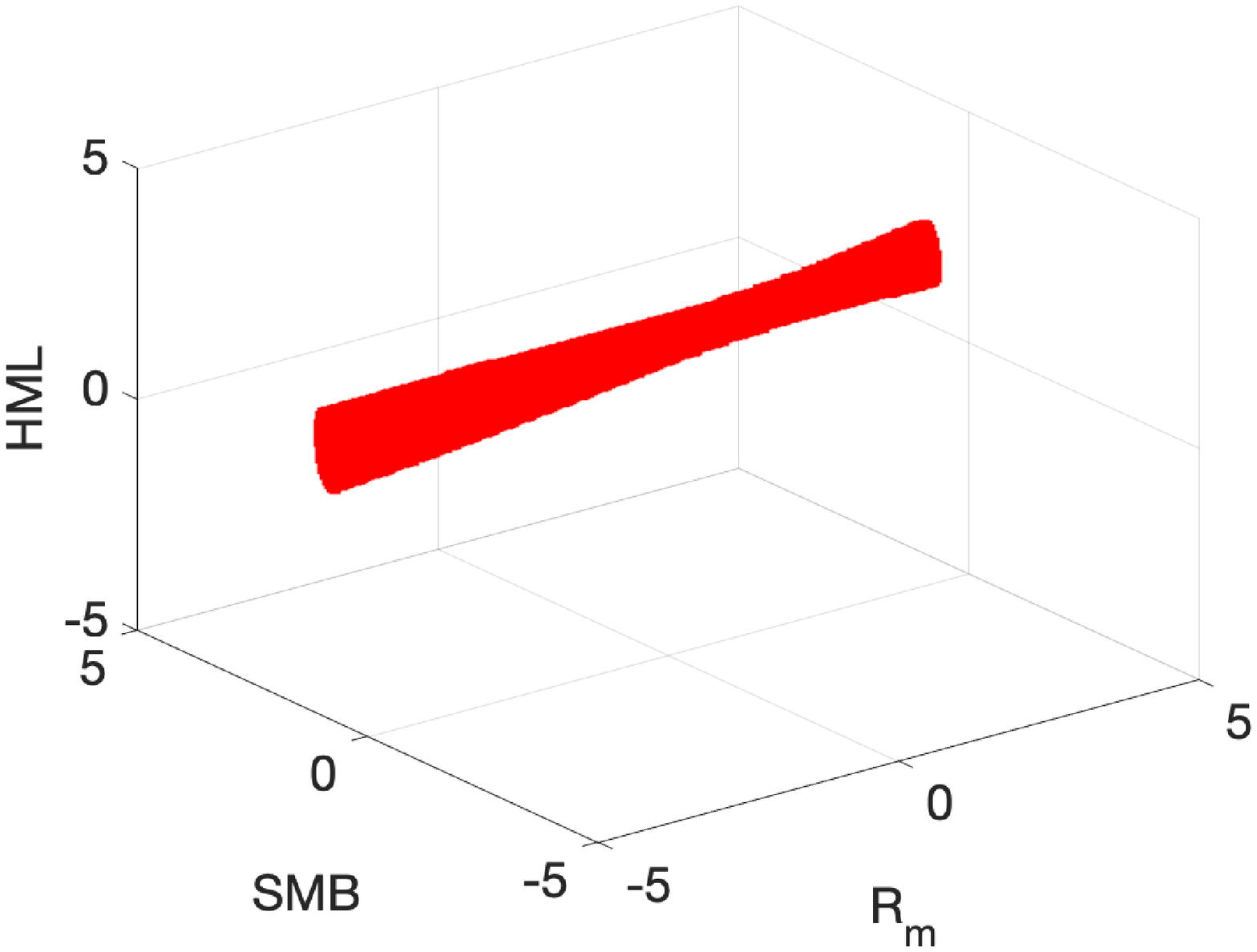}}
\subfigure[Data: Lettau and Ludvigson (2001)]{\includegraphics[width=200 pt,  height=150 pt]{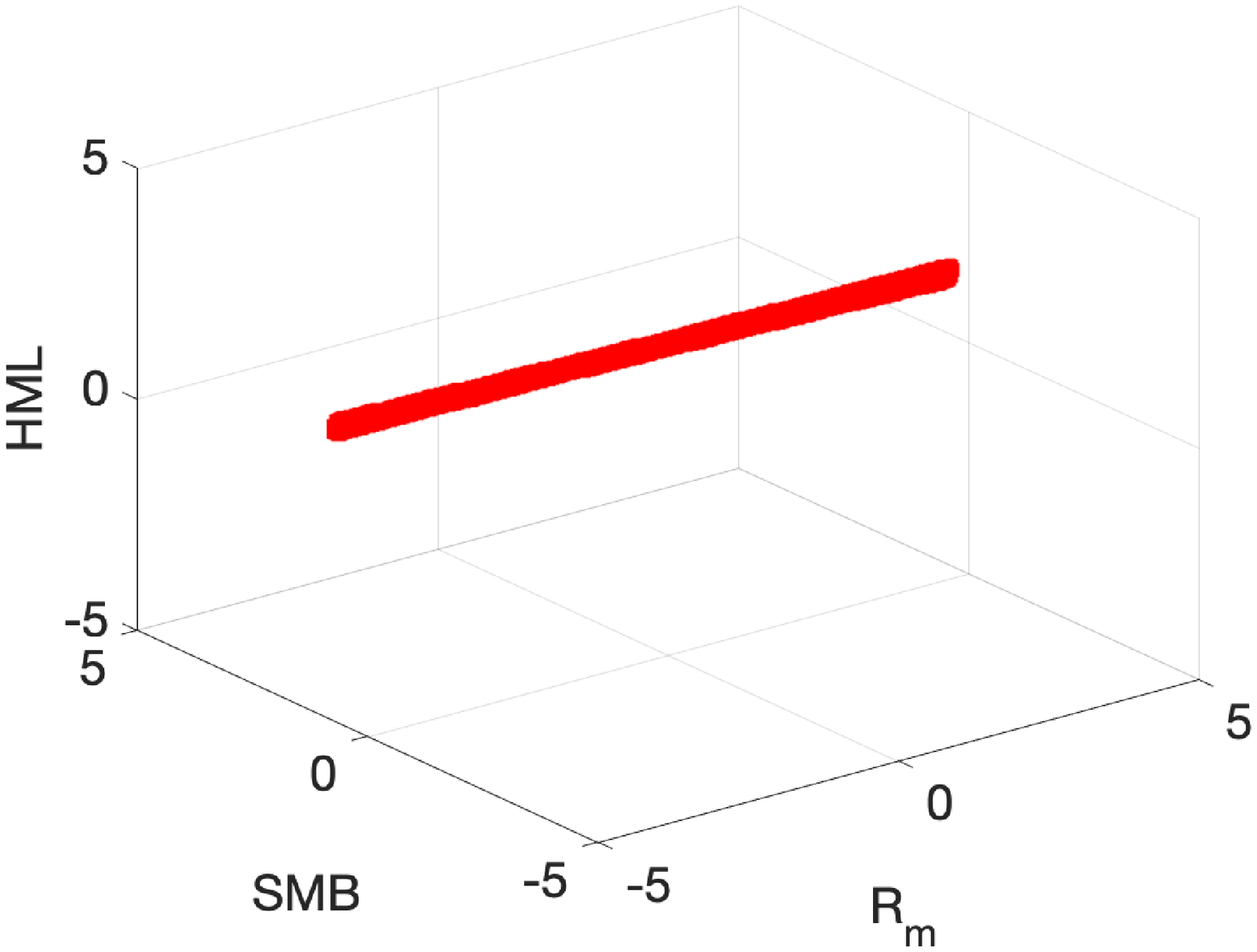}}
\subfigure[Data: Lettau and Ludvigson (2001)]{\includegraphics[width=200 pt,  height=150 pt]{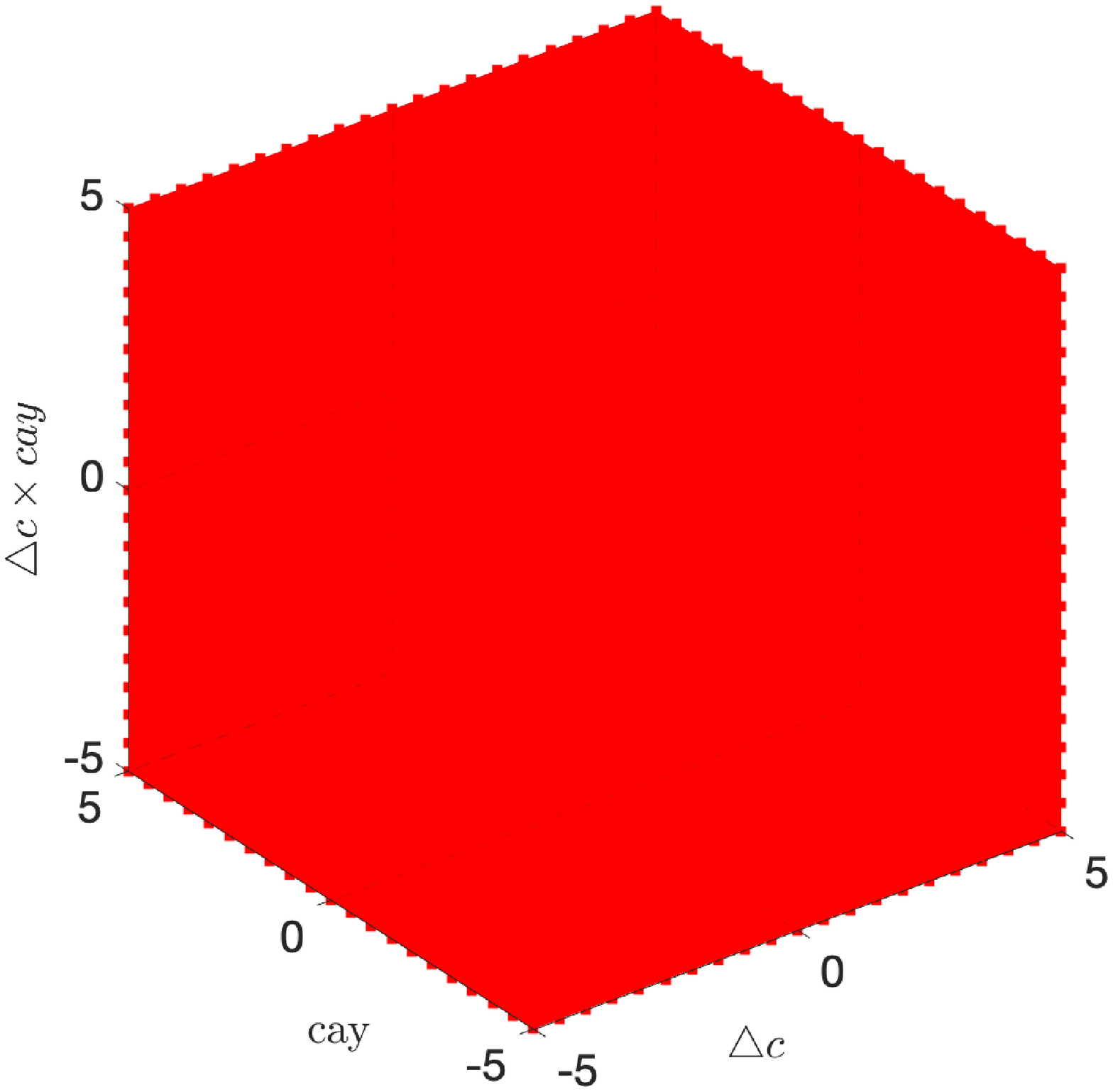}}
\subfigure[Data: Monthly data, July 1963 - June 2021]{\includegraphics[width=200 pt,  height=150 pt]{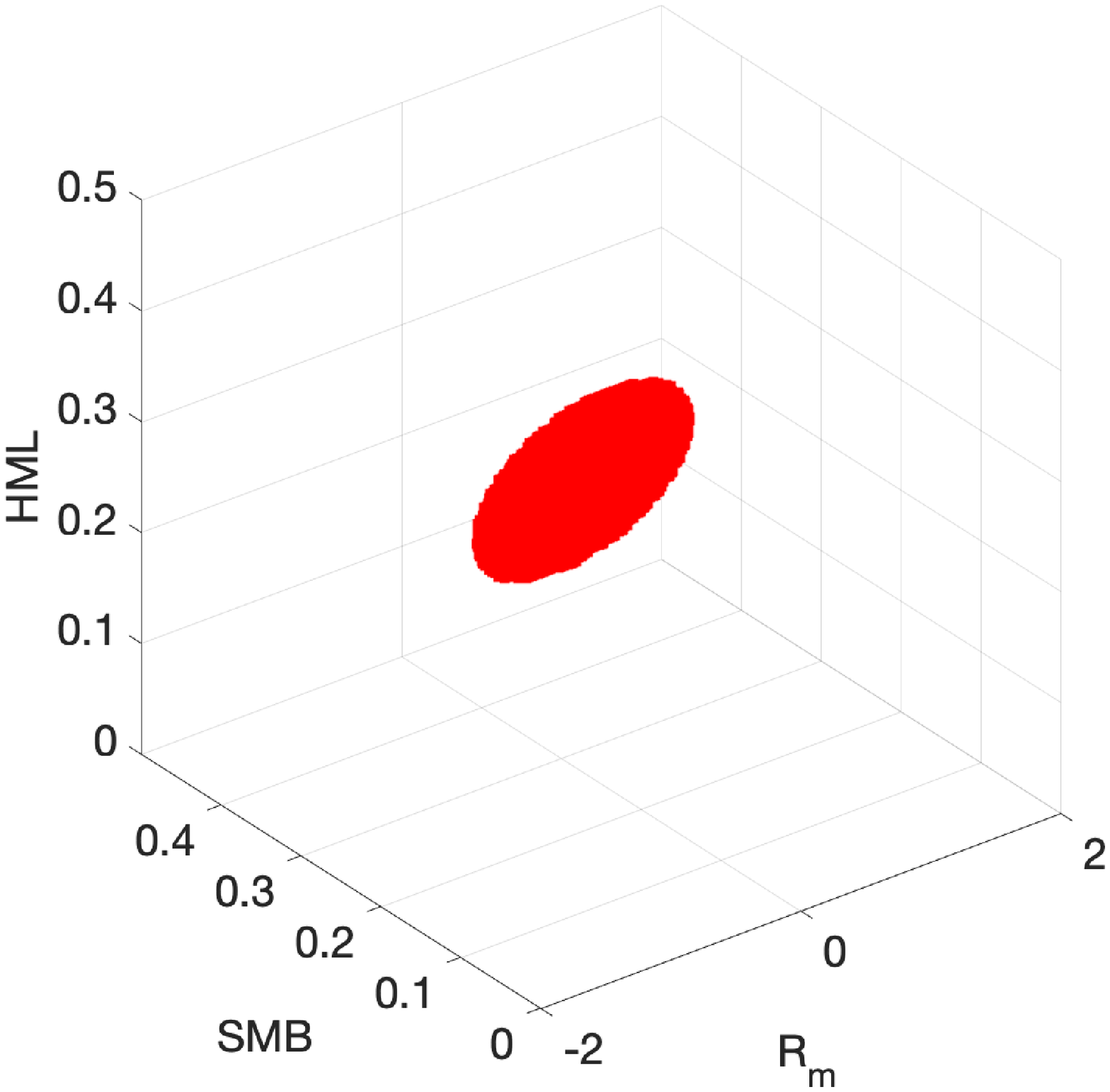}}
\caption{\textbf{Joint 95\% confidence sets of risk premia from the DRLM test.}}
\justify{\small Notes: The red region consists of risk premia values that are
not  rejected by the DRLM test at the 5\% significance level. The
test assets are the twenty-five Fama-French portfolios  taken from Lettau, Ludvigson, and Ma (2019) over 1963Q3 -
2013Q4 for (a), Lettau and Ludvigson (2001) over  1963Q3 - 1998Q3 for (b) and (c), and monthly data from July
1963 - June 2021 downloaded from French's website for  (d), respectively. (a)(b)(c)(d) correspond to Panels A, B, C,
D of Table 3. The zero-$\beta$ return is incorporated.}
\end{figure}

\doublespace

\newpage

To help answer this question, we employ the DRLM test, since this test is
robust to misspecification as well as the strength of identification.
Specifically, we test every risk premium value between -5 and 5, since Table %
\ref{lambdaf_app} Panel A and Panel B indicate that risk premia on all three
factors likely lie within this range. The resulting joint confidence sets at
the 5\% level are presented in Figure 8a-b, using the data from Lettau,
Ludvigson, and Ma (2019) and Lettau and Ludvigson (2001), respectively.

Interestingly, Figure 8a-b show that the risk premia on SMB and HML are
strongly identified, while the risk premium on $R_{m}$ is not. These
findings remain similar no matter whether we adopt the data from Lettau,
Ludvigson, and Ma (2019) or Lettau and Ludvigson (2001). In both Figure
8a-b, we can not reject any risk premium value on $R_{m}$ between -5 and 5,
which thus helps reconcile the difference in the confidence intervals for
the risk premium on $R_{m}$ reported in Table \ref{lambdaf_app} Panel A and
Panel B.

For SMB and HML, however, the DRLM test yields tight 95\% confidence sets
for their risk premia in Figure 8. For example, Figure 8a implies that the
range of the risk premium on SMB is (0.59, 1.06), while the range of the
risk premium on HML is (0.04, 2.00). Similar ranges can be derived from
Figure 8b. All these sets thus largely overlap with those resulting from the
FM $t$-test as reported in Table \ref{lambdaf_app} Panel A and Panel B.


\begin{table}[tph]
\caption{\textbf{$\protect\beta $ with 25 portfolio returns}}
\label{beta}\justify The test assets are the 25 Fama-French portfolios. For $%
R_{m}$, SMB, and HML, we use the data from 1963Q3 - 2013Q4 used by Lettau,
Ludvigson, and Ma (2019). For $\triangle c$, $cay$, $\triangle c\times cay$,
we use the data from 1963Q3 - 1998Q3 used by Lettau and Ludvigson (2001).
The reported beta estimates and their associated $t$-statistics result from
the first pass time-series regression of the Fama and MacBeth (1973)
methodology; see also Kleibergen, Kong, and Zhan (2020) for the same results
reported for Lettau and Ludvigson (2001).\nocite{kkz19}
\par
\bigskip \centering
{\small \centering%
\begin{tabular}{lrrcccc|cccccr}
\hline
& \multicolumn{2}{c}{$R_{m}$} & \multicolumn{2}{c}{SMB} & \multicolumn{2}{c}{
HML} & \multicolumn{2}{c}{$\triangle c$} & \multicolumn{2}{c}{$cay$} & 
\multicolumn{2}{c}{$\triangle c\times cay$} \\ \cline{2-13}
& $\hat{\beta}$ & $t$-stat & $\hat{\beta}$ & $t$-stat & $\hat{\beta}$ & $t$%
-stat & $\hat{\beta}$ & $t$-stat & $\hat{\beta}$ & $t$-stat & $\hat{\beta}$
& $t$-stat \\ \hline
{\footnotesize (1)} & {\footnotesize 1.08} & {\footnotesize 26.50} & 
{\footnotesize 1.50} & {\footnotesize 25.11} & {\footnotesize -0.31} & 
{\footnotesize -6.36} & {\footnotesize 6.35} & {\footnotesize 2.32} & 
{\footnotesize 4.25} & {\footnotesize 2.96} & {\footnotesize -4.56} & 
{\footnotesize -0.22} \\ 
{\footnotesize (2)} & {\footnotesize 0.94} & {\footnotesize 32.73} & 
{\footnotesize 1.33} & {\footnotesize 31.52} & {\footnotesize 0.02} & 
{\footnotesize 0.69} & {\footnotesize 6.30} & {\footnotesize 2.61} & 
{\footnotesize 3.66} & {\footnotesize 2.90} & {\footnotesize 2.25} & 
{\footnotesize 0.12} \\ 
{\footnotesize (3)} & {\footnotesize 0.86} & {\footnotesize 29.98} & 
{\footnotesize 1.14} & {\footnotesize 27.08} & {\footnotesize 0.19} & 
{\footnotesize 5.56} & {\footnotesize 5.11} & {\footnotesize 2.28} & 
{\footnotesize 3.22} & {\footnotesize 2.75} & {\footnotesize 3.22} & 
{\footnotesize 0.19} \\ 
{\footnotesize (4)} & {\footnotesize 0.81} & {\footnotesize 26.16} & 
{\footnotesize 1.10} & {\footnotesize 24.32} & {\footnotesize 0.29} & 
{\footnotesize 8.02} & {\footnotesize 5.50} & {\footnotesize 2.58} & 
{\footnotesize 2.96} & {\footnotesize 2.65} & {\footnotesize 5.12} & 
{\footnotesize 0.32} \\ 
{\footnotesize (5)} & {\footnotesize 0.94} & {\footnotesize 28.29} & 
{\footnotesize 1.17} & {\footnotesize 24.04} & {\footnotesize 0.57} & 
{\footnotesize 14.37} & {\footnotesize 5.79} & {\footnotesize 2.53} & 
{\footnotesize 2.65} & {\footnotesize 2.21} & {\footnotesize 11.36} & 
{\footnotesize 0.65} \\ 
{\footnotesize (6)} & {\footnotesize 1.11} & {\footnotesize 36.22} & 
{\footnotesize 1.05} & {\footnotesize 23.49} & {\footnotesize -0.33} & 
{\footnotesize -8.97} & {\footnotesize 4.38} & {\footnotesize 1.76} & 
{\footnotesize 4.55} & {\footnotesize 3.49} & {\footnotesize -15.82} & 
{\footnotesize -0.84} \\ 
{\footnotesize (7)} & {\footnotesize 0.94} & {\footnotesize 36.04} & 
{\footnotesize 0.97} & {\footnotesize 25.32} & {\footnotesize 0.05} & 
{\footnotesize 1.48} & {\footnotesize 3.63} & {\footnotesize 1.65} & 
{\footnotesize 3.25} & {\footnotesize 2.83} & {\footnotesize 0.02} & 
{\footnotesize 0.00} \\ 
{\footnotesize (8)} & {\footnotesize 0.92} & {\footnotesize 33.18} & 
{\footnotesize 0.76} & {\footnotesize 18.80} & {\footnotesize 0.22} & 
{\footnotesize 6.81} & {\footnotesize 3.92 } & {\footnotesize 1.97} & 
{\footnotesize 3.07} & {\footnotesize 2.95} & {\footnotesize -1.08} & 
{\footnotesize -0.07} \\ 
{\footnotesize (9)} & {\footnotesize 0.91} & {\footnotesize 29.94} & 
{\footnotesize 0.68} & {\footnotesize 15.32} & {\footnotesize 0.42} & 
{\footnotesize 11.67} & {\footnotesize 3.52} & {\footnotesize 1.91} & 
{\footnotesize 2.66} & {\footnotesize 2.76} & {\footnotesize 6.93} & 
{\footnotesize 0.50} \\ 
{\footnotesize (10)} & {\footnotesize 0.98} & {\footnotesize 30.95} & 
{\footnotesize 0.81} & {\footnotesize 17.55} & {\footnotesize 0.66} & 
{\footnotesize 17.47} & {\footnotesize 4.83} & {\footnotesize 2.41} & 
{\footnotesize 2.16} & {\footnotesize 2.06} & {\footnotesize 8.48} & 
{\footnotesize 0.56} \\ 
{\footnotesize (11)} & {\footnotesize 1.08} & {\footnotesize 39.19} & 
{\footnotesize 0.76} & {\footnotesize 18.65} & {\footnotesize -0.39} & 
{\footnotesize -11.96} & {\footnotesize 2.70} & {\footnotesize 1.21} & 
{\footnotesize 4.53} & {\footnotesize 3.88} & {\footnotesize -22.20} & 
{\footnotesize -1.31} \\ 
{\footnotesize (12)} & {\footnotesize 0.99} & {\footnotesize 35.98} & 
{\footnotesize 0.60} & {\footnotesize 14.70} & {\footnotesize 0.07} & 
{\footnotesize 2.16} & {\footnotesize 2.76} & {\footnotesize 1.47} & 
{\footnotesize 3.64} & {\footnotesize 3.69} & {\footnotesize -3.35} & 
{\footnotesize -0.23} \\ 
{\footnotesize (13)} & {\footnotesize 0.91} & {\footnotesize 29.02} & 
{\footnotesize 0.49} & {\footnotesize 10.71} & {\footnotesize 0.28} & 
{\footnotesize 7.51} & {\footnotesize 2.92} & {\footnotesize 1.69} & 
{\footnotesize 2.63} & {\footnotesize 2.91} & {\footnotesize 4.40} & 
{\footnotesize 0.34} \\ 
{\footnotesize (14)} & {\footnotesize 0.95} & {\footnotesize 28.29} & 
{\footnotesize 0.40} & {\footnotesize 8.19} & {\footnotesize 0.47} & 
{\footnotesize 11.66} & {\footnotesize 2.58} & {\footnotesize 1.57} & 
{\footnotesize 2.73} & {\footnotesize 3.18} & {\footnotesize 0.02} & 
{\footnotesize 0.00} \\ 
{\footnotesize (15)} & {\footnotesize 0.91} & {\footnotesize 23.16} & 
{\footnotesize 0.62} & {\footnotesize 10.77} & {\footnotesize 0.57} & 
{\footnotesize 12.20} & {\footnotesize 3.71} & {\footnotesize 1.98} & 
{\footnotesize 2.15} & {\footnotesize 2.19} & {\footnotesize 4.78} & 
{\footnotesize 0.34} \\ 
{\footnotesize (16)} & {\footnotesize 1.07} & {\footnotesize 38.14} & 
{\footnotesize 0.43} & {\footnotesize 10.37} & {\footnotesize -0.42} & 
{\footnotesize -12.67} & {\footnotesize 1.97} & {\footnotesize 1.02} & 
{\footnotesize 4.22} & {\footnotesize 4.20} & {\footnotesize -20.28} & 
{\footnotesize -1.39} \\ 
{\footnotesize (17)} & {\footnotesize 1.01} & {\footnotesize 31.06} & 
{\footnotesize 0.32} & {\footnotesize 6.63} & {\footnotesize 0.13} & 
{\footnotesize 3.31} & {\footnotesize 2.62} & {\footnotesize 1.49} & 
{\footnotesize 3.28} & {\footnotesize 3.58} & {\footnotesize -10.24} & 
{\footnotesize -0.77} \\ 
{\footnotesize (18)} & {\footnotesize 1.00} & {\footnotesize 32.59} & 
{\footnotesize 0.22} & {\footnotesize 4.94} & {\footnotesize 0.32} & 
{\footnotesize 8.74} & {\footnotesize 1.94} & {\footnotesize 1.21} & 
{\footnotesize 2.57} & {\footnotesize 3.06} & {\footnotesize -4.98} & 
{\footnotesize -0.41} \\ 
{\footnotesize (19)} & {\footnotesize 0.98} & {\footnotesize 30.21} & 
{\footnotesize 0.22} & {\footnotesize 4.56} & {\footnotesize 0.40} & 
{\footnotesize 10.44} & {\footnotesize 2.50} & {\footnotesize 1.56} & 
{\footnotesize 2.32} & {\footnotesize 2.76} & {\footnotesize 0.60} & 
{\footnotesize 0.05} \\ 
{\footnotesize (20)} & {\footnotesize 1.05} & {\footnotesize 27.69} & 
{\footnotesize 0.36} & {\footnotesize 6.41} & {\footnotesize 0.61} & 
{\footnotesize 13.46} & {\footnotesize 3.78} & {\footnotesize 2.05} & 
{\footnotesize 2.26} & {\footnotesize 2.34} & {\footnotesize 3.28} & 
{\footnotesize 0.23} \\ 
{\footnotesize (21)} & {\footnotesize 1.02} & {\footnotesize 46.70} & 
{\footnotesize -0.20} & {\footnotesize -6.30} & {\footnotesize -0.27} & 
{\footnotesize -10.39} & {\footnotesize 1.61} & {\footnotesize 1.01} & 
{\footnotesize 2.77} & {\footnotesize 3.33} & {\footnotesize -21.22} & 
{\footnotesize -1.76} \\ 
{\footnotesize (22)} & {\footnotesize 0.99} & {\footnotesize 39.05} & 
{\footnotesize -0.19} & {\footnotesize -5.16} & {\footnotesize 0.07} & 
{\footnotesize 2.25} & {\footnotesize 1.16} & {\footnotesize 0.80} & 
{\footnotesize 2.53} & {\footnotesize 3.33} & {\footnotesize -4.20} & 
{\footnotesize -0.38} \\ 
{\footnotesize (23)} & {\footnotesize 0.93} & {\footnotesize 32.10} & 
{\footnotesize -0.23} & {\footnotesize -5.34} & {\footnotesize 0.27} & 
{\footnotesize 7.93} & {\footnotesize 2.32} & {\footnotesize 1.88} & 
{\footnotesize 2.46} & {\footnotesize 3.80} & {\footnotesize -8.27} & 
{\footnotesize -0.88} \\ 
{\footnotesize (24)} & {\footnotesize 0.94} & {\footnotesize 36.21} & 
{\footnotesize -0.16} & {\footnotesize -4.29} & {\footnotesize 0.45} & 
{\footnotesize 14.52} & {\footnotesize 1.24} & {\footnotesize 0.94} & 
{\footnotesize 2.24} & {\footnotesize 3.25} & {\footnotesize -10.49} & 
{\footnotesize -1.05} \\ 
{\footnotesize (25)} & {\footnotesize 0.99} & {\footnotesize 26.11} & 
{\footnotesize -0.13} & {\footnotesize -2.40} & {\footnotesize 0.56} & 
{\footnotesize 12.31} & {\footnotesize 3.07} & {\footnotesize 2.13} & 
{\footnotesize 1.59} & {\footnotesize 2.11} & {\footnotesize -1.25} & 
{\footnotesize -0.11} \\ \hline
\end{tabular}
}
\end{table}

One might wonder what causes the large difference in the risk premia between 
$R_{m}$ and SMB, HML. To illustrate, we present their $\beta $ estimates and
associated $t$-statistics in Table \ref{beta}. It is clear in Table \ref%
{beta} that the three Fama and French (1993) factors are all closely related
to the test asset returns as reflected by the significant $t$-statistics.
There is, however, little cross-sectional variation in the estimated betas
of $R_{m}$, i.e. the $\beta $ estimates on $R_{m}$ are all close to 1. Thus,
if the zero-$\beta $ return is incorporated, we have near-multicollinearity
in the $(\iota _{N}$ $\vdots $ $\beta )$ matrix for the cross-sectional
regression, causing the market risk premium to be weakly identified.
Consequently, we observe in Table \ref{lambdaf_app} Panels A and B the
seemingly conflicting risk premium values on $R_{m}$, but not for SMB and
HML. For the same reason, we observe in Figure 8a-b the wide range for the
risk premium on $R_{m}$, but narrower ones for SMB, HML.

\subsection{Lettau and Ludvigson (2001)}

To compare with Fama and French (1993), we consider the conditional
consumption capital asset pricing model from Lettau and Ludvigson (2001)
where the three factors are consumption growth, $\triangle c$,
consumption-wealth ratio, $cay$, and their interaction, $\triangle c\times
cay$. The significant FM $t$ and Shanken $t$-statistics on the interaction $%
\triangle c\times cay$, the relatively small, insignificant (at the 5\%
level) $J$-statistic, together with the large cross-sectional $R^{2}$
reported in Panel C of Table \ref{lambdaf_app} provide considerable
motivation for this model for asset pricing.

The small $IS$-statistic in Panel C of Table \ref{lambdaf_app}, however,
shows that the three-factor model of Lettau and Ludvigson (2001) is just
weakly identified. Furthermore, given its proximity to the $J$-statistic,
the small value of the $J$-statistic also results from it, since the $J$%
-statistic is at most equal to the $IS$-statistic. The risk premia can
therefore not be identified. Consequently, the significant FM $t$-statistic
on $\triangle c\times cay$ should be interpreted with caution, since the FM $%
t$-test is now unreliable and this reasoning similarly applies to the KRS $t$%
-test. Also, as warned by Kleibergen and Zhan (2015)\nocite{kz15}, weakly
identified models can yield spuriously large cross-sectional $R^{2}$'s,
which should then not be taken as the evidence in support of asset pricing.

Next, we apply the DRLM test to the three-factor model of Lettau and
Ludvigson (2001). As shown by Figure 8c, we can not reject any hypothesized
value for risk premia in the range of [-5, 5], reflecting that the risk
premia can not be identified for the conditional consumption capital asset
pricing model in Lettau and Ludvigson (2001), which is in line with the
proximity of the $J$ and $IS$ statistics.

The challenge to identify the risk premia in Lettau and Ludvigson (2001) is
further explained by Table \ref{beta}, which contains the $\beta $ estimates
and their associated $t$-statistics. Table \ref{beta} shows that $\triangle
c\times cay$ is poorly correlated with asset returns, which makes a column
of $\beta $ statistically close to zero (i.e. tiny $t$-statistics in the
last column). Thus, the full rank condition of $\beta $ is problematic,
which leads to the weak identification problem in Lettau and Ludvigson
(2001). Consequently, we observe a small $IS$-statistic, which tests for a
full rank value of $\beta ,$ in Panel C of Table \ref{lambdaf_app}, and
uninformative confidence sets in Figure 8c.

In light of the findings in Figure 8a-c, one might wonder when the DRLM test
could yield an informative confidence set for the risk premia. We show that
a bounded confidence set of the risk premia is feasible if we just have more
time series observations, or if we remove the zero-$\beta $ return, so $%
\lambda _{0}=0,$ as presented in the next two subsections, respectively.

\begin{table}[tph]
\caption{\textbf{$\protect\beta $ with 25 portfolio returns using
alternative data for Fama and French (1993)}}
\label{beta_2}\justify The test assets are the 25 Fama-French portfolios.
For $R_{m}$, SMB, and HML in the left panel, we use the data from 1963Q3 -
1998Q3 as in Lettau and Ludvigson (2001); and from July 1963 - June 2021,
downloaded from French's website, for the right panel, respectively. The
reported beta estimates and their associated $t$-statistics result from the
first pass time-series regression of the Fama and MacBeth (1973) methodology.
\par
\bigskip \centering
{\small \centering%
\begin{tabular}{lrrcccc|cccccr}
\hline
& \multicolumn{2}{c}{$R_{m}$} & \multicolumn{2}{c}{SMB} & \multicolumn{2}{c}{
HML} & \multicolumn{2}{c}{$R_{m}$} & \multicolumn{2}{c}{SMB} & 
\multicolumn{2}{c}{HML} \\ \cline{2-13}
& $\hat{\beta}$ & $t$-stat & $\hat{\beta}$ & $t$-stat & $\hat{\beta}$ & $t$%
-stat & $\hat{\beta}$ & $t$-stat & $\hat{\beta}$ & $t$-stat & $\hat{\beta}$
& $t$-stat \\ \hline
{\footnotesize (1)} & {\footnotesize 1.00} & {\footnotesize 21.40} & 
{\footnotesize 1.53} & {\footnotesize 25.43} & {\footnotesize -0.29} & 
{\footnotesize -4.44} & {\footnotesize 1.09} & {\footnotesize 49.32} & 
{\footnotesize 1.38} & {\footnotesize 43.79} & {\footnotesize -0.48} & 
{\footnotesize -14.87} \\ 
{\footnotesize (2)} & {\footnotesize 0.98} & {\footnotesize 30.05} & 
{\footnotesize 1.37} & {\footnotesize 32.47} & {\footnotesize 0.11} & 
{\footnotesize 2.49} & {\footnotesize 0.95} & {\footnotesize 53.68} & 
{\footnotesize 1.31} & {\footnotesize 51.69} & {\footnotesize -0.17} & 
{\footnotesize -6.58} \\ 
{\footnotesize (3)} & {\footnotesize 0.97} & {\footnotesize 30.66} & 
{\footnotesize 1.21} & {\footnotesize 29.54} & {\footnotesize 0.28} & 
{\footnotesize 6.39} & {\footnotesize 0.92} & {\footnotesize 75.00} & 
{\footnotesize 1.09} & {\footnotesize 61.84} & {\footnotesize 0.15} & 
{\footnotesize 8.16} \\ 
{\footnotesize (4)} & {\footnotesize 0.96} & {\footnotesize 34.84} & 
{\footnotesize 1.15} & {\footnotesize 32.46} & {\footnotesize 0.43} & 
{\footnotesize 11.38} & {\footnotesize 0.87} & {\footnotesize 72.34} & 
{\footnotesize 1.07} & {\footnotesize 61.78} & {\footnotesize 0.32} & 
{\footnotesize 17.81} \\ 
{\footnotesize (5)} & {\footnotesize 1.03} & {\footnotesize 32.38} & 
{\footnotesize 1.24} & {\footnotesize 30.15} & {\footnotesize 0.76} & 
{\footnotesize 17.24} & {\footnotesize 0.93} & {\footnotesize 52.58} & 
{\footnotesize 1.08} & {\footnotesize 42.39} & {\footnotesize 0.53} & 
{\footnotesize 20.15} \\ 
{\footnotesize (6)} & {\footnotesize 1.04} & {\footnotesize 27.17} & 
{\footnotesize 1.08} & {\footnotesize 21.77} & {\footnotesize -0.49} & 
{\footnotesize -9.31} & {\footnotesize 1.12} & {\footnotesize 76.44} & 
{\footnotesize 1.03} & {\footnotesize 48.82} & {\footnotesize -0.51} & 
{\footnotesize -23.70} \\ 
{\footnotesize (7)} & {\footnotesize 1.01} & {\footnotesize 31.82} & 
{\footnotesize 1.00} & {\footnotesize 24.45} & {\footnotesize 0.00} & 
{\footnotesize 0.10} & {\footnotesize 1.00} & {\footnotesize 82.03} & 
{\footnotesize 0.92} & {\footnotesize 52.68} & {\footnotesize -0.01} & 
{\footnotesize -0.73} \\ 
{\footnotesize (8)} & {\footnotesize 1.02} & {\footnotesize 35.96} & 
{\footnotesize 0.83} & {\footnotesize 22.65} & {\footnotesize 0.23} & 
{\footnotesize 5.96} & {\footnotesize 0.96} & {\footnotesize 75.66} & 
{\footnotesize 0.76} & {\footnotesize 41.93} & {\footnotesize 0.28} & 
{\footnotesize 14.97} \\ 
{\footnotesize (9)} & {\footnotesize 1.02} & {\footnotesize 36.41} & 
{\footnotesize 0.69} & {\footnotesize 19.10} & {\footnotesize 0.47} & 
{\footnotesize 12.23} & {\footnotesize 0.94} & {\footnotesize 80.20} & 
{\footnotesize 0.72} & {\footnotesize 42.72} & {\footnotesize 0.47} & 
{\footnotesize 27.34} \\ 
{\footnotesize (10)} & {\footnotesize 1.07} & {\footnotesize 36.32} & 
{\footnotesize 0.83} & {\footnotesize 21.88} & {\footnotesize 0.76} & 
{\footnotesize 18.66} & {\footnotesize 1.07} & {\footnotesize 84.69} & 
{\footnotesize 0.88} & {\footnotesize 48.20} & {\footnotesize 0.65} & 
{\footnotesize 35.21} \\ 
{\footnotesize (11)} & {\footnotesize 1.06} & {\footnotesize 31.54} & 
{\footnotesize 0.75} & {\footnotesize 17.17} & {\footnotesize -0.46} & 
{\footnotesize -10.02} & {\footnotesize 1.10} & {\footnotesize 78.20} & 
{\footnotesize 0.74} & {\footnotesize 36.81} & {\footnotesize -0.51} & 
{\footnotesize -24.97} \\ 
{\footnotesize (12)} & {\footnotesize 1.01} & {\footnotesize 31.29} & 
{\footnotesize 0.65} & {\footnotesize 15.68} & {\footnotesize 0.00} & 
{\footnotesize 0.05} & {\footnotesize 1.01} & {\footnotesize 73.17} & 
{\footnotesize 0.60} & {\footnotesize 26.93} & {\footnotesize 0.06} & 
{\footnotesize 2.91} \\ 
{\footnotesize (13)} & {\footnotesize 1.00} & {\footnotesize 34.68} & 
{\footnotesize 0.54} & {\footnotesize 14.47} & {\footnotesize 0.31} & 
{\footnotesize 7.83} & {\footnotesize 0.97} & {\footnotesize 68.10} & 
{\footnotesize 0.44} & {\footnotesize 21.75} & {\footnotesize 0.35} & 
{\footnotesize 16.85} \\ 
{\footnotesize (14)} & {\footnotesize 1.02} & {\footnotesize 35.48} & 
{\footnotesize 0.43} & {\footnotesize 11.56} & {\footnotesize 0.50} & 
{\footnotesize 12.58} & {\footnotesize 0.98} & {\footnotesize 72.25} & 
{\footnotesize 0.44} & {\footnotesize 22.51} & {\footnotesize 0.55} & 
{\footnotesize 27.59} \\ 
{\footnotesize (15)} & {\footnotesize 1.07} & {\footnotesize 30.24} & 
{\footnotesize 0.61} & {\footnotesize 13.28} & {\footnotesize 0.78} & 
{\footnotesize 15.87} & {\footnotesize 1.07} & {\footnotesize 63.75} & 
{\footnotesize 0.57} & {\footnotesize 23.75} & {\footnotesize 0.74} & 
{\footnotesize 29.92} \\ 
{\footnotesize (16)} & {\footnotesize 1.02} & {\footnotesize 30.70} & 
{\footnotesize 0.39} & {\footnotesize 9.04} & {\footnotesize -0.48} & 
{\footnotesize -10.42} & {\footnotesize 1.07} & {\footnotesize 74.82} & 
{\footnotesize 0.39} & {\footnotesize 19.12} & {\footnotesize -0.45} & 
{\footnotesize -21.72} \\ 
{\footnotesize (17)} & {\footnotesize 1.06} & {\footnotesize 31.66} & 
{\footnotesize 0.32} & {\footnotesize 7.48} & {\footnotesize 0.02} & 
{\footnotesize 0.49} & {\footnotesize 1.06} & {\footnotesize 69.04} & 
{\footnotesize 0.23} & {\footnotesize 10.45} & {\footnotesize 0.17} & 
{\footnotesize 7.52} \\ 
{\footnotesize (18)} & {\footnotesize 1.05} & {\footnotesize 34.63} & 
{\footnotesize 0.22} & {\footnotesize 5.50} & {\footnotesize 0.32} & 
{\footnotesize 7.54} & {\footnotesize 1.03} & {\footnotesize 65.57} & 
{\footnotesize 0.18} & {\footnotesize 8.07} & {\footnotesize 0.40} & 
{\footnotesize 17.55} \\ 
{\footnotesize (19)} & {\footnotesize 1.07} & {\footnotesize 31.94} & 
{\footnotesize 0.22} & {\footnotesize 5.03} & {\footnotesize 0.50} & 
{\footnotesize 10.92} & {\footnotesize 1.02} & {\footnotesize 66.30} & 
{\footnotesize 0.23} & {\footnotesize 10.28} & {\footnotesize 0.56} & 
{\footnotesize 24.57} \\ 
{\footnotesize (20)} & {\footnotesize 1.12} & {\footnotesize 25.63} & 
{\footnotesize 0.44} & {\footnotesize 7.86} & {\footnotesize 0.70} & 
{\footnotesize 11.67} & {\footnotesize 1.15} & {\footnotesize 61.12} & 
{\footnotesize 0.30} & {\footnotesize 11.00} & {\footnotesize 0.78} & 
{\footnotesize 28.16} \\ 
{\footnotesize (21)} & {\footnotesize 1.00} & {\footnotesize 34.39} & 
{\footnotesize -0.23} & {\footnotesize -6.20} & {\footnotesize -0.39} & 
{\footnotesize -9.61} & {\footnotesize 0.98} & {\footnotesize 95.17} & 
{\footnotesize -0.23} & {\footnotesize -15.89} & {\footnotesize -0.31} & 
{\footnotesize -20.80} \\ 
{\footnotesize (22)} & {\footnotesize 1.02} & {\footnotesize 33.17} & 
{\footnotesize -0.21} & {\footnotesize -5.19} & {\footnotesize -0.05} & 
{\footnotesize 1.16} & {\footnotesize 0.97} & {\footnotesize 75.34} & 
{\footnotesize -0.18} & {\footnotesize -9.65} & {\footnotesize 0.11} & 
{\footnotesize 5.68} \\ 
{\footnotesize (23)} & {\footnotesize 0.91} & {\footnotesize 24.75} & 
{\footnotesize -0.23} & {\footnotesize -4.87} & {\footnotesize 0.16} & 
{\footnotesize 3.09} & {\footnotesize 1.05} & {\footnotesize 87.14} & 
{\footnotesize -0.13} & {\footnotesize -7.54} & {\footnotesize 0.35} & 
{\footnotesize 20.06} \\ 
{\footnotesize (24)} & {\footnotesize 1.00} & {\footnotesize 35.20} & 
{\footnotesize -0.16} & {\footnotesize -4.28} & {\footnotesize 0.44} & 
{\footnotesize 11.26} & {\footnotesize 1.05} & {\footnotesize 82.15} & 
{\footnotesize -0.20} & {\footnotesize -10.70} & {\footnotesize 0.37} & 
{\footnotesize 19.72} \\ 
{\footnotesize (25)} & {\footnotesize 0.99} & {\footnotesize 20.47} & 
{\footnotesize -0.09} & {\footnotesize -1.39} & {\footnotesize 0.69} & 
{\footnotesize 10.30} & {\footnotesize 1.06} & {\footnotesize 77.70} & 
{\footnotesize -0.26} & {\footnotesize -13.49} & {\footnotesize 0.39} & 
{\footnotesize 19.42} \\ \hline
\end{tabular}
}
\end{table}

\subsection{More time series observations}

With more time series observations, we expect more information in the data
and consequently, stronger identification of the risk premia. Figure 8d
therefore uses monthly data from July 1963 to June 2021, so $T=696,$ for the
Fama and French (1993) three factors and the twenty-five size and
book-to-market sorted portfolios, which are downloaded from French's online
data library. Table \ref{beta_2} (right panel) shows the resulting $\beta $
estimates from which it is clear that the standard errors of the estimates
have almost halved, which doubles the $t$-statistics testing their
significance, while there is also more variation in the $\beta $ estimates
for $R_{m}$ when compared to the left panel of Table \ref{beta_2} with fewer
observations. The estimate of $\beta $ is consequently distant from a
reduced rank value as shown by the large $IS$-statistic, 425.55, in Panel D
of Table \ref{lambdaf_app}. It considerably exceeds the resulting $J$%
-statistic, 52.25, indicating strong identification of the risk premia.

In line with such strong identification, Figure 8d shows that the joint 95\%
confidence set of the risk premia on the Fama and French (1993) factors
resulting from the DRLM\ test lies in a tight region of the 3-dimensional
space. Specifically, this confidence set contains all values of the risk
premia that are not rejected by the DRLM test at the 5\% level. Combining
all these values results in the red joint confidence set plotted in Figure
8d. Moreover, projecting the joint confidence set to each axis yields the
95\% confidence intervals for each risk premium: (-1.22, 0.18) for $R_{m}$,
(0.12, 0.26) for SMB, and (0.24, 0.41) for HML, respectively. These 95\%
confidence intervals are also comparable to, but often smaller than those
reported in Panel D of Table \ref{lambdaf_app} by inverting the FM $t$,
Shanken $t$ and KRS $t$ tests.

\subsection{Remove the zero-$\protect\beta $ return, impose $\protect\lambda %
_{0}=0$}

Instead of using richer data, we can also remove the zero-$\beta $ return,
so impose $\lambda _{0}=0,$ to improve the identification of the risk
premia. When we do so by using the data from Lettau, Ludvigson, and Ma
(2019), the $J$-statistic equals 87.47 and the $IS$-statistic becomes
974.39. Hence, the $IS$-statistic has increased ninefold while the $J$%
-statistic remains almost the same compared with the specification including
the zero-$\beta $ return in Panel A of Table \ref{lambdaf_app}. It indicates
that the risk premia are now well identified.

To illustrate further, we use DRLM to construct the joint 95\% confidence
set in Figure 9, where $\lambda _{0}=0$ is imposed for the Fama and French
(1993) model. Figure 9 is to be compared with Figure 8a. They use the same
data, but $\lambda _{0}=0$ is imposed for Figure 9 but not for Figure 8a.
Clearly, the $\lambda _{0}=0$ restriction strongly improves the
identification of the risk premia, so we observe a bounded 95\% confidence
set in Figure 9. Projecting the joint confidence set to each axis leads to
95\% confidence intervals for the risk premia on $R_{m}$, SMB, and HML,
respectively: $(1.52,1.84)$ for $R_{m}$, $(0.66,1.06)$ for SMB, and $%
(0.59,1.93)$ for HML.

Since the three factors are traded, we can also infer the 95\%\ confidence
sets on their risk premia by using the average value of the respective
factor $\pm $ 1.96 $\times $ standard deviation, which leads to $(0.41,2.80)$
for $R_{m}$, $(0.18,1.78)$ for SMB and $(0.05,1.81)$ for HML. These
confidence intervals are comparable to those derived above by the projection
method, but are considerably wider. All these empirical findings therefore
lend credibility to the proposed DRLM test.

\begin{figure}[htp]
\centering
\includegraphics[width=220 pt,  height=180 pt]{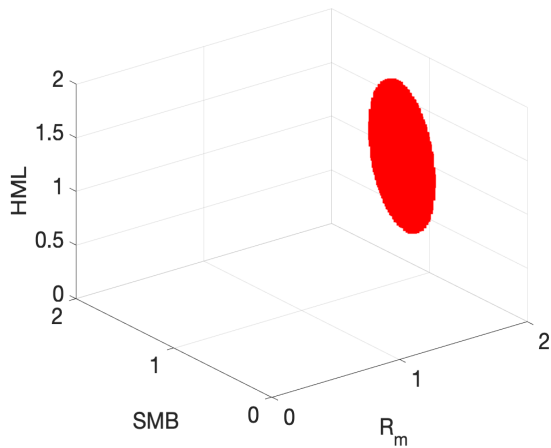}
\caption{\textbf{Joint 95\% confidence set of risk premia by the DRLM test for the  Fama and French (1993) model with the $\lambda _{0}=0$ restriction. }  }
\justify{\small Notes:\ The red region consists of risk premia values that
are not rejected by the DRLM test at the 5\% significance level. The test
assets are the twenty-five Fama-French portfolios taken from Lettau,
Ludvigson, and Ma (2019) over 1963Q3 - 2013Q4.}
\end{figure}

\doublespace

\subsection{Further discussion}

Though misspecified, the Fama and French (1993) three-factor model has been
well documented to exhibit relatively strong correlations with asset
returns. Indeed, the model leads to large $IS$-statistics as we report in
Table \ref{lambdaf_app}. Astonishingly, even when the $IS$-statistic is
significant and exceeds 100 (see Panel A of Table \ref{lambdaf_app}), we
still find that the risk premium on $R_{m}$ is only weakly identified in the
beta representation. Given that a large number of empirical studies, as
indicated by Table \ref{j_is_models}, have weaker strength of identification
than Fama and French (1993), it is therefore unlikely that their risk premia
could be precisely identified. To achieve stronger identification, we
therefore need to adopt richer data or impose extra restrictions, as
indicated by Figure 8d and Figure 9, respectively.

The conditional consumption capital asset pricing model in Lettau and
Ludvigson (2001) serves as an example where the FM $t$-statistic, the
misspecification $J$-statistic, and the cross-sectional $R^{2}$ line up
nicely to show support for the model. Yet the model appears to lack
identification, as reflected by the tiny difference between the $J$ and $IS$
statistics, and the wide confidence set resulting from the DRLM test, since
the rank condition of $\beta $ is likely violated. These empirical findings
thus clearly show the importance of adopting the $J,$ $IS$ statistics and
the DRLM test for assessing asset pricing models.

In the Appendix, we further extend our empirical analysis to the $q$-factor
model of Hou, Xue, and Zhang (2015). \nocite{hou2015digesting} In line with
Figure 9, we find that the DRLM test leads to bounded 95\% confidence sets
for the risk premia on all four factors in the $q$-factor model. In
particular, for the first three factors (R\_MKT, market excess returns;
R\_ME, size factor; R\_IA, investment factor), we find that the 95\%
confidence sets resulting from DRLM are significant at the 5\% level so they
exclude zero. Yet for the fourth factor (R\_ROE, equity factor), the
confidence set is bounded but relatively wide. These findings are consistent
with those in Kleibergen and Zhan (2015), who show that the commonly used 25
Fama-French portfolios have a strong factor structure, i.e., the majority of
variations in these portfolios are likely captured by three factors, so it
is hard to precisely identify risk premia on all four factors in the $q$%
-factor model by using such test assets in the two-pass approach.

\section{Conclusion}

\label{conclusion}

An alarmingly large number of factors in the asset pricing literature are
able to yield significant $t$-statistics on their risk premia, together with
seemingly promising misspecification $J$-statistics and cross-sectional $%
R^{2}$'s. The credibility of these conventional statistics, however, is
threatened by misspecification and weak identification, both of which are
prevalent as we document in this paper. We show that failure to account for
both misspecification and weak identification could easily lead to erroneous
conclusions. To remedy these problems, we suggest that the $J$-statistic for
misspecification, the $IS$-statistic for identification strength, and the
DRLM test for risk premia become part of a toolkit that helps to provide
trustworthy diagnosis and inference in future studies.

\newpage \setstretch{1.45} 
\bibliographystyle{plain}
\bibliography{eqrand}

\begin{thebibliography}{10}

\bibitem{aem14}
{Adrian, T., E. Etula and T. Muir}.
\newblock {Financial Intermediaries and the Cross-Section of Asset Returns}.
\newblock {\em Journal of Finance}, {\bf 69}:2557--2596, 2014.

\bibitem{cochrane2011presidential}
{Cochrane, J. H.}
\newblock Presidential address: Discount rates.
\newblock {\em Journal of Finance}, 66(4):1047--1108, 2011.

\bibitem{cradon97}
{Cragg, J.C. and S.G. Donald}.
\newblock {Inferring the rank of a matrix}.
\newblock {\em Journal of Econometrics}, {\bf 76}:223--250, 1997.

\bibitem{duf97}
{Dufour, J.-M.}
\newblock {Some Impossibility Theorems in Econometrics with Applications to
  Structural and Dynamic Models}.
\newblock {\em Econometrica}, {\bf 65}:1365--388, 1997.

\bibitem{fm73}
{Fama, E.F. and J.D. MacBeth}.
\newblock {Risk, Return and Equillibrium: Empirical Tests}.
\newblock {\em Journal of Political Economy}, {\bf 81}:607--636, 1973.

\bibitem{ff93}
{Fama, E.F. and K.R. French}.
\newblock {Common Risk Factors in the Returns on Stocks and Bonds}.
\newblock {\em Journal of Financial Economics}, {\bf 33}:3--56, 1993.

\bibitem{feng2020taming}
{Feng, G. and Giglio, S. and Xiu, D.}
\newblock {Taming the factor zoo: A test of new factors}.
\newblock {\em Journal of Finance}, 75(3):1327--1370, 2020.

\bibitem{gkr14}
{Gospodinov, N. R. Kan and C. Robotti}.
\newblock {Misspecification-Robust Inference in Linear Asset-Pricing Models
  with Irrelevant Factors}.
\newblock {\em Review of Financial Studies}, {\bf 27}:2139--2170, 2014.

\bibitem{hl21}
{Hansen, B. E. and Lee, S.}
\newblock Inference for iterated gmm under misspecification.
\newblock {\em Econometrica}, 89(3):1419--1447, 2021.

\bibitem{han82}
{Hansen, L.P.}
\newblock {Large Sample Properties of Generalized Method Moments Estimators}.
\newblock {\em Econometrica}, {\bf 50}:1029--1054, 1982.

\bibitem{hhy96}
{Hansen, L.P., J. Heaton and A. Yaron}.
\newblock {Finite Sample Properties of Some Alternative GMM Estimators}.
\newblock {\em Journal of Business and Economic Statistics}, {\bf 14}:262--280,
  1996.

\bibitem{harvey2016and}
{Harvey, C. R. and Liu, Y. and Zhu, H.}
\newblock {... and the cross-section of expected returns}.
\newblock {\em Review of Financial Studies}, 29(1):5--68, 2016.

\bibitem{hkm17}
{He, Z., B. Kelly and A. Manela}.
\newblock {Intermediary Asset Pricing: New Evidence from Many Asset Classes}.
\newblock {\em Journal of Financial Economics}, {\bf 126}:1--35, 2017.

\bibitem{hou2015digesting}
{Hou, K. and Xue, C. and Zhang, L.}
\newblock Digesting anomalies: An investment approach.
\newblock {\em Review of Financial Studies}, 28(3):650--705, 2015.

\bibitem{jw96}
{Jagannathan, R. and Z. Wang}.
\newblock {The Conditional CAPM and the Cross-Section of Expected Returns}.
\newblock {\em Journal of Finance}, {\bf 51}:3--53, 1996.

\bibitem{kz99}
{Kan, R. and C. Zhang}.
\newblock {Two-Pass Tests of Asset Pricing Models with Useless Factors}.
\newblock {\em Journal of Finance}, {\bf 54}:203--235, 1999.

\bibitem{krs13}
{Kan, R., C. Robotti and J. Shanken}.
\newblock {Pricing Model Performance and the Two-Pass Cross-Sectional
  Regression Methodology}.
\newblock {\em Journal of Finance}, {\bf 68}:2617--2649, 2013.

\bibitem{kanstam95}
{Kandel, S. and R.F. Stambaugh}.
\newblock {Portfolio Inefficiency and the Cross-section of Expected Returns}.
\newblock {\em Journal of Finance}, {\bf 50}:157--184, 1995.

\bibitem{kf00a}
{Kleibergen, F.}
\newblock {Testing Parameters in GMM without assuming that they are
  identified}.
\newblock {\em Econometrica}, {\bf 73}:1103--1124, 2005.

\bibitem{kf04}
{Kleibergen, F.}
\newblock {Generalizing weak instrument robust IV statistics towards multiple
  parameters, unrestricted covariance matrices and identification statistics}.
\newblock {\em Journal of Econometrics}, {\bf 139}:181--216, 2007.

\bibitem{kf09}
{Kleibergen, F.}
\newblock {Tests of Risk Premia in Linear Factor Models}.
\newblock {\em Journal of Econometrics}, {\bf 149}:149--173, 2009.

\bibitem{kpaap02}
{Kleibergen, F. and R. Paap}.
\newblock {Generalized Reduced Rank Tests using the Singular Value
  Decomposition}.
\newblock {\em Journal of Econometrics}, {\bf 133}:97--126, 2006.

\bibitem{kz15}
{Kleibergen F. and Z. Zhan}.
\newblock Unexplained factors and their effects on second pass r-squared's.
\newblock {\em Journal of Econometrics}, {\bf 189}:101--116, 2015.

\bibitem{kleibergen2018identification}
{Kleibergen, F. and Z. Zhan}.
\newblock Identification-robust inference on risk premia of mimicking
  portfolios of non-traded factors.
\newblock {\em Journal of Financial Econometrics}, 16(2):155--190, 2018.

\bibitem{kz20}
{Kleibergen, F. and Z. Zhan}.
\newblock {Robust Inference for Consumption-Based Asset Pricing}.
\newblock {\em Journal of Finance}, {\bf 75}:507--550, 2020.

\bibitem{kz21}
{Kleibergen F. and Z. Zhan}.
\newblock {Double Robust Inference for Continuous Updating GMM}.
\newblock Working paper, University of Amsterdam, 2021.

\bibitem{kkz19}
{Kleibergen, F., L. Kong and Z. Zhan}.
\newblock {Identification Robust Testing of Risk Premia in Finite Samples}.
\newblock Working paper, University of Amsterdam, 2020.

\bibitem{Kro17}
{Kroencke, T.A.}
\newblock {Asset Pricing without Garbage}.
\newblock {\em Journal of Finance}, {\bf 72}:47--98, 2017.

\bibitem{letlud01}
{Lettau, M. and S. Ludvigson}.
\newblock {Resurrecting the (C)CAPM: A Cross-Sectional Test When Risk Premia
  are Time-Varying}.
\newblock {\em Journal of Political Economy}, {\bf 109}:1238--1287, 2001.

\bibitem{llm19}
{Lettau, M., S. Ludvigson and S. Ma}.
\newblock {Capital Share Risk in U.S. Asset Pricing}.
\newblock {\em Journal of Finance}, {\bf 74}:1753--1792, 2019.

\bibitem{lns2009}
{Lewellen, J., S. Nagel and J. Shanken}.
\newblock {A Skeptical Appraisal of Asset-Pricing Tests}.
\newblock {\em Journal of Financial Economics}, {\bf 96}:175--194, 2010.

\bibitem{robsmit00}
{Robin, J.-M. and R.J. Smith}.
\newblock {Tests of Rank}.
\newblock {\em Econometric Theory}, {\bf 16}:151--175, 2000.

\bibitem{sav11}
{Savov, A.}
\newblock {Asset Pricing with Garbage}.
\newblock {\em Journal of Finance}, {\bf 72}:47--98, 2011.

\bibitem{sh92}
{Shanken, J.}
\newblock {On the Estimation of Beta-Pricing Models}.
\newblock {\em Review of Financial Studies}, {\bf 5}:1--33, 1992.

\bibitem{stst97}
{Staiger, D. and J.H. Stock}.
\newblock {Instrumental Variables Regression with Weak Instruments}.
\newblock {\em Econometrica}, {\bf 65}:557--586, 1997.

\bibitem{sw00}
{Stock, J.H. and J.H. Wright}.
\newblock {GMM with Weak Identification}.
\newblock {\em Econometrica}, {\bf 68}:1055--1096, 2000.

\bibitem{yogo06}
{Yogo, M.}
\newblock {A consumption-based explanation of expected stock returns}.
\newblock {\em Journal of Finance}, {\bf 61}:539--580, 2006.

\end{thebibliography}

\newpage \setstretch{1.50}

\section*{Appendix}

\paragraph{Proof of Theorem 1:}

Consider repackaging the assets to a new set of $N^{\ast }$ assets by an
invertible $N\times N$ weight matrix $A:$%
\begin{equation*}
R_{t}^{\ast }=AR_{t}.
\end{equation*}%
The pseudo-true value resulting from the FM two-pass population objective
function is:%
\begin{equation*}
\begin{array}{c}
\lambda _{F,FM}^{\ast }=\left( \beta ^{\ast \prime }\beta ^{\ast }\right)
^{-1}\beta ^{\ast \prime }\mu _{R^{\ast }}=\left( \beta ^{\prime }A^{\prime
}A\beta \right) ^{-1}\beta ^{\prime }A^{\prime }A\mu _{R},%
\end{array}%
\end{equation*}%
with $\beta ^{\ast }=A\beta ,$ $\mu _{R^{\ast }}=A\mu _{R},$ which is not
equal to the pseudo-true value resulting from the orginal set of assets
unless $A$ is orthogonal. Since an orthogonal matrix $A$ does not lead to a
set of portfolios, the FM\ pseudo-true value is not invariant under
repackaging. Under correct specification, $\mu _{R}=\beta \lambda _{F},$ so
the risk premium is invariant to repackaging.

The pseudo-true value resulting from the CUE\ objective function: 
\begin{equation*}
\begin{array}{rl}
\lambda _{F,CUE}^{\ast }= & \arg \min_{\lambda _{F}}(\mu _{R}-\beta \lambda
_{F})^{\prime }\left[ \text{var(}\hat{\mu}_{R}-\hat{\beta}\lambda _{F})%
\right] ^{-1}(\mu _{R}-\beta \lambda _{F}) \\ 
= & \arg \min_{\lambda _{F}}(A\mu _{R}-A\beta \lambda _{F})^{\prime }\left[ 
\text{var(}A\hat{\mu}_{R}-A\hat{\beta}\lambda _{F})\right] ^{-1}(A\mu
_{R}-A\beta \lambda _{F}) \\ 
= & \arg \min_{\lambda _{F}}(\mu _{R^{\ast }}-\beta ^{\ast }\lambda
_{F})^{\prime }\left[ \text{var(}\hat{\mu}_{R^{\ast }}-\hat{\beta}^{\ast
}\lambda _{F})\right] ^{-1}(\mu _{R^{\ast }}-\beta ^{\ast }\lambda _{F}),%
\end{array}%
\end{equation*}%
is clearly invariant to repackaging by an invertible matrix $A$, but not to
repackaging by an $N^{\ast }\times N$ matrix with $N^{\ast }$ smaller than $%
N.$

\paragraph{Proof of Theorem 2:}

The FM two-pass estimator is given by:%
\begin{equation*}
\hat{\lambda}_{F}=(\hat{\beta}^{\prime }\hat{\beta})^{-1}\hat{\beta}^{\prime
}\hat{\mu}_{R}.
\end{equation*}%
We characterize the limit behavior of the FM two-pass estimator for the one
factor setting, so $\hat{\lambda}_{F}=\frac{\hat{\beta}^{\prime }\hat{\mu}%
_{R}}{\hat{\beta}^{\prime }\hat{\beta}}.$ It results from the joint limit
behavior of its two different elements:%
\begin{equation*}
\begin{array}{c}
\sqrt{T}\left( 
\begin{array}{c}
\hat{\mu}_{R}-\mu _{R} \\ 
\hat{\beta}-\beta%
\end{array}%
\right) \underset{d}{\rightarrow }\left( 
\begin{array}{c}
\psi _{\mu } \\ 
\psi _{\beta}%
\end{array}%
\right) ,%
\end{array}%
\end{equation*}%
with $\psi _{\mu }\sim N(0,\Omega +\beta Q\beta ^{\prime })$ and $\psi
_{\beta}\sim N(0,\Omega Q^{-1})$ independently distributed which corresponds
with Shanken (1992, Lemma 1). To focus on a setting where both the
misspecification and betas are small and perhaps just borderline
significant, we use the weak factor/small $\beta $ and misspecification
assumption (\ref{small beta})$:$%
\begin{equation*}
\begin{array}{c}
\beta =\beta _{T}=\frac{b}{\sqrt{T}},\text{ }\mu _{R}-\beta \lambda _{F}=%
\frac{a}{\sqrt{T}},\text{ }\lambda _{F}^{\ast }=\lambda _{F}+(b^{\prime
}b)^{-1}b^{\prime }a,%
\end{array}%
\end{equation*}%
with $b$ and $a$ $N$-dimensional vectors of constants. Under the small
misspecification and $\beta $ assumption, the limit behavior of the least
squares estimator $\hat{\beta}$ and $\hat{\mu}_{R}$ are characterized by:%
\begin{equation*}
\begin{array}{c}
\sqrt{T}\hat{\beta}\underset{d}{\rightarrow }b+\psi _{\beta},\text{ }\sqrt{T}%
\hat{\mu}_{R}\underset{d}{\rightarrow }b\lambda _{F}+a+\psi _{\mu },%
\end{array}%
\end{equation*}%
which we use to characterize the behavior of the FM\ risk premia estimator%
\begin{equation*}
\begin{array}{rl}
\hat{\lambda}_{F}= & \frac{\hat{\mu}_{R}^{\prime }\hat{\beta}}{\hat{\beta}%
^{\prime }\hat{\beta}}=\frac{(\hat{\mu}_{R}-\hat{\beta}\lambda _{F}^{\ast }+%
\hat{\beta}\lambda _{F}^{\ast })^{\prime }\hat{\beta}}{\hat{\beta}^{\prime }%
\hat{\beta}}=\lambda _{F}^{\ast }+\frac{(\hat{\mu}_{R}-\hat{\beta}\lambda
_{F}^{\ast })^{\prime }\hat{\beta}}{\hat{\beta}^{\prime }\hat{\beta}} \\ 
= & \lambda _{F}^{\ast }+\frac{[\hat{\mu}_{R}-\mu _{R}+(\mu _{R}-\beta
\lambda _{F}^{\ast })-(\hat{\beta}-\beta )\lambda _{F}^{\ast }]^{\prime }%
\hat{\beta}}{\hat{\beta}^{\prime }\hat{\beta}}%
\end{array}%
\end{equation*}%
with $\lambda _{F}^{\ast }=\frac{\beta ^{\prime }\mu _{R}}{\beta ^{\prime
}\beta }=\lambda _{F}+(b^{\prime }b)^{-1}b^{\prime }a,$ the pseudo-true
value, so for small values of the betas and misspecification:%
\begin{equation*}
\begin{array}{rl}
\hat{\lambda}_{F}= & \lambda _{F}^{\ast }+\frac{\left[ \sqrt{T}(\hat{\mu}%
-\mu _{R})+\sqrt{T}(\mu _{R}-\beta \lambda _{F}^{\ast })-\sqrt{T}(\hat{\beta}%
-\beta )\lambda _{F}^{\ast })\right] ^{\prime }(\sqrt{T}\hat{\beta})}{(\sqrt{%
T}\hat{\beta})^{\prime }(\sqrt{T}\hat{\beta})} \\ 
\underset{d}{\rightarrow } & \lambda _{F}^{\ast }+\frac{[\psi _{\mu
}+b\lambda _{F}+a-b(\lambda _{F}+(b^{\prime }b)^{-1}b^{\prime }a)-\psi
_{\beta}\lambda _{F}^{\ast }]^{\prime }(b+\psi _{\beta})}{(b+\psi
_{\beta})^{\prime }(b+\psi _{\beta})} \\ 
= & \lambda _{F}^{\ast }+\frac{(\psi _{\mu }+e-\psi _{\beta}\lambda
_{F}^{\ast })^{\prime }(b+\psi _{\beta})}{(b+\psi _{\beta})^{\prime }(b+\psi
_{\beta})} \\ 
= & \lambda _{F}^{\ast }\left( 1-\frac{\psi _{\beta}^{\prime }(b+\psi
_{\beta})}{(b+\psi _{\beta})^{\prime }(b+\psi _{\beta})}\right) +\frac{\psi
_{\mu }^{\prime }(b+\psi _{\beta})}{(b+\psi _{\beta})^{\prime }(b+\psi
_{\beta})}+\frac{e^{\prime }(b+\psi _{\beta})}{(b+\psi _{\beta})^{\prime
}(b+\psi _{\beta})},%
\end{array}%
\end{equation*}%
with $e=a-b(b^{\prime }b)^{-1}b^{\prime }a,$ which shows that the limit
behavior of the FM\ two-pass estimator consists of four components.

\paragraph{The DRLM\ test:}

Kleibergen and Zhan (2021) propose the double robust Lagrange multiplier
(DRLM)\ statistic for testing hypotheses on the pseudo-true value of the
CUE. They show that under the hypothesis of interest, H$_{0}:\lambda
_{F,CUE}^{\ast }=\lambda _{F,CUE,0}^{\ast },$ the limiting distribution of
the DRLM\ statistic is bounded by a $\chi _{K}^{2}$ distribution for general
values of the identification and misspecification strengths under weak
conditions. The DRLM\ statistic involves a recentered estimator of the $%
\beta $'s that depends on the hypothesized value of the pseudo-true value,
which we indicate by the $K$-dimensional vector $l:$%
\begin{equation*}
\begin{array}{rl}
\hat{D}(l)= & -\hat{\beta}-\left[ \hat{V}_{\beta _{1}(\mu _{R}-\beta l)}(l)%
\hat{V}_{(\mu _{R}-\beta l)}(l)^{-1}(\hat{\mu}_{R}-\hat{\beta}l)\ldots
\right. \\ 
& \qquad \left. \hat{V}_{\beta _{K}(\mu _{R}-\beta l)}(l)\hat{V}_{(\mu
_{R}-\beta l)}(l)^{-1}(\hat{\mu}_{R}-\hat{\beta}l)\right] \\ 
= & -\hat{\beta}-(\bar{R}-\hat{\beta}l)(1+l^{\prime }\hat{Q}_{\bar{F}\bar{F}%
}^{-1}l)^{-1}l^{\prime }\hat{Q}_{\bar{F}\bar{F}}^{-1} \\ 
= & -\frac{1}{T}\sum_{t=1}^{T}R_{t}(\bar{F}_{t}+l)^{\prime }\left[ \frac{1}{T%
}\sum_{t=1}^{T}(\bar{F}_{t}+l)(\bar{F}_{t}+l)^{\prime }\right] ^{-1},%
\end{array}%
\end{equation*}%
where $\hat{V}_{\beta _{i}(\mu _{R}-\beta l)}$ is the estimator of the
covariance between $\hat{\beta}_{i}$ and $\hat{\mu}_{R}-\hat{\beta}l,$ for $%
i=1,\ldots ,K,$ $\hat{\beta}=(\hat{\beta}_{1}\ldots \hat{\beta}_{K})$ and $%
\hat{V}_{(\mu _{R}-\beta \lambda _{f})}(l)$ is the covariance matrix
estimator of the sample pricing error $\hat{\mu}_{R}-\hat{\beta}l.$ The
identical expressions on the last two lines are for a setting of i.i.d.
errors with $\hat{Q}_{\bar{F}\bar{F}}$ the estimator of the covariance
matrix of the factors, $\hat{Q}_{\bar{F}\bar{F}}=\frac{1}{T}\sum_{t=1}^{T}%
\bar{F}_{t}\bar{F}_{t}^{\prime },$ $\bar{F}_{t}=F_{t}-\bar{F},$ $\bar{F}=%
\frac{1}{T}\sum_{t=1}^{T}F_{t}.$ The DRLM\ statistic for testing H$%
_{0}:\lambda _{F,CUE}^{\ast }=l$ then reads:%
\begin{equation*}
\begin{array}{rl}
DRLM(l)= & T\times (\hat{\mu}_{R}-\hat{\beta}l)^{\prime }\hat{V}_{(\mu
_{R}-\beta l)}(l)^{-1}\hat{D}(l)\left[ \hat{D}(l)^{\prime }\hat{V}_{(\mu
_{R}-\beta l)}(l)^{-1}\hat{D}(l)+\right. \\ 
& \left. \left( I_{N}\otimes \hat{V}_{(\mu _{R}-\beta l)}(l)^{-1}(\hat{\mu}%
_{R}-\hat{\beta}l)\right) ^{\prime }\hat{V}_{\hat{D}(l)}(l)\left(
I_{N}\otimes \hat{V}_{(\mu _{R}-\beta l)}(l)^{-1}(\hat{\mu}_{R}-\hat{\beta}%
l)\right) \right] ^{-1} \\ 
& \hat{D}(l)^{\prime }\hat{V}_{(\mu _{R}-\beta l)}(l)^{-1}(\hat{\mu}_{R}-%
\hat{\beta}l),%
\end{array}%
\end{equation*}%
where $\hat{V}_{\hat{D}(l)}$ is the estimator of the covariance matrix of
vec($\hat{D}(l)).$ For the i.i.d. setting, it simplifies to:%
\begin{equation*}
\begin{array}{rl}
DRLM(l)= & \hat{\mu}(l)^{\ast \prime }\hat{D}(l)^{\ast }\left[ \hat{\mu}%
(l)^{\ast \prime }\hat{\mu}(l)^{\ast }I_{N}+\hat{D}(l)^{\ast \prime }\hat{D}%
(l)^{\ast }\right] ^{-1}\hat{D}(l)^{\ast \prime }\hat{\mu}(l)^{\ast },%
\end{array}%
\end{equation*}%
with $\hat{\mu}(l)^{\ast }=\sqrt{T}\hat{\Omega}^{-\frac{1}{2}}(\bar{R}-\hat{%
\beta}l)(1+l^{\prime }\hat{Q}_{\bar{F}\bar{F}}^{-1}l)^{-\frac{1}{2}},$ and $%
\hat{D}(l)^{\ast }=\sqrt{T}\hat{\Omega}^{-\frac{1}{2}}\hat{D}(l)(\hat{Q}_{%
\bar{F}\bar{F}}+ll^{\prime })^{\frac{1}{2}}.$

The $100\times (1-\alpha )\%$ confidence set for $\lambda _{F, CUE}^{\ast }$
(denoted by $\text{CS}_{\lambda _{F, CUE}^{\ast }}\text{(}\alpha )$ below)
that results from the DRLM test consists of all values of $l$ for which the
DRLM\ test does not reject using the $100\times (1-\alpha )\%$ critical
value that results from the $\chi_K ^{2}$-distribution:%
\begin{equation*}
\begin{array}{c}
\text{CS}_{\lambda _{F, CUE}^{\ast }}\text{(}\alpha )=\left\{ l:\text{DRLM(}%
l)\leq \chi _{K}^{2}(\alpha )\right\} ,%
\end{array}%
\end{equation*}%
where $\chi _{K}^{2}(\alpha )$ is the upper $\alpha $-th quantile of the $%
\chi _{K}^{2}$ distribution. DRLM($l)$ is not a quadratic function of $l$,
so it cannot directly be inverted to obtain the confidence set. The
confidence set does therefore not have the usual expression of an estimator
plus or minus a multiple of the standard error. Instead, we have to specify
a $K$-dimensional grid of values for $l$, and compute the DRLM statistic for
each value of $l$ on the $K$-dimensional grid to determine if it does not
exceed the appropriate critical value so $l$ is part of the confidence set.

The DRLM\ statistic is a quadratic form of the derivative of the sample CUE
objective function with respect to $l.$ It is therefore equal to zero at all
stationary points of the sample CUE\ objective function. Since the DRLM\
statistic is not a quadratic function of $l,$ there can be multiple
stationary points where it equals zero. This affects the discriminatory
power of the DRLM test. Kleibergen and Zhan (2021) therefore propose a power
improvement rule which rejects values of $l$ at the $\alpha \%$ significance
level, alongside values of $l$ where the DRLM\ statistic is significant at
the $\alpha \%$ level also, when there are significant values of the DRLM\
statistic on every line going from the hypothesized value to the CUE.
Kleibergen and Zhan (2021) show that the power improvement rule does not
affect the size of the DRLM\ test and improves power considerably.

Since the DRLM\ test is size correct, the coverage of a $100\times (1-\alpha
)\%$ confidence set is at least $100\times (1-\alpha )\%.$ By projecting
these confidence sets on the $K$ different axes, we obtain $100\times
(1-\alpha )\%$ univariate confidence sets for the individual risk premium
whose coverage is also at least $100\times (1-\alpha )\%.$ When we plug in
estimators for some of the risk premia, the coverage of the resulting
confidence sets is not guaranteed nor is the size of the resulting subset
DRLM\ test.

When using the power improvement rule, the confidence sets resulting from
the DRLM\ test can have two distinct shapes.

\begin{enumerate}
\item Bounded and convex: there is a closed compact set of values of $l$ for
which the DRLM\ statistic does not exceed the critical value.

\item Unbounded: this occurs either when there are no values of $l$ for
which the DRLM\ statistic exceeds the critical value (unbounded and convex),
or when there is a bounded set of values of $l$ for which the DRLM statistic
exceeds the critical value (unbounded and disjoint).
\end{enumerate}

Bounded and convex confidence sets occur when the risk premia are well
identified. Unbounded confidence sets are indicative of identification
failure. Dufour (1997, Theorems 3.3 and 3.6) \nocite{duf97} formally proves
that a size correct test on a parameter which is potentially not identified
must have a positive probability of producing an unbounded 95\% confidence
set. Conversely, any test procedure, such as the FM $t$-test, that cannot
generate an unbounded 95\% confidence set, cannot be a size correct test
procedure when the tested parameter can be non-identified.

\paragraph{$J$ and $IS$ statistics:}

\begin{enumerate}
\item If $\lambda_0=0$ is not imposed: consider $\mathcal{R}_t=(\mathcal{R}%
_{1,t}\ldots \mathcal{R}_{N+1,t})^{\prime }$: $(N+1)\times 1$ vector of
returns; $F_{t}$: $K\times 1$ vector of risk factors, $t=1,...,T$. By
subtracting the $(N+1)$-th asset return, we obtain the $N\times 1$ column
vector $R_{t}$: 
\begin{equation*}
R_{t}=(\mathcal{R}_{1,t}\ldots \mathcal{R}_{N,t})^{\prime }-\iota _{N}%
\mathcal{R}_{N+1,t}.
\end{equation*}

\item If $\lambda_0=0$ is imposed: consider $R_t$ as the observed $N\times 1$
vector of returns, and $F_t$ as the $K\times 1$ vector of risk factors.
\end{enumerate}

Estimation of the auxiliary linear factor model $R_{t}=\alpha +\beta
F_{t}+u_{t}$ yields 
\begin{equation*}
\hat{\beta}=\sum_{t=1}^{T}\bar{R}_{t}\bar{F}_{t}^{\prime }\left(
\sum_{t=1}^{T}\bar{F}_{t}\bar{F}_{t}^{\prime }\right) ^{-1},\ \hat{\Omega}=%
\frac{1}{T}\sum_{t=1}^{T}\hat{u}_{t}\hat{u}_{t}^{\prime },\text{ \ and \ \ }%
\hat{Q}_{FF}=\frac{1}{T}\sum_{t=1}^{T}\bar{F}_{t}\bar{F}_{t}^{\prime },
\end{equation*}%
where $\bar{F}_{t}=F_{t}-\bar{F},$ $\bar{F}=\frac{1}{T}\sum_{t=1}^{T}F_{t},$ 
$\bar{R}_{t}=R_{t}-\bar{R}$, $\bar{R}=\frac{1}{T}\sum_{t=1}^{T}R_{t}$, and $%
\hat{u}_{t}=(R_{t}-\bar{R})-\hat{\beta}(F_{t}-\bar{F})$ is the residual at
time $t$.

\hspace{\parindent} Let $rk$ be the smallest root of 
\begin{equation*}
\begin{array}{c}
\left\vert \mu \hat{Q}_{FF}^{-1}-\hat{\beta}^{\prime }\hat{\Omega}^{-1}\hat{%
\beta}\right\vert =0,%
\end{array}%
\end{equation*}%
which is identical to the smallest eigenvalue of the matrix $\hat{Q}_{FF}%
\hat{\beta}^{\prime }\hat{\Omega}^{-1}\hat{\beta}.$ The $IS$-statistic for H$%
_{0}:$ rank($\beta)=K-1,$ reads: 
\begin{equation*}
\begin{array}{c}
IS=T\times \text{rk}\preceq \ \chi_{N-K+1}^2.%
\end{array}%
\end{equation*}

\hspace{\parindent} Let ``$\text{Eigen}_{min}$" be the smallest root of 
\begin{equation*}
\begin{array}{c}
\left\vert \mu \left( 
\begin{array}{cc}
1 & 0 \\ 
0 & \hat{Q}_{FF}^{-1}%
\end{array}%
\right) -\left( 
\begin{array}{cc}
\bar{R} & \hat{\beta}%
\end{array}%
\right) ^{\prime }\hat{\Omega}^{-1}\left( 
\begin{array}{cc}
\bar{R} & \hat{\beta}%
\end{array}%
\right) \right\vert =0,%
\end{array}%
\end{equation*}%
which is equal to the smallest eigenvalue of $\left( 
\begin{array}{cc}
1 & 0 \\ 
0 & \hat{Q}_{FF}%
\end{array}%
\right) \left( 
\begin{array}{cc}
\bar{R} & \hat{\beta}%
\end{array}%
\right) ^{\prime }\hat{\Omega}^{-1}\left( 
\begin{array}{cc}
\bar{R} & \hat{\beta}%
\end{array}%
\right) . $ The misspecification $J$-statistic for testing H$_{0}:$ $%
E(R_t)=\beta \lambda_F$ reads: 
\begin{equation*}
J=T\times \text{Eigen}_{min}\overset{d}{\to} \chi_{N-K}^2.
\end{equation*}

\paragraph{DRLM for the $q$-factor model:}

See Table \ref{lambdaf_q}.

\begin{table}[htp]
\caption{\textbf{Application to the Hou, Xue, and Zhang (2015) $q$-factor
model} }
\label{lambdaf_q}\justify The four factors (R\_MKT, R\_ME, R\_IA, R\_ROE) of
Hou, Xue, and Zhang (2015) are downloaded from \url{https://global-q.org}.
The test assets are the 25 Fama-French size and book-to-market portfolios
from Jan 1967 - Dec 2020. The zero-$\beta $ return, $\lambda_0=0$
restriction is imposed for the DRLM test. The resulting $J$-statistic is
72.92, and $IS$-statistic is 194.64. The table reports the mean of factors,
together with their 95\% confidence interval (C.I.) of risk premia by
inverting the DRLM test.
\par
\bigskip
\par
\centering{\small 
\begin{tabular}{lcccc}
\hline
& R\_MKT & R\_ME & R\_IA & R\_ROE \\ \hline
mean & 0.58 & 0.27 & 0.33 & 0.51 \\ 
&  &  &  &  \\ 
95\% C.I. of risk premia by DRLM & {\footnotesize (0.53, 0.65)} & 
{\footnotesize (0.20, 0.67)} & {\footnotesize (0.02, 0.75)} & {\footnotesize %
(-0.24, 2.07)} \\ \hline
\end{tabular}
}
\end{table}

\end{document}